\documentclass[11pt]{article}
\usepackage{amsmath}
\usepackage{amsmath}
\usepackage{amsthm}
\usepackage{amssymb}
\usepackage{epsfig}
\usepackage{epstopdf}

\usepackage{calc}
\usepackage{xspace}
\usepackage{comment}
\usepackage{paralist}
\usepackage{cancel}

\usepackage{url}
\usepackage[normalem]{ulem}
\usepackage{setspace}
\usepackage{rotating}
\usepackage{amsmath,amssymb}
\usepackage{subfigure}
\usepackage{array,arydshln}	
\usepackage{multirow}
\usepackage{booktabs}
\usepackage{color,colortbl}
\usepackage{authblk}
\usepackage{hyperref}
\usepackage{enumitem}
\usepackage{tabularx}	
\usepackage[table]{xcolor}
\usepackage{adjustbox}
\usepackage{float}
\usepackage{makecell}
\usepackage{cite}
\usepackage{bm}
\usepackage{mathrsfs}
\usepackage{indentfirst}
\usepackage{soul}
\usepackage{threeparttable}
\usepackage{placeins}

\usepackage[authoryear]{natbib}
\usepackage{bibunits}
\DeclareMathOperator*{\argmax}{argmax}

\newtheorem{corollary}{Corollary}
\newtheorem{thm}{Theorem}
\newtheorem{prop}{Proposition}

\newtheorem{lemma}{Lemma}

\newtheorem{@remark}{\bf Remark}
\newenvironment{remark}{\begin{@remark}\rm}{\end{@remark}}
\newtheorem{@exam}{\bf Example}
\newenvironment{exam}{\begin{@exam}\rm}{\end{@exam}}

\newcommand {\uP}{\mathrm{P}}

\newcommand {\E} {{\rm E}}
\newcommand {\Var} {{\rm Var}}
\newcommand {\A} {\alpha}
\newcommand {\pa} {\partial }

\newcommand {\T} {\theta}
\newcommand{\Gt}{\widetilde G_n}
\newcommand{\Gto}{\Gt^\infty}
\newcommand{\GtM}{\Gt^{M}}

\newcommand{\s}{\scriptscriptstyle}
\newcommand{\cD}{\mathcal D}   
\newcommand{\cV}{\mathcal V}   
\newcommand{\cA}{\mathcal A}  
\newcommand{\revR}{R}
\newcommand{\TR}{\hat{\theta}_n^{\revR}}

\newcommand{\ep}{\epsilon}
\newcommand{\te}{\tilde e_t}
\newcommand{\les}{\lesssim}

\newcommand{\tcr}{}
\newcommand{\tco}{}
\newcommand{\tcb}{}

\title{Robust tests for parameter change in conditionally heteroscedastic time series models}
\author{Junmo Song}
\affil{Department of Statistics,  Kyungpook National University}

\begin{document}
\begin{bibunit}[apalike_noissue]
\maketitle

\begin{abstract}
Structural changes and outliers often coexist, complicating statistical inference. This paper addresses the problem of testing for parameter changes in conditionally heteroscedastic time series models, particularly in the presence of outliers. To mitigate the impact of outliers, we introduce a two-step procedure comprising robust estimation and residual truncation. Based on this procedure,  we propose a residual-based robust CUSUM test and its self-normalized counterpart.  We derive the limiting null distributions of the proposed robust tests and establish their consistency. We also investigate the breakdown of the naive tests under additive and innovation outlier contamination and provide theoretical support for the robustness of the proposed tests. Simulation results demonstrate the strong robustness of our tests against outliers. Finally, we analyze Bitcoin data to illustrate the practical application.
\end{abstract}

\section{Introduction}
In practical analysis, we often encounter events that affect underlying dynamics. In finance, for instance, major events such as changes in monetary policy, critical social events, or economic crises are common examples. Following shocks from such events, the underlying dynamics may either shift to a new state or remain unchanged. When changes occur, they are typically represented through structural breaks or parameter changes in a fitted model. The statistical analysis for identifying and testing these changes is referred to as change point analysis. On the other hand, when the dynamics retain their original state after such events, the data often contain aberrant observations, such as outliers or extreme values, which can adversely affect statistical inferences. So-called robust inference methods have been developed to mitigate the impact of these outlying observations. Since ignoring structural changes or outlying observations can lead to unreliable results, change point analysis and robust inference have attracted considerable attention. For historical background and a general review of change point analysis, see, for example, \cite{aue:horvath:2013} and \cite{horvath:rice:2014}. For robust inference, see \cite{maronna2019robust}.

It is important to recognize, however, that these events often cause both outlying observations and structural changes simultaneously, or that outliers and structural changes can coexist throughout the observation period. Many previous studies have addressed these problems separately. While much of the literature has dealt with structural changes in the absence of outliers, it should be noted that the presence of outliers can significantly influence the outcomes of change point analysis. Specifically, when atypical observations are present in a dataset suspected of having structural changes, it can be difficult to determine whether the results of change point analysis are due to genuine change or the influence of outliers.

Although limited, there are several studies that address this issue. For example, \cite{tsay:1988} investigated a procedure for detecting outliers, level shifts, and variance changes in univariate time series. \cite{lee:na:2005} and \cite{kang:song:2015} introduced CUSUM tests based on robust estimators. More recently, \cite{fearnhead:rigaill:2019} proposed a robust penalized cost function for detecting changes in the location parameter. Additionally, \cite{song2021test} and \cite{song2021} introduced robust tests based on a divergence  in retrospective and sequential frameworks, respectively. 

In this study, we focus on testing for parameter changes in conditionally heteroscedastic time series models, particularly in the presence of outliers. While various change point tests have been developed, we specifically consider the CUSUM and self-normalized tests based on residuals, and propose their robust versions. Since \cite{brown1975techniques} introduced the CUSUM test, it has been widely used to detect mean or variance changes across various statistical models. In particular, the residual-based CUSUM tests have been actively applied to testing for parameter change in time series models because of their simple implementation. See, for example, \cite{kulperger:yu2005}, \cite{song:kang:2018}, and \cite{oh2019modified}. The self-normalized test for parameter change, introduced by \cite{shao2010testing}, has also received particular attention because it effectively avoids issues with the estimation of long-run variance. See, for example, \cite{betken2016testing} and \cite{choi2020self}. While this issue is less pronounced for residual-based tests in the absence of outliers, it should be noted that the variance estimation in residual-based CUSUM tests can be strongly affected by outliers.  Our simulation study indicates that additive outliers can lead to an underestimation of the variance term in the test statistics, resulting in size distortions for the robust version of the residual-based CUSUM test. To address this issue, we propose a robust self-normalized test.

 \begin{table}[t]
 \renewcommand\arraystretch{1.12}
\tabcolsep=7pt
 \centering
 \caption{Empirical sizes and powers of $T_n$ and $T_n^R$}
 \begin{tabular}{cccccccc}
\hline
                             &        & &\multicolumn{2}{c}{no outliers}&     &\multicolumn{2}{c}{$s=5$, $p=1\%$}\\ \cline{4-5} \cline{7-8}
                             &    $n$ & & $T_n$    & $T_n^R $  & & $T_n$      & $T_n^R $ \\ \hline
size                         &100     & & 0.033    &   0.034   & &  0.013      & 0.028         \\
$\sigma^2=1$                 &300     & & 0.041    &   0.042   & & 0.021 &   0.040 \\
                             &500     & & 0.045    &   0.044   & & 0.030 &   0.048 \\
                           \hline
power                        &100     & & 0.440    &   0.469   & &  0.218          &0.428          \\
$\sigma^2: 1 \rightarrow 2$  &300     & & 0.958    &   0.959   & & 0.472 &0.916 \\
                             &500     & & 0.999    &   0.998   & & 0.698 &  0.996 \\
                             \hline
\end{tabular}\label{tab:intro}
\end{table}
To demonstrate the impact of outliers and provide the intuition behind our proposed approach, we present a simple simulation example for the variance-change problem in an i.i.d. setting. In this example, truncation is directly applied to the squared observations as the main tool for reducing the impact of outliers. This idea is later extended to squared residuals in conditionally heteroscedastic time series models, together with robust estimation.
 
Let $\{ X_t\}$ be a sequence of i.i.d. random variables generated by $X_t = X_t^o + s \cdot \text{sign}(X_t^o) P_t$, where $X_t^o$ are i.i.d. random variables from $N(0, \sigma^2)$, and $P_t$ are i.i.d. Bernoulli random variables with a success probability of $p$. This setup represents a scenario in which the series ${ X_t^o }$ is contaminated by outliers. We consider the case where the parameter $\sigma^2$ changes from 1 to 2 at the midpoint of the series, under contamination with $s = 5$ and $p = 0.01$. We now test for the constancy of the variance $\sigma^2$ using the following CUSUM statistic: 
\begin{eqnarray*}
T_n:=\frac{1}{\sqrt{n}\hat{\tau}_n} \max_{1\leq k\leq n } \Big|\sum_{t=1}^k X^2_{t} -\frac{k}{n}\sum_{t=1}^n X_{t}^2\Big|,
\end{eqnarray*}
where $\hat{\tau}^2_n$ is the sample variance of $\{X_t^2\}_{t=1}^n$.
In the case of $p = 0$, it is well known that $T_n$ converges in distribution to $\sup_{0 \leq t \leq 1} |B^o_t|$, where $\{B^o_t|0\leq t\leq 1\}$ denotes a standard Brownian bridge. In the presence of outliers, the test procedure is expected to be unduly affected. A natural approach to mitigate the impact of outliers is to truncate the deviating observations using a suitable truncation function and then construct the CUSUM statistic based on the truncated observations. Specifically, for a suitable truncation function $f_{\s M}^{tr}$ with a threshold value $M>0$, one may consider the following modified statistic:
\begin{eqnarray*}
T_n^R:=\frac{1}{\sqrt{n}\hat{\tau}_{n,{\s R}}} \max_{1\leq k\leq n } \Big|\sum_{t=1}^k f^{tr}_{\s M}(X_{t}^2) -\frac{k}{n}\sum_{t=1}^n f^{tr}_{\s M}(X_{t}^2)\Big|,
\end{eqnarray*}
where  $\hat{\tau}^2_{n, {\s R}}$ is the sample variance of $\{f^{tr}_{\s M}(X_t^2)\}_{t=1}^n$. In this example, the following truncation function
\begin{eqnarray}\label{f.tr}
    f^{tr}_{\s M}(x) 
    = 
    \begin{cases}
    x &\text{if } 0\leq x\leq M,\\ 
    M &\text{if } x> M, 
    \end{cases}
\end{eqnarray}
with $M=3^2$ is used. Since ${f^{tr}_{\s M}(X_1^2), \cdots, f^{tr}_{\s M}(X_n^2)}$ are i.i.d. bounded  random variables, $T_n^R$ has the same limiting null distribution as $T_n$. We implement simulations at the 5\% significance level. The results are presented in Table \ref{tab:intro}. As seen in the table, the naive CUSUM test $T_n$ suffers from power loss, with empirical sizes lower than the 5\% significance level in contaminated data. In contrast, the truncated version, $T_n^R$, performs well in both uncontaminated and contaminated cases, yielding sizes substantially closer to the nominal level  and much higher power than the naive test.

This simple example shows that truncating the terms entering a CUSUM statistic can substantially reduce the influence of outliers while preserving the power to detect a genuine change. This approach was introduced by \citet{song:2020}, who proposed a robust CUSUM test for detecting changes in the dispersion parameter of diffusion processes. In this study, we extend this truncation-based approach to conditionally heteroscedastic time series models.

The main contributions of this study are threefold. First, we develop robust CUSUM and self-normalized tests for parameter changes by combining residual truncation with robust parameter estimation. Second, we derive the limiting null distributions of the proposed tests and establish their consistency under fixed alternatives. Third, we theoretically investigate the behavior of both the naive and proposed tests in the presence of outliers, explaining how the naive tests can break down and deriving conditions under which the proposed tests remain asymptotically unaffected by contamination.

The rest of the paper is organized as follows. In Section \ref{sec:robust_tests}, we introduce
the robust CUSUM and self-normalized tests and establish their asymptotic
properties. In Section \ref{sec:outlier_behavior}, we investigate the breakdown of the naive tests and the robustness of the proposed tests under outlier contamination. Section \ref{sec:simul} presents the simulation results, and Section \ref{sec:real_data} provides a real data analysis. Section \ref{sec:conclude} concludes the paper.

\section{Robust CUSUM and self-normalized tests}\label{sec:robust_tests}
Consider the following time series model with  parameter $\T$:
 \begin{eqnarray}\label{ts.model}
X_t=\sigma_t(\theta)\epsilon_t,
\end{eqnarray}
where $\sigma_t^2(\T)=\rm{Var}(X_t|\mathcal{F}_{t-1})$ and  $\mathcal{F}_t=\sigma(X_s|s\leq t)$. The sequence $\{\epsilon_t|t\in\mathbb{Z}\}$ consists of i.i.d. random variables  with zero mean and unit variance. We assume that the process  $\{X_t | t\in\mathbb{Z}\}$  defined by the model above is strictly stationary and ergodic.  We further assume that the parameter space $\Theta$ is a compact subset of $\mathbb{R}^d$ with the true parameter $\T_0$ lying in its interior. The model encompasses a broad class of scale time series models, including standard GARCH models as well as nonlinear and asymmetric models, such as power-transformed and threshold GARCH (PTT-GARCH) models.

When estimating the model above, $\{\sigma_t(\T)\}_{t=1}^n$ is often not explicitly obtained due to the initial value issue. In such cases, a proxy for $\{\sigma_t(\T)\}_{t=1}^n$, denoted by $\{\tilde{\sigma}_t(\T)\}_{t=1}^n$, is usually computed in one of two ways. The first method uses the fact that $\sigma_t(\T)$ can be expressed as a measurable function of $\{X_{t-1}, X_{t-2}, \cdots\}$ and the parameter $\T$ (cf. Theorem 20.1 in \cite{billingsley:1995}). Using this expression, $\{\tilde{\sigma}_t(\T)\}_{t=1}^n$ can be obtained with appropriate initial values. For instance, see \cite{berkes.et.al:2003} and \cite{pan2008estimation} for the standard GARCH models and PTT-GARCH models, respectively. The second approach is through recursion. $\sigma_t(\T)$ is usually defined by a recurrence equation. From this equation,  $\{\tilde \sigma_t(\T)\}_{t=1}^n$ can be obtained recursively by inserting suitable initial values. For the case of the GARCH models and PTT-GARCH models, see \cite{francq:zakoian:2004} and \cite{hamadeh2011asymptotic}, respectively.

\subsection{Robust CUSUM of squares test}
Let $\{X_1,\cdots, X_n\}$ be observations from the model (\ref{ts.model}). Based on these, we aim to test the following hypotheses in the potential presence of outliers:
\begin{eqnarray*}
&& H_0:\ \text{The true parameter }\theta_0\text{ does not change over }X_1,\cdots, X_n.\quad \textrm{vs}.\quad H_1:\ \textrm{not}\ \ H_0\,.
\end{eqnarray*}
To this end,  we first consider the residual-based CUSUM of squares test (cf. \citet{kulperger:yu2005}). To be more explicit, let $\hat\theta_n$ be an estimator of $\theta$. Then, the residuals for the model above are typically given as $\tilde e_t(\hat\theta_n):=\frac{X_t}{\tilde \sigma_t(\hat\T_n)}$
and, based on these residuals, the naive CUSUM of squares test is defined by
 \begin{eqnarray*}
T_n:=\frac{1}{\sqrt{n}\hat{\tau}_n} \max_{1\leq k\leq n } \Big|\sum_{t=1}^k \tilde e^2_t(\hat\theta_n) -\frac{k}{n}\sum_{t=1}^n \tilde e^2_t(\hat\theta_n)\Big|,
\end{eqnarray*}
where $\hat\tau_n^2$ is a consistent estimator for the variance of $\ep_t^2$.

As in the simulation study mentioned in the introduction, the test is likely to be significantly influenced by outliers. One reason is that outliers may bias the model estimates, causing the residuals to deviate from the behavior of ideal residuals. Therefore, using a robust estimator for $\T$ is a natural choice to reduce the influence of outliers on model estimation. Another reason is that, even with a robust estimator, residuals may still exhibit abnormal values at outlying observations. In other words, residuals computed at these observations may themselves be outliers. Consequently, it is necessary to truncate such residuals using an appropriate truncation function. In summary, to mitigate the impact of outliers on the test procedure, a two-step robust procedure involving robust estimation and residual truncation is required.

In this study, to avoid technical difficulties in proving the main theorems below, we consider the following truncation function:
\begin{eqnarray*} 
f_{{\s M},\delta}(x)=
\begin{cases} 
~~~x & \text{if } x \in [0, M - \delta), \\[2pt]
\displaystyle -\frac{1}{4\delta} (x - M - \delta)^2 + M  & \text{if } x \in [M - \delta, M+\delta),\\[2pt] 
 ~~~M & \text{if } x \in [M+\delta, \infty),\end{cases} \end{eqnarray*}
 where $M>0$ and $0<\delta<M$.
For sufficiently small $\delta>0$, this function approximates the truncation function $f^{tr}_{\s M}$ introduced in the introduction. Furthermore, it is continuously differentiable and Lipschitz continuous with a Lipschitz constant of one. These two properties play a key role in the proofs below, making it easier to handle. 
Hereafter, we denote $f_{{\s M},\delta}$ as $f_{\s M}$ for notational convenience. 

Meanwhile, rather than using the squares of truncated residuals, we truncate the squared residuals since this makes the proofs of the results below more tractable. Specifically, $\{ f_{\s M}\big(\tilde e^2_t(\hat{\theta}_n^{\revR})\big) \}_{t=1}^n$ serves as our building block for constructing  robust test statistics, where $\hat{\theta}_n^{\revR}$ is a robust estimator. 
Using these truncated squared residuals, we first propose the following robust CUSUM test:
\begin{eqnarray*}
T_{n}^M(\TR):=\begin{cases}
\displaystyle\ \frac{1}{\sqrt{n}\hat\tau_{\s M}} \max_{1\leq k\leq n } \Big|\sum_{t=1}^k f_{\s M}\big(\tilde e^2_t(\TR)\big) -\frac{k}{n}\sum_{t=1}^n f_{\s M}\big(\tilde e^2_t(\TR)\big)\Big| &\text{on } \cD_n^c,\\[12pt]
\ 0 &\text{on } \cD_n,
\end{cases}
\end{eqnarray*}
where 
\[\hat{\tau}_{\s M}^2=\frac{1}{n}\sum_{t=1}^nf_{\s M}^2\big(\te^2(\TR)\big)-\Big(\frac{1}{n}\sum_{t=1}^nf_{\s M}\big(\te^2(\TR)\big)\Big)^2\]
and $\cD_n$ denotes the event given by
\[
\cD_n=\Big\{\,f_{\s M}\big(\tilde e_1^2(\TR)\big)=\cdots=f_{\s M}\big(\tilde e_n^2(\TR)\big)\,\Big\}.
\]

\begin{remark}\label{rmk:zero}
Observe that $\hat{\tau}_{\s M}=0$ on $\cD_n$. Actually, these two events $\cD_n$ and $\{\hat{\tau}_{\s M}=0\}$ are the same. Since the equality of all truncated squared residuals provides no evidence against $H_0$, it is natural to set  $T_n^M(\hat\theta_n^{\revR})=0$ on $\cD_n$. This event can occur, for example, when $M$ is chosen so small that all
squared residuals are truncated to $M$. 

\end{remark}

The following are the conditions required to obtain the limiting null
distribution of $T_n^M(\hat{\theta}_n^{\revR})$. Throughout this paper,
$\|\cdot\|$ denotes the Euclidean norm.
\begin{enumerate}
\item[\bf A1.] $\sigma^2_t(\T)$ is  continuously differentiable with respect to $\T$.
\item[\bf A2.] For some constant $\underline{c}>0$, $\displaystyle \inf_{\T\in\Theta} \big\{\sigma^2_t(\T)\wedge \tilde \sigma^2_t(\T)\big\} \ge \underline{c}\quad a.s.$
\item[\bf A3.] For some positive random variables $V$,  $W_t$ satisfying $\E \log^+ W_t <\infty$, and a generic constant $0<\rho<1$,
\[\sup_{\T\in\Theta}   | \sigma^2_t(\T)-\tilde\sigma^2_t(\T)|  \leq V W_t\, \rho^t\quad a.s.\]
\item[\bf A4.] $\displaystyle\E \sup_{\T\in\Theta} \Big\|\frac{1}{\sigma_t^2(\T)}\frac{\pa}{\pa \T}\sigma^2_t(\T)\Big\|^2 <\infty$ and $\E X_t^4<\infty$.
\item[\bf A5.] $\{\sigma_t^2(\T)\}$ is strictly stationary and ergodic for each $\T\in\Theta$.
\item[\bf A6.] Under $H_0$, $\TR$ converges almost surely to $\T_0$ and $\sqrt{n}\|\TR -\T_0\|=O_{\uP}(1).$
\end{enumerate}
These assumptions are not restrictive. Assumptions {\bf A2} -- {\bf A4} are typically established to derive the asymptotic properties of estimators.   In the proof below, assumption {\bf A4} is required to show that $\E \sup_{\T\in\Theta} \big\| \frac{X_t^2}{\sigma_t^2(\T)}\frac{\pa}{\pa \T}\sigma^2_t(\T)\big\|<\infty$. Hence, if $\sup_{\T\in\Theta} \big\| \frac{1}{\sigma_t^2(\T)}\frac{\pa}{\pa \T}\sigma^2_t(\T)\big\|$ has a higher-order moment, the moment condition on $X_t$ can be weakened. Necessary and sufficient conditions for the existence of moments for the GARCH process can be found, for example, in \cite{chen:an:1998}.  Assumption {\bf A5} is usually deduced by the stationarity and ergodicity of $\{X_t\}$. 

\begin{remark}\label{rmk:A5}
Our robust test statistics require only that the plugged-in estimator satisfies
assumption {\bf A6}, that is, strong consistency and $\sqrt{n}$-consistency.
Hence, any estimator with these two properties can be used, and the limiting null
distributions in Theorems \ref{Theorem.main} and \ref{Theorem.main.SN} below remain
valid. In particular, one may plug in a non-robust estimator such as the QMLE.
Even in this case, the resulting test still gains some robustness from the residual
truncation. Indeed, our simulation study below shows that the tests using truncation
only with the QMLE are reasonably robust against outliers.  See also Subsection \ref{subsec:rob_cont}, where the effect of
the truncation on the tests is studied theoretically.
\end{remark}
We examine whether assumptions {\bf A1}--{\bf A5} hold for the GARCH and PTT-GARCH models in the following examples.

\begin{exam}
 For the standard GARCH models, the assumptions above are established, for example, in \cite{francq:zakoian:2004}. Notably, it holds that $\E\sup_{\T\in\Theta^*} \big\| \frac{1}{\sigma_t^2(\T)}\frac{\pa}{\pa \T}\sigma^2_t(\T)\big\|^d<\infty$ for all $d>0$, where $\Theta^*$ is a compact subset of $\Theta$ with $\T_0\in \Theta^* \subset \Theta^i$. Hence, it suffices that $\E |X_t|^\tau<\infty$ for some $\tau>0$.
\end{exam}

\begin{exam} The conditional variance of PTT-GARCH$(p,q)$ model satisfies the following equation:
\[\sigma^{2\delta}_t(\T) = \omega + \sum_{i=1}^p \A_{1i}(X^+_{t-i})^{2\delta} + \sum_{i=1}^p \A_{2i}(X^-_{t-i})^{2\delta} + \sum_{j=1}^q \beta_j \sigma^{2\delta}_{t-j}(\T),\]
where the parameter $\T$ is $(\delta,\omega,\A_{11},\cdots,\A_{1p},\A_{21},\cdots,\A_{2,p},\beta_{1},\cdots,\beta_{q}).$ The parameter space $\Theta$ is assumed to be compact with $0<\underline{c}\leq \delta,\omega\leq \bar{c}$ for any $\T\in\Theta$. We also assume that $\{X_t\}$ from the PTT-GARCH($p,q$) model is strictly stationary and ergodic. For detailed conditions, see the appendix of \cite{pan2008estimation}. We can easily see that assumption {\bf A1} holds from the derivatives of $\sigma^{2\delta}_t(\T)$ provided in \cite{pan2008estimation}. Assumption {\bf A2} is easy to check by the compactness of $\Theta$ with $\omega \geq \underline{c} $. 
As an approximation $\{\tilde \sigma_t(\T)\}_{t=1}^n$ to $\{\sigma_t(\T)\}_{t=1}^n$, one can use the formula given in (2.8) in \cite{pan2008estimation}. They showed that 
\[\sup_{\T\in\Theta}   | \sigma^{2\delta}_t(\T)-\tilde\sigma^{2\delta}_t(\T)|  \leq V_0 \rho^t\quad\text{and}\quad
\sup_{\T\in\Theta}   \sigma^{2\delta}_t(\T)\leq V_{1t},\]
where $V_0=C\sum_{j=0}^\infty \rho^j ( |X_{-j}|^{2\bar{c}} +1)$ and $V_{1t}=C\sum_{j=1}^\infty \rho^j ( |X_{t-j}|^{2\bar{c}} +1)$ with some constants $C>0$ and $0<\rho<1$. Since $V_0\rho^t$ and $V_{1t}$ serve as upper bounds, we can assume that $C>1$, and hence $V_0>1$. Using these facts  and the  mean value theorem, we have
\begin{eqnarray*}
    | \sigma^2_t(\T)-\tilde\sigma^2_t(\T)| &\leq&\begin{cases} 
\frac{1}{\delta} \big(\frac{1}{\omega}\big)^{1-1/\delta} | \sigma^{2\delta}_t(\T)-\tilde\sigma^{2\delta}_t(\T)| &\text{if } \delta\ge1,\\[10pt]
\frac{1}{\delta} \big(V_0+2\sigma_t^{2\delta}(\T)\big)^{1/\delta-1} | \sigma^{2\delta}_t(\T)-\tilde\sigma^{2\delta}_t(\T)| &\text{if }  0<\delta<1
    \end{cases}\\[10pt]
    &\leq&
    \frac{2^{2/\underline{c}}}{\underline{c}^2}\big( 1+ V_0^{1/\underline{c}}+V_{1t}^{1/\underline{c}}\big) V_0 \rho^t,
\end{eqnarray*}
where we assume  without loss of generality that  $\underline{c}<1$.  Let $W_t= 1+ V_0^{1/\underline{c}}+V_{1t}^{1/\underline{c}}$. Under the assumption {\bf (A1)} of \cite{pan2008estimation}, $\E |X_t|^\tau <\infty$ for some $\tau>0$. Taking $s>0$ such that $s/\underline{c}<1$ and $2s<\tau$, and using the fact that $(a+b)^s\leq a^s+b^s$ for $a,b\geq 0$ and $0<s\leq 1$, it can be shown that $\E W_t^s <\infty$, implying that $\E \log^+ W_t <\infty$. Hence, setting $V=\frac{2^{2/\underline{c}}}{\underline{c}^2} V_0$, assumption {\bf A3} holds.  
Note that
\[
\frac{1}{\sigma_t^2(\T)} \frac{\pa \sigma_t^2(\T)}{\pa \T} 
= 
\frac{1}{\delta \sigma_t^{2\delta}(\T)} \frac{\pa \sigma_t^{2\delta}(\T)}{\pa\T} 
- \frac{1}{\delta^2} \ln \sigma_t^{2\delta}(\T) \, \mathbf{e}_1,
\]
where $\mathbf{e}_1=(1,0,\cdots,0)'\in \mathbb{R}^{2+2p+q}$.
Using equations (5.3)--(5.5) in \cite{pan2008estimation}, we can see that $\sup_{\T\in\Theta}\big\|\frac{1}{\sigma_t^2(\T)}\frac{\pa}{\pa \T} \sigma_t^2(\T)\big\|$ has all finite moments. Hence, instead of requiring $\E X_t^4<\infty$, it suffices for assumption {\bf A4} that $\E |X_t|^\tau <\infty$ for some $\tau>0$, as stated above, which is ensured by assumption {\bf (A1)} in \cite{pan2008estimation}. We can also see that assumption {\bf A5} holds because $\sigma^{2\delta}_t$ can be expressed as a function of $\{X_s | s\leq t\}$ and $\{X_t\}$ is strictly stationary and ergodic. 
\end{exam}

Throughout this paper, we assume that $f_{\s M}(\epsilon_1^2)$ is not degenerate, because otherwise the limiting null distributions of the test statistics proposed in this study cannot be obtained. Note that
$f_{\s M}(\epsilon_1^2)$ is degenerate only when $\epsilon_1^2$ is almost surely constant or
$\epsilon_1^2\ge M+\delta$ almost surely, and the latter cannot occur when $M\ge1$ because
$E\epsilon_1^2=1$. Hence, one may naturally consider the practical range of $M$ as $M\ge1$.
A more specific choice of $M$ is discussed in Remark \ref{rmk:choice_M} below.

In the proofs that follow, we shall use the relation $A\lesssim B$, where $A$ and $B$ are
nonnegative, to mean that $A\le CB$ for some constant $C>0$ that does not depend on $n$.
We are now ready to present our first result.
\begin{thm}\label{Theorem.main}
Suppose that assumptions {\bf A1}--{\bf A6} are satisfied. Then, under $H_0$, it holds that
\begin{eqnarray*}
T^M_{n}(\TR)
\stackrel{d}{\longrightarrow}\sup_{0\leq t \leq 1} |B^o_t|\quad \text{as } n\rightarrow \infty,
\end{eqnarray*}
where $\{B_t^o \mid 0\leq t \leq  1\}$ is a standard Brownian bridge.
\end{thm}
\begin{proof}

Since $f_{\s M}(\epsilon^2_1), \cdots, f_{\s M}(\epsilon^2_n)$ are i.i.d. bounded  random variables, it follows from
the invariance principle and the continuous mapping theorem that
\begin{eqnarray}\label{f.conv}
\frac{1}{\sqrt{n}\tau_{\s M}} \max_{1\leq k\leq n } \Big|\sum_{t=1}^k f_{\s M}(\epsilon_t^2) -\frac{k}{n}\sum_{t=1}^nf_{\s M}(\epsilon_t^2)\Big|
\stackrel{d}{\longrightarrow}\sup_{0\leq t \leq 1} |B^o_t|,
\end{eqnarray}
where $\tau_{\s M}^2=\mathrm{Var}\big(f_{\s M}(\epsilon_1^2)\big)$.
 We first show that
\begin{eqnarray}\label{main.conv}
\frac{1}{\sqrt{n}}\max_{1\leq k\leq n } \Big|\sum_{t=1}^k \big(f_{\s M}(\te^2(\TR))-f_{\s M}(\epsilon_t^2)\big) 
-\frac{k}{n}\sum_{t=1}^n \big(f_{\s M}(\te^2(\TR))-f_{\s M}(\epsilon_t^2)\big) \Big|=o_{\uP}(1)
\end{eqnarray}
and
\begin{eqnarray}\label{tau.conv}
\hat{\tau}_{\s M}^2\stackrel{\uP}{\longrightarrow} \tau_{\s M}^2.
\end{eqnarray}
Let $e_t(\T)=X_t/\sigma_t(\T)$. We first  note that $\ep_t=e_t(\T_0)$. 
Now, split $f_{\s M}(\te^2(\TR))-f_{\s M}(\epsilon_t^2)$ into the following two terms:
\begin{eqnarray}\label{fm.split}
    f_{\s M}(\te^2(\TR))-f_{\s M}(\epsilon_t^2) &=& \big\{f_{\s M}(\te^2(\TR))-f_{\s M}(e_t^2(\TR))\big\}+\big\{f_{\s M}(e_t^2(\TR))-f_{\s M}(e_t^2(\T_0))\big\}\nonumber\\
    &=:& I_{t} +II_{t}.
\end{eqnarray}
By assumptions {\bf A2} and {\bf A3}, we have
\begin{eqnarray}\label{diff.ep}
   \big|\te^2(\TR)-e_t^2(\TR) \big|
   &=&\Big|\frac{X_t^2}{\tilde\sigma_t^2(\TR)}-\frac{X_t^2}{\sigma_t^2(\TR)}\Big| \ \les\  V W_t X_t^2 \rho^t.
\end{eqnarray}
Since $\E \log^+ W_tX_t^2 \leq \E \log^+W_t +\E \log^+X_t^2 <\infty$, it follows from  Lemma 2.1 in \cite{straumann:mikosch:2006} that $\sum_{t=1}^n W_tX_t^2\rho^t=O(1)\ a.s.$
Hence, noting that $|f_{\s M}(x)-f_{\s M}(y)|\leq |x-y|$ for all $x,y\geq0$, we have 
\begin{eqnarray}\label{main.conv1}
 \frac{1}{\sqrt{n}}\max_{1\leq k\leq n } \Big|\sum_{t=1}^k I_t-\frac{k}{n}\sum_{t=1}^n I_t \Big| 
 &\les&\frac{V}{\sqrt{n}}\sum_{t=1}^n W_tX_t^2\rho^t=o(1)\quad a.s.
 \end{eqnarray}
Next, letting $g_t(\T):=\frac{\pa}{\pa x} f_{\s M}( e_t^2(\T))\, \frac{\pa}{\pa \T} e_t^2(\T)$, we have
\begin{eqnarray*}\label{II}
    II_t= (\TR-\T_0)'\, g_t(\T^*_{t,n}), 
\end{eqnarray*}
where $\T^*_{t,n}$ is an intermediate point between $\TR$ and $\T_0$.
Here, we note that  $g_t(\T)$ is continuous by assumption {\bf A1} and, for each $\T \in \Theta$,  $\{g_t(\T)\}$ is stationary and ergodic by assumption {\bf A5}. Since $|\frac{\pa}{\pa x} f_{\s M}(x)| \leq 1$ for all $x> 0$ and  $\frac{\pa}{\pa \T} e^2_t(\T)=-\frac{X_t^2}{\sigma_t^4(\T) }\frac{\pa}{\pa \T} \sigma_t^2(\T)$, it follows from the Cauchy-Schwarz inequality and assumption {\bf A4} that  $\E \sup_{\T\in\Theta} \|g_t(\T)\|<\infty$. Therefore, for any $\ep>0$, by the dominated convergence theorem and the continuity of $g_t$, we can choose a constant $r_\ep>0$ such that
\begin{eqnarray}\label{e.neighbor}
    \E \sup_{\T\in N_\ep(\T_0)} \| g_t(\T) -g_t(\T_0)\|\  \le\ \ep,
\end{eqnarray}
where $N_\ep(\T_0)=\{\T\in\Theta \mid \|\T-\T_0\| \leq r_\ep\}$. Since $\|\T^*_{t,n}-\T_0\|\leq \|\TR-\T_0\|$ and $\TR$ converges almost surely to $\T_0$ by assumption {\bf A6}, we have that  for sufficiently large $n$,
\begin{eqnarray*}
    \Big\|\frac{1}{n}\sum_{t=1}^n g_t(\T^*_{t,n})- \E g_t(\T_0)\Big\| &\leq &
    \frac{1}{n}\sum_{t=1}^n \sup_{\T\in N_\ep(\T_0)} \|g_t(\T)-g_t(\T_0)\| +\Big\|\frac{1}{n}\sum_{t=1}^n g_t(\T_0) -\E g_t(\T_0)\Big\|
    \quad a.s.,
\end{eqnarray*}
from which, together with (\ref{e.neighbor}),  we can see by the ergodic theorem that $\frac{1}{n} \sum_{t=1}^n g_t(\T^*_{t,n})$ converges almost surely to $\E g_t(\T_0)$.
Thus, it follows that
\begin{eqnarray*}
    &&\hspace{-1cm}\max_{1\leq k \leq \sqrt{n}} \frac{k}{n}\Big\| \frac{1}{k}\sum_{t=1}^k g_t(\T^*_{t,n})
    -\frac{1}{n}\sum_{t=1}^n g_t(\T^*_{t,n}) \Big\| \\
    &\leq &
    \frac{1}{\sqrt{n}} \sup_{k\geq 1} \Big\| \frac{1}{k}\sum_{t=1}^k g_t(\T^*_{t,n})\Big\|
    +\frac{1}{\sqrt{n}}\Big\|\frac{1}{n}\sum_{t=1}^n g_t(\T^*_{t,n}) \Big\| =o(1)\quad a.s.
\end{eqnarray*}
and
\begin{eqnarray*}
   \max_{\sqrt{n}\leq k \leq n} \frac{k}{n}\Big\| \frac{1}{k}\sum_{t=1}^k g_t(\T^*_{t,n})
    -\frac{1}{n}\sum_{t=1}^n g_t(\T^*_{t,n}) \Big\| 
    &\leq& 
    \max_{\sqrt{n}\leq k \leq n} \Big\| \frac{1}{k}\sum_{t=1}^k g_t(\T^*_{t,n})
    -\frac{1}{n}\sum_{t=1}^n g_t(\T^*_{t,n}) \Big\|\\
    &=&o(1)\quad a.s.,
\end{eqnarray*}
which imply 
\[\max_{1\leq k \leq n} \frac{k}{n}\Big\| \frac{1}{k}\sum_{t=1}^k g_t(\T^*_{t,n})
    -\frac{1}{n}\sum_{t=1}^n g_t(\T^*_{t,n}) \Big\| =o(1)\quad a.s.\]
Therefore, since $\sqrt{n}\|\TR-\T_0\|=O_{\uP}(1)$ by assumption {\bf A6}, it follows from the Cauchy-Schwarz inequality  that 
\begin{eqnarray*}
    \frac{1}{\sqrt{n}}\max_{1\leq k\leq n } \Big|\sum_{t=1}^k II_t-\frac{k}{n}\sum_{t=1}^n II_t \Big|
    &\leq&\sqrt{n} \| \TR-\T_0\| \max_{1\leq k \leq n} \frac{k}{n}\Big\| \frac{1}{k}\sum_{t=1}^k g_t(\T^*_{t,n})
    -\frac{1}{n}\sum_{t=1}^n g_t(\T^*_{t,n}) \Big\|\\
    &=& o_{\uP}(1).  
\end{eqnarray*}
Combining this with (\ref{main.conv1}), (\ref{main.conv}) is asserted.

Recall that  $|f_{\s M}(x)-f_{\s M}(y)| \leq |x-y|$ for all $x,y\geq0$ and $f_{\s M}$ is bounded above by $M$. Then, we have that 
\begin{eqnarray}\label{eq:cond_consi_tau}
    \frac{1}{n}\sum_{t=1}^n \big|f_{\s M}(\te^2(\TR))-f_{\s M}(\epsilon_t^2)\big| \vee \frac{1}{n}\sum_{t=1}^n \big|f_{\s M}^2(\te^2(\TR))-f_{\s M}^2(\epsilon_t^2)\big| 
\ \les\ \frac{1}{n} \sum_{t=1}^n \big|\te^2(\TR) -\ep_t^2\big|.
\end{eqnarray}
Since $\sum_{t=1}^n W_tX_t^2\rho^t=O(1)$ a.s. and $\|\TR-\T_0\|=o(1)$ a.s., we have by (\ref{diff.ep}) and  the ergodic theorem that
\begin{eqnarray*}
    \frac{1}{n}\sum_{t=1}^n \big|\te^2(\TR) -\ep_t^2\big| 
    &\leq&  \frac{1}{n}\sum_{t=1}^n \big|\te^2(\TR) -e_t^2(\TR)\big|+ \frac{1}{n}\sum_{t=1}^n \big|e_t^2(\TR) -e_t^2(\T_0)\big|
\end{eqnarray*}
\begin{eqnarray*}
    \hspace{5.55cm}
    \les\   \frac{V}{n}\sum_{t=1}^n W_tX_t^2\rho^t +\|\TR-\T_0\| \frac{1}{n}\sum_{t=1}^n \sup_{\T\in\Theta} \| \pa_\T e_t^2(\T)\| =o(1)\quad a.s.,
\end{eqnarray*}
 which together with \eqref{eq:cond_consi_tau} asserts (\ref{tau.conv}).
 
 Now, let $T_{n,\epsilon}^M$ and $\Delta_n^M$ be the terms on the left sides of (\ref{f.conv}) and (\ref{main.conv}), respectively, and put $\cA_n:=\big\{\hat\tau_{\s M}>\tau_{\s M}/2\big\}$. Since $f_{\s M}(\epsilon_1^2)$ is assumed to be non-degenerate, we have $\tau_{\s M} >0$, implying $\cA_n\subset \cD_n^c$. Thus, we have 
\begin{eqnarray*}
\big|T_n^M(\TR)-T^M_{n,\ep}\big|
\ \leq\
\frac{2}{\tau_{\s M}}\Big(\Delta^M_n+|\tau_{\s M}-\hat\tau_{\s M}|\,T^M_{n,\epsilon}\Big)=:\Lambda_n\quad\text{on }\cA_n.
\end{eqnarray*}
Note that  $\Lambda_n=o_{\uP}(1)$ by \eqref{f.conv}, \eqref{main.conv}, and \eqref{tau.conv}. Therefore, since $P(\cA_n^c)\to 0$ by (\ref{tau.conv}), we obtain that for any $\eta>0$,
\begin{eqnarray*}
\uP\big(\big|T_n^M(\TR)-T^M_{n,\epsilon}\big|>\eta\big)
&\leq&
\uP(\cA_n^c)+\uP\big(\big\{\big|T_n^M(\TR)-T^M_{n,\epsilon}\big|>\eta\big\}\cap \cA_n\big)\\
&\leq&
\uP(\cA_n^c)+\uP\big(\Lambda_n>\eta\big)\ =\ o(1),
\end{eqnarray*}
which together with (\ref{f.conv}) yields the theorem. This completes the proof.
\end{proof}

The limiting null distribution of the naive CUSUM test $T_n$ can also be readily obtained by noting that \eqref{main.conv} continues to hold when $f_{\s M}$ is replaced by any continuously differentiable function $f$ satisfying $|f(x)-f(y)|\le|x-y|$ for all $x,y\ge0$. Since the identity function $f(x)=x$ satisfies this condition, we have
\begin{eqnarray*}\label{eq:res.approx}
\frac{1}{\sqrt n}\max_{1\le k\le n}\Big|\sum_{t=1}^k\big(\te^2(\hat\T_n)-\ep_t^2\big)
-\frac kn\sum_{t=1}^n\big(\te^2(\hat\T_n)-\ep_t^2\big)\Big|=o_{\uP}(1).
\end{eqnarray*}
Note that $\E\ep_1^4<\infty$ because $\E X_t^4 =\E\sigma_t^4(\T_0)\, \E \ep_t^4<\infty$ by
assumption {\bf A4} and $\E\sigma_t^4(\T_0)>0$ by assumption {\bf A2}. Then, $\tau^2=\Var(\ep_1^2)$ is finite, and thus, \eqref{f.conv} also holds
with $f_{\s M}$ replaced by the identity function. Therefore, we obtain the following corollary.

\begin{corollary}\label{Cor:Tn}
Suppose that assumptions {\bf A1}--{\bf A5} are satisfied and that  assumption {\bf A6} holds for $\hat\T_n$ used in $T_n$.
Under $H_0$, we have
\begin{eqnarray*}
T_{n}
\stackrel{d}{\longrightarrow}\sup_{0\leq t \leq 1} |B^o_t|\quad\text{as }n\to\infty.
\end{eqnarray*}
\end{corollary}
\noindent The limiting null distribution of $T_n$ for the standard GARCH models was
also established in \cite{kulperger:yu2005}. The above corollary extends
this result to a broader class of timeseries models.
\subsection{Robust self-normalized test}
While the two-step procedure introduced above effectively prevents the truncated
squared residuals at outlier times from reaching large values, it does not fully
mitigate the propagation of outlier effects on subsequent fitted conditional variances.
Such a propagated effect can be roughly described in the case of additive outlier
contamination. When an additive outlier is exceptionally large, it may inflate the
fitted conditional variances in the subsequent periods, causing the truncated squared
residuals therein to be computed as small. This can lead to an underestimation of
$\tau_{\s M}^2$ and, consequently, a size distortion of $T_n^M(\TR)$. This indicates
that although $\hat\tau_{\s M}$ is obtained using the residuals through the two-step
procedure, it is still not free from the effect of the outliers, and thus neither is
$T_n^M(\TR)$. To address the issue related to the estimation of $\tau_{\s M}^2$ in the
presence of large outliers, we consider a self-normalized version of $T_n^M(\TR)$, in
which the estimation of $\tau_{\s M}$ is not required.

The self-normalization method has been successfully used to avoid issues related to the
estimation of the long-run variance. See, for example, \cite{lobato2001testing} and
\cite{shao2010self}. Building on this idea, \cite{shao2010testing} introduced a
self-normalized test for parameter changes. This type of test has since been widely
applied across various statistical models. For instance, \cite{betken2016testing}
explored the self-normalized test in long-range dependent time series, and
\cite{choi2020self} examined its use in detecting correlation breaks. Although the
purpose of using the self-normalization method in these studies is different from ours,
the idea of excluding the estimation of the variance term can serve as a promising
remedy for the underestimation issue of $\hat\tau_{\s M}^2$.

We now introduce the self-normalized version of $ T_n^M(\TR) $. To proceed, for a sequence $x=(x_1,\cdots,x_n)$, we define
\vspace{-0.3cm}
\begin{eqnarray*}
D_{n,k}(x):=\sum_{t=1}^k x_t-\frac{k}{n}\sum_{t=1}^n x_t
\end{eqnarray*}
and
\vspace{-0.3cm}
\begin{eqnarray*}
V_{n,k}(x):=
\sum_{t=1}^k \Big(\sum_{j=1}^tx_j -\frac{t}{k}\sum_{j=1}^k x_j\Big)^2 +
\sum_{t=k+1}^n\Big(\sum_{j=t}^nx_j -\frac{n-t+1}{n-k}\sum_{j=k+1}^n x_j\Big)^2. 
\end{eqnarray*}
Let $f_{\s M}\big(\tilde e^2(\TR)\big):=\big(f_{\s M}\big(\tilde e_1^2(\TR)\big),\cdots,f_{\s M}\big(\tilde e_n^2(\TR)\big)\big)$. Then, the robust self-normalized test is given as follows:
\begin{eqnarray*}
SN_n^M(\TR)
:=
\begin{cases}
\displaystyle \max_{1\leq k \leq n-1}
\frac{\frac{1}{n} D^2_{n,k}\big(f_{\s M}\big(\tilde e^2(\TR)\big)\big)}
{\frac{1}{n^2}V_{n,k}\big(f_{\s M}\big(\tilde e^2(\TR)\big)\big) } & \text{on } \cV_n,\\[10pt]
\ 0 &\text{on } \cD_n,\\[3pt]
\ \infty &\text{on }(\cV_n\cup\cD_n)^c,
\end{cases}
\end{eqnarray*}
where $\mathcal{V}_n$ denotes the event that $V_{n,k}\big(f_{\s M}\big(\tilde e^2(\TR)\big)\big)>0$ for all $1\leq k\leq n-1$.

\begin{remark}\label{rmk:sndeg}
Note that, for each $1\leq k\leq n-1$, $V_{n,k}\big(f_{\s M}\big(\tilde e^2(\TR)\big)\big)=0$ holds if and only if
\begin{eqnarray*}
f_{\s M}\big(\tilde e_1^2(\TR)\big)=\cdots=f_{\s M}\big(\tilde e_k^2(\TR)\big)
\quad\mbox{and}\quad
f_{\s M}\big(\tilde e_{k+1}^2(\TR)\big)=\cdots=f_{\s M}\big(\tilde e_n^2(\TR)\big).
\end{eqnarray*}
One can readily see that, on $\cD_n^c$, there exists at most one $k$ satisfying $V_{n,k}\big(f_{\s M}\big(\tilde e^2(\TR)\big)\big)=0$.
Denoting such an index by $k_0$, we have $D_{n,k_0}^2\big(f_{\s M}\big(\tilde e^2(\TR)\big)\big)>0$, and thus the ratio at $k_0$ has a positive numerator and a zero denominator. Since the truncated squared residuals are constant before and after $k_0$ and take different values on the two segments, it is natural to regard $k_0$ as a change point and to set $SN_n^M(\TR)=\infty$ in this case.
\end{remark}

The test above can also be viewed as a robust counterpart of the following naive self-normalized test:
\begin{eqnarray*}
SN_n
:=
\max_{1\leq k \leq n-1}
\frac{\frac{1}{n} D^2_{n,k}\big(\tilde e^2(\hat{\theta}_n)\big)}
{\frac{1}{n^2}V_{n,k}\big(\tilde e^2(\hat{\theta}_n)\big) },
\end{eqnarray*}
where $\tilde e^2(\hat{\theta}_n)=(\tilde e_1^2(\hat{\theta}_n),\cdots,\tilde e_n^2(\hat{\theta}_n))$. We note that $V_{n,k}\big(\tilde e^2(\hat\T_n)\big)>0$ for all $1\le k\le n-1$ almost surely since $\{\tilde e_t^2(\hat\T_n)\}_{t=1}^n$ are continuous random variables.
The two tests share the same limiting null distribution, which is established in the
following theorem for $SN_n^M(\hat\theta_n^R)$ and in the subsequent corollary for $SN_n$. The quantiles of this distribution can be found in Table 1 of \cite{shao2010testing}. The critical values for a 1-dimensional Brownian bridge are given as
$29.6$ and $40.1$ at the $10\%$ and $5\%$ levels, respectively.

To state the following theorem, we introduce a map and a functional. Let $D_0:=\{g\in D[0,1]\mid g(0)=g(1)=0\}$. For $g\in D_0$, define $V_g \in D[0,1]$ by
\begin{eqnarray*}\label{cV}
     V_g(t):=\int_0^t\Big(g(u)-\frac{u}{t}g(t)\Big)^2 du
    +\int_t^1\Big(g(u)-\frac{1-u}{1-t}g(t)\Big)^2 du,\qquad 0<t<1,
\end{eqnarray*}
and set $V_g(0)=V_g(1)=\int_0^1g^2(u)\,du$. \noindent Next, letting $D_0^+:=\{g\in D_0\mid\inf_{0\le t\le1}V_g(t)>0\}$, we define the functional $\Phi$ on  $D_0$ by
\begin{eqnarray*}
\Phi(g):=
\begin{cases}
\displaystyle \sup_{0\leq t\leq 1}\frac{g^2(t)}{V_g(t)} &\text{if } g\in D_0^+,\\[7pt]
\ 0 &\text{if } g\equiv0,\\[0pt]
\ \infty& \text{otherwise}.
\end{cases}
\end{eqnarray*}

\begin{thm}\label{Theorem.main.SN}
Suppose that assumptions {\bf A1}--{\bf A6} are satisfied.  Under $H_0$, it holds that
\begin{eqnarray*}
SN_n^M(\TR)
&\stackrel{d}{\longrightarrow}& \Phi(B^o)=\sup_{0\leq t \leq 1} \frac{\big(B_t^o\big)^2}{V_{B^o}(t)}\quad \text{as } n\rightarrow \infty,
\end{eqnarray*}
where $B^o=\{B^o_t \mid 0 \le t \le 1\}$ is  a standard Brownian bridge.
\end{thm}
\begin{proof}
Define the process $\GtM=\{\GtM(u)\mid 0\le u \le 1\}$ by
\begin{eqnarray*}\label{Gn.def}
    \GtM(u):=\frac{1}{\sqrt{n}} D_{n,\lfloor nu\rfloor} \big(f_{\s M}\big(\tilde e^2(\hat{\theta}^R_n)\big)\big).
\end{eqnarray*}
Then, by the invariance principle and \eqref{main.conv}, we have
\begin{eqnarray}\label{Gn.conv}
    \GtM \stackrel{w}{\longrightarrow} \tau_{\s M} B^o \quad \text{in}\ \ D[0,1].
\end{eqnarray}
Since we have $SN_n^M(\TR)=\Phi\big(\GtM\big)+o_{\uP}(1)$ by Lemma \ref{lem:sn} and $\Phi$ is continuous at $\tau_{\s M} B^o$ with probability one by Lemma \ref{lem:cmt}, it follows from (\ref{Gn.conv}) and the continuous mapping theorem that
\begin{eqnarray*}
    SN_n^M(\TR) \stackrel{d}{\longrightarrow}
    \Phi(\tau_{\s M} B^o).
\end{eqnarray*}
Since  $V_{cg}=c^2V_g$ for all $c\neq0$, we have $\Phi(\tau_{\s M} B^o)=\Phi(B^o)$. This completes the proof.
\end{proof}

Since $SN_n$ is very close to $SN_n^M$ when $M$ is sufficiently large, one may naturally expect that the two statistics have the same limiting null distribution. However, verifying this requires rigorous justification. We apply Theorem 3.2 of \citet{billingsley:1999} to prove the following corollary. The proof is provided in the Appendix.

\begin{corollary}\label{Cor:SN}
Suppose that assumptions {\bf A1}--{\bf A5} are satisfied and that  assumption {\bf A6} holds for $\hat\T_n$ used in $SN_n$.  Under $H_0$, we have
\begin{eqnarray*}
SN_{n}
\stackrel{d}{\longrightarrow}\sup_{0\leq t \leq 1} \frac{\big(B_t^o)^2}{V_{B^o}(t)}
\quad\text{as }n\to\infty.
\end{eqnarray*}
\end{corollary}

\subsection{Consistency of the robust tests}
We now investigate the consistency of $T^M_n(\TR)$ and $SN^M_n(\TR)$ under the alternative hypothesis below. To be more specific, let $\{X_{0,t} | t\in\mathbb{Z}\}$  and  $\{X_{1,t} | t\in\mathbb{Z}\}$ be the strictly stationary and ergodic processes from the model (\ref{ts.model}) with the parameters $\theta_0$ and $\theta_1(\neq \theta_0)$, respectively. The alternative hypothesis under consideration is as follows:
\[
H_1: X_t = \begin{cases}
X_{0,t}, \quad & t=1,\cdots, k^*, \\
X_{1,t}, \quad & t=k^*+1,\cdots, n,
\end{cases}
\]
where the change point $k^*$ is assumed to be $\lfloor n\lambda \rfloor$ for some $ 0 <\lambda < 1 $.

To establish the asymptotic property of $T_n^M(\TR)$ and $SN_n^M(\TR)$ under $H_1$, we require additional conditions on the estimator plugged in the robust tests. An estimator is usually defined as the optimizer of an objective function. Let $\tilde L_n(\T):=\frac{1}{n}\sum_{t=1}^n \tilde l (\T; X_t)$ be the objective function for the estimator $\TR$, where $\tilde l(\T;X_t)$ is actually a function of $X_t$ and $\tilde \sigma_t^2(\T)$, and let $l(\T;X_t)$ denote its counterpart of $\tilde l(\T;X_t)$ obtained by replacing $\tilde \sigma_t^2(\T)$ with $\sigma_t^2(\T)$.
We first assume that the objective function converges uniformly as follows:
\begin{eqnarray}\label{obj.H1}
   \tilde L_n(\T)&=&\frac{1}{n}\sum_{t=1}^{k^*} \tilde l(\T;X_{0,t}) + \frac{1}{n}\sum_{t=k^*+1}^n \tilde l(\T;X_{1,t}) \nonumber\\
&\rightarrow& L(\T):=\lambda \E \, l(\T;X_{0,t}) + (1-\lambda) \E \, l(\T;X_{1,t})\quad a.s.
\end{eqnarray} 
This can be obtained, for example, by showing that, for each $i=0,1$,  $\E \sup_{\T\in \Theta} | l(\T;X_{i,t})| <\infty$; $\frac{1}{k^*} \sum_{t=1}^{k^*} \sup_{\T\in\Theta} | \tilde l(\T;X_{0,t})- l(\T;X_{0,t})|=o(1)$ a.s. and $\frac{1}{n-k^*} \sum_{t=k^*+1}^{n} \sup_{\T\in\Theta} | \tilde l(\T;X_{1,t})- l(\T;X_{1,t})|=o(1)$ a.s. 
Let $\bar\T$ be the optimizer of $L(\T)$. Under $H_1$, we make the following assumptions to ensure the consistency of $T_n^M(\TR)$ and $SN_n^M(\TR)$:
\begin{enumerate}
\item[\bf A7.] $\bar\T$ is the unique optimizer of $L(\T)$.
\item[\bf A8.] $C_{\s M}:=\big|\E f_{\s M}\big(e^2_{0,t}(\bar\T)\big)-\E f_{\s M}\big( e^2_{1,t}(\bar\T)\big)\big| > 0$, where $e_{i,t}(\T)=\frac{X_{i,t}}{\sigma_t(\T)}$ for $i=0,1$.
\end{enumerate}

\begin{remark}\label{rmk:CM}
The proposed tests are based on the truncated squared residuals, and thus their ability to detect a change inevitably depends on whether the corresponding truncated second moment changes.
Note that $C_{\s M}$ represents the difference between the truncated second moments before and after the change. 
As shown in Theorems \ref{Thm:cons.T} and \ref{Thm:cons.SN} below, $C_{\s M}$ directly governs the divergence of $T_n^M(\TR)$ and $SN_n^M(\TR)$. When $C_{\s M}>0$, both robust tests are consistent, whereas consistency is no longer guaranteed if $C_{\s M}=0$. Hence, assumption {\bf A8} serves as a sufficient condition for the consistency of the proposed tests. Further, we emphasize that there can be cases in which a parameter change
exists but $C_{\s M}=0$. In such cases, or when  $C_{\s M}$ is close to zero,  the tests are very likely to suffer from lower power. 
This limitation also applies to the naive tests when the difference between the corresponding second moments before and after the change is small.  
\end{remark}

We note that, by the uniform convergence of (\ref{obj.H1}) and assumption {\bf A7}, $\TR$ converges almost surely to $\bar\T$. We first present the result establishing the consistency of $T_n^M(\TR)$.
\begin{thm}\label{Thm:cons.T}
Suppose that assumptions {\bf A1}--{\bf A3} and {\bf A5} still hold for $\{X_{0,t}\}$ and $\{X_{1,t}\}$. If assumptions {\bf A7} and {\bf A8} are satisfied, then it holds that  under $H_1$,
\begin{eqnarray*}
\frac1n \argmax_{1\leq k\leq n } \big|D_{n,k}\big(f_{\s M}\big(\tilde e^2(\TR)\big)\big)\big| \stackrel{a.s.}{\longrightarrow} \lambda\quad\text{as }n\to\infty.
\end{eqnarray*}
Further, we have
\begin{eqnarray*}
\frac{1}{\sqrt{n}}\,T_n^M(\hat\T_n^{\revR}) \stackrel{a.s.}{\longrightarrow} \frac{\lambda(1-\lambda)C_{\s M}}{\tau_{\s M}^*}\quad\text{as }n\to\infty,
\end{eqnarray*}
where $\tau_{\s M}^*$ is a finite positive constant, and thus
$T_n^M(\hat\T_n^{\revR})\rightarrow\infty$ almost surely.
\end{thm}
\begin{proof} 

Let $n'$ and $n''$ be integers such that $1\le n'<n''\le k^*$ for $i=0$ and
$k^*+1\le n'<n''\le n$ for $i=1$, respectively, and that $n''-n'+1\ge c\,n$ for some constant $c>0$. Note that $n''/(n''-n'+1)$ and $(n'-1)/(n''-n'+1)$ are bounded by $1/c$. Then, for each $i=0,1$, since $\{f_{\s M}(e^2_{i,t}(\theta))\}$ is
stationary and ergodic and $\sup_{\theta\in\Theta}f_{\s M}(e^2_{i,t}(\theta))\le M$,
we have by Theorem 2.7 in \cite{straumann:mikosch:2006} that
\vspace{-0.3cm}
\begin{eqnarray}\label{cons.1}
   &&\hspace{-1cm}
   \sup_{\T\in\Theta}\Big| \frac{1}{n''-n'+1}\sum_{t=n'}^{n''} f_{\s M}\big(e^2_{i,t}(\T)\big) - \E\,f_{\s M}\big(e_{i,t}^2(\T)\big) \Big|\notag\\
   &\le&
   \frac{n''}{n''-n'+1}\sup_{\T\in\Theta}\Big|\frac{1}{n''}\sum_{t=1}^{n''}f_{\s M}\big(e^2_{i,t}(\T)\big)-\E\,f_{\s M}\big(e_{i,t}^2(\T)\big)\Big| \notag\\
   &&
   +
   \frac{n'-1}{n''-n'+1}\sup_{\T\in\Theta}\Big|\frac{1}{n'-1}\sum_{t=1}^{n'-1}f_{\s M}\big(e^2_{i,t}(\T)\big)-\E\,f_{\s M}\big(e_{i,t}^2(\T)\big)\Big|\notag\\
   &=&
   o(1)\quad a.s.
\end{eqnarray}
Furthermore, since $\TR$ converges almost surely to $\bar\T$ and $\E\,f_{\s M}(e_{i,t}^2(\T))$  is continuous in $\T$, we have by (\ref{cons.1}) that
\begin{eqnarray}\label{cons.2}
   \Big| \frac{1}{n''-n'+1}\sum_{t=n'}^{n''} f_{\s M}\big(e^2_{i,t}(\TR)\big) - \E\,f_{\s M}\big(e_{i,t}^2(\bar\T)\big) \Big| =o(1)\quad a.s.
\end{eqnarray}
Recall that  $|f_{\s M}(x)-f_{\s M}(y)|\leq |x-y|$ for all $x,y\geq 0$. Then, similarly as in the proof of (\ref{main.conv1}),  we have
\begin{eqnarray*}
    \frac{1}{n''-n'+1}\sum_{t=n'}^{n''} \big| f_{\s M}(\tilde e^2_{i,t}(\TR))-  f_{\s M}\big(e^2_{i,t}(\TR)\big)\big|
    &\les& \frac{1}{n''-n'+1} \sum_{t=n'}^{n''} \big|\tilde e_{i,t}^2(\TR) -e^2_{i,t}(\TR)\big|\\[5pt]
    &=& O\Big(\frac{1}{n''-n'+1}\Big)\quad a.s.,
\end{eqnarray*}
where $\tilde e_{i,t}^2(\T)=X_{i,t}^2/\tilde \sigma_t^2(\T)$ for $i=0,1$. 
Thus, we have by (\ref{cons.2}) that
\begin{equation}\label{cons.3}
  \frac{1}{n''-n'+1}\sum_{t=n'}^{n''} f_{\s M}(\tilde e^2_{i,t}(\TR))
  \stackrel{a.s.}{\longrightarrow}
  \E\,f_{\s M}\big(e_{i,t}^2(\bar\T)\big). 
\end{equation}
Using this, we can show the following. 
In the case of $k=\lfloor ns\rfloor<k^*$,
\begin{eqnarray*}\label{eq:def_Tnk}
  \frac{1}{n}T_{n,k} &:=&  \frac{1}{n} \big|D_{n,k}\big(f_{\s M}\big(\tilde e^2(\TR)\big)\big)\big| \\
   &=&
   \frac{1}{n}\Big|\sum_{t=1}^k f_{\s M}\big(\tilde e^2_t(\hat{\theta}_n^{\revR})\big) -\frac{k}{n}\sum_{t=1}^n f_{\s M}\big(\tilde e^2_t(\hat{\theta}_n^{\revR})\big)\Big|\\
  &=&  \Big|\frac{k}{n}\frac{n-k}{n}\frac{1}{k} \sum_{t=1}^k  f_{\s M}\big(\tilde e^2_{0,t}(\hat{\theta}_n^{\revR})\big) 
  -\frac{k}{n}\frac{k^*-k}{n}\frac{1}{k^*-k} \sum_{t=k+1}^{k^*}  f_{\s M}\big(\tilde e^2_{0,t}(\hat{\theta}_n^{\revR})\big)
   \end{eqnarray*}
\begin{eqnarray}\label{eq:def_Tnk}  
  &&\quad\quad -\frac{k}{n}\frac{n-k^*}{n}\frac{1}{n-k^*} \sum_{t=k^*+1}^n  f_{\s M}\big(\tilde e^2_{1,t}(\hat{\theta}_n^{\revR})\big)\Big| \notag\\
  && \hspace{-2cm}
  \stackrel{a.s.}{\longrightarrow}\ s(1-\lambda) C_{\s M}.
\end{eqnarray}
Similarly, it can be shown  that $\frac{1}{n}T_{n,k}$ converges almost surely to $\lambda(1-\lambda) C_{\s M} $ when $k=k^*$, and  to $\lambda(1-s) C_{\s M} $ when $k=\lfloor ns \rfloor>k^*$, respectively.\\
Without loss of generality, assume that $t>s$. Then, we have
\begin{eqnarray*}
   \limsup_{n\rightarrow \infty}  \Big|\frac{1}{n} T_{n,\lfloor nt \rfloor}-\frac{1}{n} T_{n,\lfloor ns \rfloor} \Big|
    &\leq&
   \limsup_{n\rightarrow \infty}\frac{1}{n}\Big|\sum_{i=\lfloor ns \rfloor+1}^{\lfloor nt \rfloor} f_{\s M}\big(\tilde e^2_i(\hat{\theta}_n^{\revR})\big) -\frac{\lfloor nt \rfloor-\lfloor ns \rfloor}{n}\sum_{i=1}^n f_{\s M}\big(\tilde e^2_i(\hat{\theta}_n^{\revR})\big)\Big|\\
    &\leq&
    2M\limsup_{n\rightarrow \infty} \Big|\frac{\lfloor nt \rfloor-\lfloor ns \rfloor}{n}\Big| =2M|t-s|.
\end{eqnarray*}
Hence,  $\big\{\frac{1}{n}T_{n,k}\big\}$ is asymptotically equicontinuous, and consequently, we obtain the following uniform convergence:
\begin{eqnarray}\label{conv.Tnk}
    \sup_{0\leq s \leq 1} \Big| \frac{1}{n} T_{n,\lfloor ns \rfloor} - T(s) \Big|=o(1)\quad a.s.,
\end{eqnarray}
where
\[
T(s) = \begin{cases}
s(1-\lambda) C_{M} &\text{if } s\leq\lambda, \\
\lambda(1-s) C_{M} &\text{if } s>\lambda.
\end{cases}
\]
Observing that $\lambda$ is the unique maximizer of $T(s)$, the first result in the theorem follows.  

Next, to prove the second assertion, put $\mu_{i,{\s M}} = \E f_{\s M} \big(e_{i,t}^2(\bar\T)\big)$ and
$\nu_{i,{\s M}}=\E f_{\s M}^2\big(e_{i,t}^2(\bar\T)\big)$, where $i=0,1$.
Applying (\ref{cons.3}) with $n'=1$ and $n''=k^*$ for $i=0$, and with $n'=k^*+1$ and $n''=n$ for $i=1$, we obtain
\begin{eqnarray}\label{cons.4}
\frac1n\sum_{t=1}^n
f_{\s M}\big(\tilde e_t^2(\hat\T_n^{\revR})\big)
\stackrel{a.s.}{\longrightarrow}
\lambda\mu_{0,{\s M}}+(1-\lambda)\mu_{1,{\s M}}.
\end{eqnarray}
Furthermore,  since $f_{\s M}^2$ is bounded and $|f_{\s M}^2(x)-f_{\s M}^2(y)|\leq 2M|x-y|$, the same argument used to
establish (\ref{cons.3}) and \eqref{cons.4} yields
\begin{eqnarray*}
\frac1n\sum_{t=1}^n
f_{\s M}^2\big(\tilde e_t^2(\hat\T_n^{\revR})\big)
\stackrel{a.s.}{\longrightarrow}
\lambda\nu_{0,{\s M}}+(1-\lambda)\nu_{1,{\s M}}.
\end{eqnarray*}
Thus, we have
\begin{eqnarray}\label{eq:tau_H1}
\hat\tau_{\s M}^2
\stackrel{a.s.}{\longrightarrow}
(\tau_{\s M}^*)^2:=\lambda\nu_{0,{\s M}}+(1-\lambda)\nu_{1,{\s M}}
-
\big(\lambda\mu_{0,{\s M}}+(1-\lambda)\mu_{1,{\s M}}\big)^2.
\end{eqnarray}
Here, we note by assumption {\bf A8} that
\begin{eqnarray*}
(\tau_{\s M}^*)^2
&=&
\lambda\Var\big[f_{\s M}\big(e_{0,t}^2(\bar\T)\big)\big]
+
(1-\lambda)\Var\big[f_{\s M}\big(e_{1,t}^2(\bar\T)\big)\big]
+
\lambda(1-\lambda)(\mu_{0,{\s M}}-\mu_{1,{\s M}})^2
\\
&\geq&
\lambda(1-\lambda)C_{\s M}^2
\ >\ 0.
\end{eqnarray*}
Moreover, we have by (\ref{conv.Tnk}) that
\begin{eqnarray*}
\frac1n\max_{1\leq k\leq n}T_{n,k}
\stackrel{a.s.}{\longrightarrow}
\max_{0\leq s\leq1}T(s)
=
\lambda(1-\lambda)C_{\s M}.
\end{eqnarray*}
Combining this with  \eqref{eq:tau_H1} yields the second assertion.
\end{proof}

As stated in \cite{shao2010testing}, proving
\begin{eqnarray*}
\frac{1}{n}\argmax_{1\leq k \leq n-1}\frac{\frac{1}{n}  D^2_{n,k}\big(f_{\s M}\big(\tilde e^2(\hat{\theta}_n^{\revR})\big)\big)}{\frac{1}{n^2}V_{n,k}\big(f_{\s M}\big(\tilde e^2(\hat{\theta}_n^{\revR})\big)\big) }\ \rightarrow \ \lambda
\end{eqnarray*}
appears challenging. 
However, it is noteworthy that, as shown in Theorem \ref{Thm:cons.T}, the value of $k$ that maximizes the numerator converges to the change point $k^*$. Hence, when the robust self-normalized test $SN_n^M(\TR)$ rejects the null hypothesis, the change point can be located as the argmax of the numerator. The following theorem establishes the consistency of $SN_n^M(\TR)$.
\begin{thm}\label{Thm:cons.SN}
Suppose that assumptions {\bf A1}--{\bf A5} hold for $\{X_{0,t}\}$ and $\{X_{1,t}\}$. If assumptions {\bf A7} and {\bf A8} are satisfied, then it holds that  under $H_1$,\begin{eqnarray*}
 SN_n^M(\TR) \stackrel{a.s.}{\longrightarrow}  \infty\quad\text{as }n\to\infty.
\end{eqnarray*}
\end{thm}
\begin{proof}
We first show that
\begin{eqnarray}\label{eq:con_Vnks}
&&\hspace{-1cm}
\frac{1}{n^3}V_{n,k^*}\big(f_{\s M}(\tilde e^2(\TR))\big)\notag\\
&=&
\frac{1}{n^3}\sum_{t=1}^{k^*} \Big(\sum_{j=1}^t \tilde f_{0,j} -\frac{t}{k^*}\sum_{j=1}^{k^*} \tilde f_{0,j}\Big)^2 +
\frac{1}{n^3}\sum_{t=k^*+1}^n\Big(\sum_{j=t}^n \tilde f_{1,j} -\frac{n-t+1}{n-k^*}\sum_{j=k^*+1}^n  \tilde f_{1,j}\Big)^2\notag\\
&\stackrel{a.s.}{\longrightarrow}& 0,
\end{eqnarray}
where $\tilde f_{i,j}=f_{\s M}(X^2_{i,j}/\tilde \sigma_j^2(\TR))$ for $i\in\{0,1\}$.
We just deal with the first term. The convergence result of the second term can be shown similarly. 

Now, let $f_{0,j}:=f_{\s M}(X^2_{0,j}/\sigma_j^2(\bar\T))$. Since $\{X_{0,t}\}$ and $\{\sigma_t^2(\bar\T)\}$ are strictly stationary and ergodic, $\{f_{0,t}\}$ is also stationary and ergodic. Hence, $\frac{1}{t}\sum_{j=1}^t f_{0,j}$ converges almost surely. Thus, one can see that for any $\eta>0$ and sufficiently large $n$,
\begin{eqnarray*}
\Big|\frac{1}{t}\sum_{j=1}^{t}f_{0,j}-\frac{1}{k^*}\sum_{j=1}^{k^*}f_{0,j}\Big|
\ \le\ \eta
\quad a.s.,
\end{eqnarray*}
where $t\geq\sqrt n$. Observe also that
$\big|\frac{1}{t}\sum_{j=1}^{t}f_{0,j}-\frac{1}{k^*}\sum_{j=1}^{k^*}f_{0,j}\big|
\le 2M.$ Using these, we obtain that for sufficiently large $n$,
\begin{eqnarray*}
&&\hspace{-1cm}
\frac{1}{n^3}\sum_{t=1}^{k^*}t^2\Big(\frac{1}{t}\sum_{j=1}^{t}f_{0,j}
-\frac{1}{k^*}\sum_{j=1}^{k^*}f_{0,j}\Big)^2\\
&=&
\frac{1}{n^3}\sum_{t=1}^{\sqrt n}t^2\Big(\frac{1}{t}\sum_{j=1}^{t}f_{0,j}
-\frac{1}{k^*}\sum_{j=1}^{k^*}f_{0,j}\Big)^2
+
\frac{1}{n^3}\sum_{t=\sqrt n+1}^{k^*}t^2\Big(\frac{1}{t}\sum_{j=1}^{t}f_{0,j}
-\frac{1}{k^*}\sum_{j=1}^{k^*}f_{0,j}\Big)^2\\
&\les&
\frac{1}{n^3}\sum_{t=1}^{\sqrt n}t^2+\frac{\eta}{n^3}\sum_{t=1}^{n}t^2
\ =\ O(n^{-3/2})+\eta\,O(1)\quad a.s.,
\end{eqnarray*}
which implies 
\begin{eqnarray}\label{conv.V3}
 \frac{1}{n^3}\sum_{t=1}^{k^*} \Big(\sum_{j=1}^t  f_{0,j} -\frac{t}{k^*}\sum_{j=1}^{k^*}  f_{0,j}\Big)^2 
 &\stackrel{a.s.}{\longrightarrow}& 0.
\end{eqnarray}

Next, putting $h_j(\T):=\frac{\pa}{\pa \T}f_{\s M}(X_{0,j}^2/\sigma_j^2(\T))$, we have
\begin{eqnarray}\label{split}
    \big|\tilde f_{0,j}-f_{0,j}\big|
    &\leq&\Big|f_{\s M}\Big( \frac{X^2_{0,j}}{\tilde \sigma_j^2(\TR)}\Big) -f_{\s M}\Big( \frac{X^2_{0,j}}{ \sigma_j^2(\TR)} \Big)\Big|
    +\Big|f_{\s M}\Big( \frac{X^2_{0,j}}{\sigma_j^2(\TR)}\Big)- f_{\s M}\Big(\frac{X^2_{0,j}}{ \sigma_j^2(\bar\T)}\Big)\Big| \nonumber\\
    &\leq&\Big|\frac{X^2_{0,j}}{\tilde \sigma_j^2(\TR)} - \frac{X^2_{0,j}}{\sigma_j^2(\TR)}\Big|
    + \big\|\TR-\bar\T\,\big\| \big\| h_j(\T^*_{j,n})\big\|\nonumber\\[3pt]
    &\les& V_0W_{0,j} X^2_{0,j}\rho^j
    + \big\|\TR-\bar\T\,\big\| \big\| h_j(\T^*_{j,n})\big\|,
\end{eqnarray}
where $V_0$ and $W_{0,j}$ are such that $\sup_{\T\in\Theta}|\tilde \sigma_t^2(\T)-\sigma_t^2(\T)|\leq V_0 W_{0,j}\rho^j$ for $\{X_{0,t}\}$ in assumption {\bf A3}, and  $\T^*_{j,n}$ is an intermediate point between $\TR$ and $\bar\T$. Using the same approach as in proving the convergence of $\frac{1}{n}\sum_{t=1}^n g_t(\T^*_{t,n})$ in Theorem \ref{Theorem.main}, it can be shown that $\frac{1}{n}\sum_{j=1}^n h_j(\T^*_{j,n})$ converges almost surely to $\E h_j(\bar\T)$. Furthermore, noting that $\tilde f_{0,j}$ and $f_{0,j}$ are bounded by $M$, we have, from elementary calculations and (\ref{split}), that 
\begin{align*}
    &\hspace{-1cm}
    \frac{1}{n^3}\sum_{t=1}^{k^*}\Big| \Big(\sum_{j=1}^t  \tilde f_{0,j} -\frac{t}{k^*}\sum_{j=1}^{k^*}  \tilde f_{0,j}\Big)^2 
    -\Big(\sum_{j=1}^t  f_{0,j} -\frac{t}{k^*}\sum_{j=1}^{k^*}  f_{0,j}\Big)^2 \Big|\\   
    &\les\
    \frac{1}{n^3}\sum_{t=1}^{n} \sum_{j=1}^n t \big|\tilde f_{0,j}-f_{0,j}\big|
     \displaybreak[4]\\
    &\les\
    \frac{1}{n} \sum_{j=1}^n V_0W_{0,j} X^2_{0,j}\rho^j + \big\|\TR-\bar\T\,\big\| \frac{1}{n} \sum_{j=1}^n\big\| h_j(\T^*_{j,n})\big\|\\
    &= O(n^{-1})+O\big(\big\|\TR-\bar\T\,\big\|\big)\quad a.s.,
\end{align*}
which together with (\ref{conv.V3}) asserts that the first term of the right side of \eqref{eq:con_Vnks} converges almost surely to zero.  

Finally, let $T_{n,k}$ be the one defined in \eqref{eq:def_Tnk}. Since we have by \eqref{conv.Tnk} that
\begin{equation*}\label{eq:conv_Tks}
    \frac{1}{n} T_{n,k^*}\stackrel{a.s.}{\longrightarrow} \lambda(1-\lambda)C_{\s M}\ >\ 0,
\end{equation*}
one can see that $T_{n,k^*}>0$ almost surely for all sufficiently large $n$. 
Observe also that $\{T_{n,k^*}>0\}\subset\cD_n^c$. Then, we have $SN_n^M(\TR)=\infty$ on $\{T_{n,k^*}>0\}\cap\cV_n^c$ and
\begin{eqnarray}\label{eq:SN_lbd}
    SN_n^M(\TR)
   \ \geq\ 
   \frac{\frac{1}{n^2} T^2_{n,k^*}} {\frac{1}{n^3}V_{n,k^*}\big(f_{\s M}\big(\tilde e^2(\TR)\big)\big)}
   \quad\text{on }\{T_{n,k^*}>0\}\cap \cV_n.
\end{eqnarray}
Since $\frac{1}{n^2}T^2_{n,k^*}$ converges almost surely to $\lambda^2(1-\lambda)^2C_{\s M}^2>0$, we
have by \eqref{eq:con_Vnks} that for any $K>0$ and all sufficiently
large $n$,
\[
\frac{1}{n^2}T^2_{n,k^*}>
K\cdot\frac{1}{n^3}V_{n,k^*}\big(f_{\s M}\big(\tilde e^2(\TR)\big)\big)\quad a.s.,
\]
and thus $SN_n^M(\TR)>K$ on $\{T_{n,k^*}>0\}$, which completes the proof.
\end{proof}

\begin{remark}\label{rmk:choice_M}
The choice of $M$ controls a trade-off between robustness and power. For any fixed $M$
with $\tau_{\s M}^2=\Var\big(f_{\s M}(\epsilon_1^2)\big)>0$, the truncated tests have the
same limiting null distributions as their non-truncated counterparts, and hence $M$ does
not affect their asymptotic size. Under $H_1$, however, the power depends on $M$.
Intuitively, a smaller $M$ provides stronger protection against outliers, but excessive
truncation may also remove information useful for detecting a genuine change, whereas a
larger $M$ preserves more of the original residuals but weakens the robustness.
Particularly for the CUSUM test, the relation between the power and $M$ can be seen more
precisely from Theorem \ref{Thm:cons.T}, where it is shown that $T_n^{\s M}(\TR)/\sqrt{n}$
converges almost surely to $\lambda(1-\lambda)C_{\s M}/\tau_{\s M}^*$. That is, since
$\lambda$ is fixed, the larger the ratio $C_{\s M}/\tau_{\s M}^*$ is, the higher the power
becomes. The optimal $M$ necessarily depends on the degree of contamination,
which is generally unknown in practice, and thus the theoretical identification of an optimal $M$ appears difficult and lies beyond the scope of this paper.
Instead, we provide some practical guidance. Since the innovations have unit variance, it is likely that  most of the residuals in the uncontaminated case are distributed over $[-3,3]$ and values outside $[-4,4]$ are quite rare. Hence, truncation with $M \le 4^2$ seems reasonable. This also explains why, in Table \ref{tab:no.out}, the proposed tests with $M=3^2$ or $4^2$ perform almost identically to the non-truncated ones. Under contamination, the tests with $M=3^2$ generally exhibit higher power than those with $M=4^2$. See Tables \ref{tab:IO.mild}--\ref{tab:AO.sev}. However, we also note that for the CUSUM test, the larger value $M=4^2$ can reduce the size distortion under severe AO contamination, at the cost of lower power. Although one should be careful in applying these findings to general cases, our limited simulation results suggest that choosing $M$ between $3^2$ and $4^2$ provides a reasonable balance between robustness and power. Accordingly, we recommend setting $M$ close to $3^2$ when no prior information on the contamination is available.
\end{remark}

\begin{remark}\label{rmk:test_choice}
As shown in Table \ref{tab:no.out}, in the absence of contamination, the robust CUSUM test $T_n^M(\hat\T_n^R)$ generally exhibits higher power
than the robust self-normalized test $SN_n^M(\hat\T_n^R)$ when the sample size is large, whereas the latter tends to perform slightly better for
smaller sample sizes. Similar behavior is also observed under IO contamination in Tables \ref{tab:IO.mild} and \ref{tab:IO.sev}. However, the situation is different under AO contamination. As shown in Tables
\ref{tab:AO.mild} and \ref{tab:AO.sev}, particularly when the process is
highly persistent, $T_n^M(\hat\T_n^R)$ can exhibit size distortions, which
become more pronounced under severe contamination, whereas
$SN_n^M(\hat\T_n^R)$ maintains relatively stable sizes and reasonable
power. Therefore, when data exhibit high persistence and are suspected to be
severely contaminated by additive outliers, we recommend using
$SN_n^M(\hat\T_n^R)$.
\end{remark}

\begin{remark}\label{rmk:multi.chg}
For both robust tests $T_n^M(\TR)$ and $SN_n^M(\TR)$, the change point is
estimated as
\[
\hat k_n
=
\argmax_{1\leq k \leq n}
\big|D_{n,k}\big(f_{\s M}\big(\tilde e^2(\TR)\big)\big)\big|.
\]
As noted in \cite{zhao2022}, a global self-normalized test can suffer
substantial power loss in the presence of multiple change points, because
other changes can make the self-normalizer large (cf. \cite{zhao2022}). Our additional simulations, not reported here, also show that such a power loss can occur, although its degree may depend on the pattern of the changes and can be particularly severe in certain cases. Since $SN_n^M(\TR)$ has the same self-normalization structure, it also appears to have this limitation. Therefore, when multiple changes are suspected, we recommend using $T_n^M(\TR)$ together with the binary segmentation procedure for detecting multiple change points. Specifically:
\begin{enumerate}
    \item[1.] Perform the test $T_n^M(\TR)$ on the whole series
    $\{X_1,\cdots,X_n\}$. If $H_0$ is rejected, split the series at
    $\hat k_n$ into two subseries
    $\{X_1,\cdots,X_{\hat k_n}\}$ and
    $\{X_{\hat k_n+1},\cdots,X_n\}$.
    
    \item[2.] Repeat this procedure for each subseries until no further
    change points are detected.
\end{enumerate}
For more details on the binary segmentation procedure with CUSUM-type tests, see \cite{aue:horvath:2013} and references therein.
\end{remark}

\section{Behavior of naive and robust tests in the presence of outliers}
\label{sec:outlier_behavior}

In this section, we provide an explanation for why the naive tests
break down in the presence of outliers and how our procedure, particularly
truncation, helps control size distortions and mitigate power loss.

We consider additive outliers (AOs) and innovation outliers (IOs) as
contamination mechanisms, which have been widely used in the time series
literature. See, for example, \cite{chang1988estimation},
\cite{tsay:1988}, and \cite{tsay2000outliers}. The main difference
between the two mechanisms is that an AO directly contaminates only the observation at the outlier time, whereas the effect of an IO propagates to subsequent observations.

Let $\{X_{t,0}\}$ be the uncontaminated process that would be obtained
from model (\ref{ts.model}) with the true parameter $\T_0$ if no outliers
occurred. Denote its conditional variance process by
$\{\sigma_{t,0}^2(\T_0)\}$. Then, we have $X_{t,0}=\sigma_{t,0}(\T_0)\epsilon_t$. Define the outlier indicator
$I_t=\mathbf{1}(t\in O_n)$, where $\mathbf 1(\cdot)$ is the indicator function and $O_n=\{r_1,\cdots,r_m\}, 1\le r_1<\cdots<r_m \le n,$ denotes the set of times at which outliers occur. Then, the AO- and
IO-contaminated observations  can be formulated as follows:
\begin{equation}\label{eq:AOIO}
\begin{aligned}
\mathrm{AO}:&\quad
X_t(s)
=
X_{t,0}
+s\zeta_t\,\mathrm{sign}(X_{t,0})I_t,
\\[3pt]
\mathrm{IO}:&\quad
X_t(s)
=
\sigma_t(\T_0;s)
\big\{
\epsilon_t
+s\zeta_t\,\mathrm{sign}(\epsilon_t)I_t
\big\},
\end{aligned}
\end{equation}
where $s>0$ is a constant controlling the overall magnitude of the outliers
and $\{\zeta_t\}$ is a sequence of positive random variables representing
their relative magnitudes. Note that $s\zeta_t$ is the magnitude of the
outlier shock at time $t$. The conditional variance
$\sigma_t^2(\T_0;s)$ is given by the same conditional variance function as
$\sigma_{t,0}^2(\T_0)$, but is evaluated using the contaminated past
observations. We note that
$X_t(0)=X_{t,0}$ and $\sigma_t^2(\T_0;0)=\sigma_{t,0}^2(\T_0)$.

Let $\hat\T_n(s)$ and $\hat\T_n^R(s)$ denote the usual and robust estimators,
respectively, computed using the contaminated sample
$\{X_t(s)\}_{t=1}^n$. To treat both types of estimators in a unified manner,
we write $\hat\T_n^\circ(s)$ for either $\hat\T_n(s)$ or
$\hat\T_n^R(s)$. We similarly use  $\hat\T_n^\circ(0)$ to denote the corresponding estimator based on the uncontaminated sample $\{X_t(0)\}_{t=1}^n$. Furthermore, we define
\begin{eqnarray*}
\tilde Z_t(s;\T):=\frac{X_t^2(s)}{\tilde\sigma_t^2(\T;s)}.
\end{eqnarray*}
Then, the corresponding squared residual is given by $\tilde Z_t(s):=\tilde Z_t(s;\hat\T_n^\circ(s))$.
Since $\tilde Z_t(0)$ is the squared residual from the uncontaminated sample, we can express
the distortion of the squared residual as
\[
u_t(s)=\tilde Z_t(s)-\tilde Z_t(0).
\]
Write $\tilde Z(s):=(\tilde Z_1(s),\cdots,\tilde Z_n(s))$ and $f_{\s M}(\tilde Z(s)):=\big(f_{\s M}(\tilde Z_1(s)),\cdots,f_{\s M}(\tilde Z_n(s))\big)$.
Then, the non-truncated CUSUM and self-normalized statistics computed
from the contaminated sample can be written as
\[
T_n(\hat\T_n^\circ(s))
=
\frac{1}
{\sqrt n\,\hat\tau_n(s)}
\max_{1\leq k\leq n}
|D_{n,k}(\tilde Z(s))|
\quad\text{and}\quad
SN_n(\hat\T_n^\circ(s))
=
\max_{1\leq k\leq n-1}
\frac{nD_{n,k}^2(\tilde Z(s))}
{V_{n,k}(\tilde Z(s))},
\]
where $\hat\tau_n^2(s)$ is the sample variance of
$\{\tilde Z_t(s)\}_{t=1}^n$. The corresponding truncated statistics
$T_n^M(\hat\T_n^\circ(s))$ and $SN_n^M(\hat\T_n^\circ(s))$ are obtained by replacing $\tilde Z(s)$ with
$f_{\s M}(\tilde Z(s))$.
\subsection{Breakdown of the naive tests}
\label{subsec:naive_breakdown}

Throughout this subsection, the sample size $n$, the uncontaminated sample $\{X_t(0)\}_{t=1}^n$, and the outlier
set $O_n$ are fixed, and only the outlier magnitude $s$ is allowed to
increase. We assume that, for each $t=1,\cdots,n$,
\begin{eqnarray}\label{eq:ut}
u_t(s)=s^2\xi_t+r_t(s),
\end{eqnarray}
where $\{\xi_t\}_{t=1}^n$ is a sequence of nonnegative random
variables and $r_t(s)=o(s^2)$ as $s\to\infty$. Although it is difficult to verify this condition in general,
we impose it in order to explain, in a simple manner, the breakdown of the
naive tests as the outlier magnitude $s$ increases. A sufficient condition
for (\ref{eq:ut}) under the AO and IO cases is given in the following
remark.

\begin{remark}\label{rmk:AOIO}
Let
\[
a_{t,{\s A}}(s)
:=
\frac{1}
{\tilde\sigma_t^2(\hat\T_n^\circ(s);s)}
\qquad\text{and}\qquad
a_{t,{\s I}}(s)
:=
\frac{\sigma_t^2(\T_0;s)}
{\tilde\sigma_t^2(\hat\T_n^\circ(s);s)},
\]
and suppose that for each $t$, $a_{t,{\s A}}(s)$ and $a_{t,{\s I}}(s)$ converge to
finite limits $a_{t,{\s A}}$ and $a_{t,{\s I}}$, respectively, as
$s\rightarrow\infty$. For the AO-contaminated sample
$\{X_t(s)\}_{t=1}^n$ in (\ref{eq:AOIO}), since 
\[
\tilde Z_t(s)
=
\frac{1}
{\tilde\sigma_t^2(\hat\T_n^\circ(s);s)}
\big(|X_{t,0}|+s\zeta_t I_t\big)^2,
\]
we have
\begin{eqnarray*}
\frac{u_t(s)}{s^2}
=
\frac{1}
{\tilde\sigma_t^2(\hat\T_n^\circ(s);s)}
\Big(
\frac{|X_{t,0}|}{s}+\zeta_t I_t
\Big)^2
-\frac{\tilde Z_t(0)}{s^2}
\ \rightarrow\ 
a_{t,{\s A}}\zeta_t^2 I_t
\quad\text{as }s\rightarrow\infty.
\end{eqnarray*}
Similarly, for the IO-contaminated case, we have
\begin{eqnarray*}
\frac{u_t(s)}{s^2}
=
\frac{\sigma_t^2(\T_0;s)}
{\tilde\sigma_t^2(\hat\T_n^\circ(s);s)}
\Big(
\frac{|\epsilon_t|}{s}+\zeta_t I_t
\Big)^2
-\frac{\tilde Z_t(0)}{s^2}
\ \rightarrow\ 
a_{t,{\s I}}\zeta_t^2 I_t
\quad\text{as }s\rightarrow\infty.
\end{eqnarray*}
Hence, we can see that (\ref{eq:ut}) holds with
$\xi_t=a_{t,{\s A}}\zeta_t^2I_t$ and $\xi_t=a_{t,{\s I}}\zeta_t^2I_t$
for the AO- and IO-contaminated cases, respectively.
\end{remark}

\begin{remark}
For the GARCH(1,1) model, one can derive explicit forms of $\xi_t$
under the assumption that
\[
\hat\T_n^\circ(s)
\ \rightarrow\
\T_\infty=(\omega_\infty,\alpha_\infty,\beta_\infty)
\quad\text{as }s\to\infty,
\]
where $\alpha_\infty>0$ and $0<\beta_\infty<1$. Although the compactness
of $\Theta$ guarantees that $\hat\T_n^\circ(s)$ has convergent
subsequences as $s\to\infty$, establishing convergence of the full
sequence is difficult because its limiting behavior may depend on the model
and the contamination pattern. We therefore impose the above convergence
condition to obtain explicit expressions for $\xi_t$. In the AO and IO
cases, $\xi_t$ is given by
\begin{equation}\label{eq:G_AOIO}
\xi_t=c_{\s A}\mathbf{1}(t=r_1)
\qquad\text{and}\qquad
\xi_t=c_{t,{\s I}}\mathbf{1}(t\in O_n),
\end{equation}
respectively.
The detailed forms of $c_{\s A}$ and $c_{t,{\s I}}$ are given in Section S3 of the
supplement.
\end{remark}

Now, let $v_{\xi }^2$ denote the sample variance of $\xi_1,\cdots,\xi_n$ and put $\xi:=(\xi_1,\cdots,\xi_n)$. We first describe the behavior of the naive tests.  The proof of the following proposition is provided in the supplementary material.

\begin{prop}\label{prop:naive-breakdown}
Suppose that (\ref{eq:ut}) holds and that $v_\xi^2>0$. Then, we have
\[
T_n(\hat\T_n^\circ(s))
\ \rightarrow\
R_n(\xi)
:=
\frac{1}
{\sqrt n\,v_{\xi}}
\max_{1\leq k\leq n}
|D_{n,k}(\xi)|
\quad\text{as }s\rightarrow\infty.
\]
Moreover, provided that $V_{n,k}(\xi)>0$ for every
$1\leq k\leq n-1$, we have
\[
SN_n(\hat\T_n^\circ(s))
\ \rightarrow\
Q_n(\xi)
:=
\max_{1\leq k\leq n-1}
\frac{nD_{n,k}^2(\xi)}
{V_{n,k}(\xi)}
\quad\text{as }s\rightarrow\infty.
\]
\end{prop}

The proposition states that, as the outlier magnitude becomes large, the
decisions of the non-truncated tests are eventually determined by
$\{\xi_t\}_{t=1}^n$, regardless of whether there is a parameter change in the
uncontaminated sample $\{X_t(0)\}_{t=1}^n$. More specifically, if $R_n(\xi)$ (resp. $Q_n(\xi)$) is below the critical value for $T_n$ (resp. $SN_n$) from Corollary \ref{Cor:Tn} (resp. Corollary \ref{Cor:SN}), then the test becomes unlikely to reject $H_0$ as $s$ increases, resulting in a conservative size under $H_0$ and a loss of power under $H_1$. In contrast, if $R_n(\xi)$ (resp. $Q_n(\xi)$) exceeds its corresponding critical value, $T_n$ (resp. $SN_n$) becomes more likely to reject even in the absence of a parameter change, leading to size inflation under $H_0$.

For the GARCH(1,1) model, $\xi_t$ under AO contamination is given by
(\ref{eq:G_AOIO}). It is noteworthy that only $r_1$ appears in $\xi_t$. Thus,
when $s$ is large, the behavior of the test is governed only by the first
outlier. In this case, it can be shown that
\[
R_n(\xi)
=
\frac{\max\{r_1-1,n-r_1\}}{\sqrt{n(n-1)}}
\quad\text{and}\quad
Q_n(\xi)
=\begin{cases}
    \infty &\text{if }r_1 \in \{1,n\},\\
  \displaystyle  \max_{1\leq k\leq n-1}q_{n,k}(r_1) &\text{if }1<r_1<n,
\end{cases}
\]
where 
\[
q_{n,k}(r_1)
=
\begin{cases}
\displaystyle
\frac{k^2(n-k)^2}
{n (J_{r_1-k-1}+J_{n-r_1})}
&\text{if }1\leq k<r_1,\\[12pt]
\displaystyle
\frac{k^2(n-k)^2}
{n(J_{r_1-1}+J_{k-r_1})}
&\text{if }r_1\leq k\leq n-1
\end{cases}
\]
and $J_\ell=\sum_{j=1}^{\ell}j^2$ with $J_0=0$.
Since $R_n(\xi)<1$ and the critical value of $T_n$ at the 10\% significance level is 1.224, the non-truncated CUSUM test is likely to produce a value below 1 as $s$ increases, regardless of the location of the first AO and the number of subsequent AOs. This explains the undersized values and substantial power losses of the naive CUSUM test $T_n$ observed in Tables \ref{tab:AO.mild} and \ref{tab:AO.sev} below. 

Unlike $R_n(\xi)$, however, $Q_n(\xi)$ can be strongly affected by the location
of the first AO. By some calculations, it can be shown that $Q_n(\xi)$ is minimized when the first AO is located near the center of the sample, with the minimum converging to 1.5 as $n\to\infty$, and increases rapidly as $r_1$ approaches either boundary of the sample. Recall that the critical value of $SN_n$ at the 10\% significance level is 29.6. Thus, one can expect that the non-truncated self-normalized test is unlikely to reject when the first AO is located near the center of the sample, and vice versa when it is located near a boundary.

Under IO contamination, all outlier times can generally contribute to $\xi_t$, as seen in \eqref{eq:G_AOIO}. Hence, $R_n(\xi)$ and $Q_n(\xi)$ depend on both the locations of the IOs and the corresponding coefficients, making their behavior more complicated than in the AO case. Hence, we focus only on the effect of the IO locations by simplifying the coefficients as $c_{r_j,{\s I}}=c_{\s I}$ for all $j$. See Subsection S3.2.2 of the supplement for detailed results.

\subsection{Robustness of the proposed tests}
\label{subsec:rob_cont}
 For notational simplicity, put
\[
T_n^M(s)
:=
T_n^M(\hat\T_n^\circ(s))
\qquad\text{and}\qquad
SN_n^M(s)
:=
SN_n^M(\hat\T_n^\circ(s))
\]
without causing confusion with $T_n^M(\TR)$ and $SN_n^M(\TR)$ in Section \ref{sec:robust_tests},
and let
\[
B_n^M(s)
:=
\sum_{t=1}^n\big|f_{\s M}(\tilde Z_t(s))-f_{\s M}(\tilde Z_t(0))\big|.
\]
Further, $\hat\tau^2_{\s M}(s)$ denotes the sample variance of $f_{\s M}(\tilde Z_1(s)),\cdots,f_{\s M}(\tilde Z_n(s))$, and we consider the case in which $\hat\tau^2_{\s M}(0)>0$. 
The following proposition provides the key bounds used to establish the robustness of the proposed tests. Its proof is given in the supplement.
\begin{prop}\label{prop:sparse}
For any $n\geq2$ and $s>0$, it holds that 
\begin{equation}\label{eq:T_uncond}
\big|T_n^M(s)-T_n^M(0)\big|
\ \leq\
\frac{1}{\hat\tau_{\s M}(0)} \Big(1+\frac{M}{\hat\tau_{\s M}(0)}\Big)
\frac{B_n^M(s)}{\sqrt{n}},
\end{equation}
and
\begin{equation}\label{eq:SN_uncond}
\Big|\sqrt{SN_n^M(s)}-\sqrt{SN_n^M(0)}\Big|
\ \leq\
\Big(1+\sqrt{SN_n^M(0)}\,\Big)
\frac{B_n^M(s)/\sqrt{n}}{\sqrt{ \underline V_n^M(0)}-B_n^M(s)/\sqrt{n}},
\end{equation}
provided that $\underline V_n^M(0)=\min_{1\leq k\leq n-1}\frac{1}{n^2}V_{n,k}\big(f_{\s M}(\tilde Z(0))\big)>0$ and $B_n^M(s)<\sqrt{ n\underline V_n^M(0)}$.
\end{prop}
In the following theorems, we allow the number and size of the outliers to depend on $n$, that is, $m=m_n$ and $s=s_n$ for some sequences $\{m_n\}$ and $\{s_n\}$.

\begin{thm}\label{thm:robust_size}
Suppose that the assumptions of Theorem \ref{Theorem.main} hold under $H_0$.  If $B_n^M(s_n)=o_{\uP}(\sqrt{n})$, then $T_n^M(s_n)$ and $SN_n^M(s_n)$ have the same limiting distributions as $T_n^M(0)$ and $SN_n^M(0)$, given in Theorems \ref{Theorem.main} and \ref{Theorem.main.SN}, respectively.
\end{thm}

\begin{proof}
Since $\hat\tau_{\s M}(0) \to \tau_{\s M}>0$ in probability by \eqref{tau.conv} and $B_n^M(s_n)=o_{\uP}(\sqrt{n})$, it follows from \eqref{eq:T_uncond} that $T_n^M(s_n)-T_n^M(0)=o_{\uP}(1)$, and thus $T_n^M(s_n)$ has the same limiting distribution as $T_n^M(0)$.

Next, recall that $\underline V_n^M$ defined in \eqref{eq:d_V_nM} is indeed $\underline V_n^M(0)$ with $\hat\T^\circ_n(0)=\hat\T_n(0)$, and note that  \eqref{eq:d_V_nM} remains valid  for $\hat\T^\circ_n(0)=\TR(0)$. 
Hence, we have 
\begin{equation*}
  \underline V_n^M(0)\ \stackrel{d}{\longrightarrow}\ \inf_{0\le t\le1}V_{\tau_{\s M} B^o}(t).
\end{equation*}
Since the limit above is positive almost surely by  \eqref{eq:inf_BB} and $SN_n^M(0)=O_{\uP}(1)$ by Theorem \ref{Theorem.main.SN}, one can see that the right side of  \eqref{eq:SN_uncond} converges in probability to zero. This completes the proof. 
\end{proof}
For consistency under $H_1$, we assume that the objective function associated with $\hat\T_n^\circ(0)$ satisfies the uniform convergence in \eqref{obj.H1}, and that  {\bf A7} holds for its limiting objective function. Consequently,
$\hat\T_n^\circ(0)$ converges almost surely to $\bar\T$, and the intermediate results
established in the proof of Theorems \ref{Thm:cons.T} and \ref{Thm:cons.SN} also hold for $\hat\T_n^\circ(0)$.
\begin{thm}\label{thm:robust_cons}
Suppose that $B_n^M(s_n)=o_{\uP}(n)$ under $H_1$. Then, it holds that
\[
T_n^M(s_n)\stackrel{\uP}{\longrightarrow}\infty
\qquad\text{and}\qquad
SN_n^M(s_n)\stackrel{\uP}{\longrightarrow}\infty
\]
under the assumptions of Theorems \ref{Thm:cons.T} and \ref{Thm:cons.SN}, respectively.
\end{thm}
\begin{proof}
By \eqref{eq:tau_H1}, $\hat\tau_{\s M}(0)$ converges in probability to $\tau_{\s M}^*>0$. Since $B_n^M(s_n)=o_{\uP}(n)$, it follows  from \eqref{eq:T_uncond} that
\[
\frac{1}{\sqrt n}
\big|T_n^M(s_n)-T_n^M(0)\big|
\ \leq\
\frac{1}{\hat\tau_{\s M}(0)} \Big(1+\frac{M}{\hat\tau_{\s M}(0)}\Big)
\frac{B_n^M(s_n)}{n}
=o_{\uP}(1).
\]
We also have by Theorem \ref{Thm:cons.T} that
\[
\frac{T_n^M(0)}{\sqrt n}\
\stackrel{\uP}{\longrightarrow}\
\frac{\lambda(1-\lambda)C_{\s M}}{\tau_{\s M}^*}>0.
\]
Combining these two results, the first assertion is yielded. 
 
Next, put $Y_s:=f_{\s M}(\tilde Z(s))$ for brevity. 
By (S2) and (S3) of the supplement, we have
\begin{equation}\label{eq:ingred}
\Big|\sqrt{V_{n,k^*}(Y_{s_n})}-\sqrt{V_{n,k^*}(Y_0)}\Big|
\ \leq\ \sqrt{n}\,B_n^M(s_n)
\quad\text{and}\quad
\big|D_{n,k^*}(Y_{s_n})-D_{n,k^*}(Y_0)\big|
\ \leq\ B_n^M(s_n),
\end{equation}
where $k^*$ is the change point under $H_1$.
Moreover, as shown in the proof of Theorems \ref{Thm:cons.T} and \ref{Thm:cons.SN}, one can see that
\[
\frac{1}{n}\big|D_{n,k^*}(Y_0)\big|\stackrel{\uP}{\longrightarrow}\lambda(1-\lambda)C_{\s M}>0
\quad\text{and}\quad
\frac{1}{n^3}V_{n,k^*}(Y_0)\stackrel{\uP}{\longrightarrow}0.
\]
Then, using \eqref{eq:ingred} and $B_n^M(s_n)=o_{\uP}(n)$, we obtain
\[
\frac{1}{n}\big|D_{n,k^*}(Y_{s_n})\big|
\stackrel{\uP}{\longrightarrow}\lambda(1-\lambda)C_{\s M}
\]
and
\[
\frac{1}{n^3}V_{n,k^*}(Y_{s_n})
\ \leq\ 
\bigg(\sqrt{\frac{V_{n,k^*}(Y_0)}{n^3}}+\frac{B_n^M(s_n)}{n}\bigg)^2
=o_{\uP}(1),
\]
from which the second assertion follows.
\end{proof}

As stated in the introduction, our two-step procedure is an intuitive way to reduce the effect of outliers, and the theorems above provide theoretical support for this under the conditions on $B_n^M(s_n)$. Since $B_n^M(s_n)=o_{\uP}(\sqrt n)$ implies $B_n^M(s_n)=o_{\uP}(n)$, we finally discuss sufficient conditions for the former. For this, note that 
\begin{eqnarray*}
B_n^M(s)
&\leq&
\sum_{t=1}^n\big|f_{\s M}\big(\tilde Z_t(s;\hat\T_n^\circ(s))\big)
-f_{\s M}\big(\tilde Z_t(0;\hat\T_n^\circ(s))\big)\big|
+
\sum_{t=1}^n\big|f_{\s M}\big(\tilde Z_t(0;\hat\T_n^\circ(s))\big)
-f_{\s M}\big(\tilde Z_t(0;\hat\T_n^\circ(0))\big)\big|\\
&=:&
\Psi_n^M(s)+\Gamma_n^M(s).
\end{eqnarray*}
Letting $ Z_t(s;\T):=\frac{X_t^2(s)}{\sigma_t^2(\T;s)}$, we have
\begin{eqnarray*}
\Gamma_n^M(s)
&\leq&
\sum_{t=1}^n\big|f_{\s M}\big(\tilde Z_t(0;\hat\T_n^\circ(s))\big)
-f_{\s M}\big(Z_t(0;\hat\T_n^\circ(s))\big)\big|
 +
\sum_{t=1}^n\big|f_{\s M}\big(Z_t(0;\hat\T_n^\circ(s))\big)
-f_{\s M}\big(Z_t(0;\hat\T_n^\circ(0))\big)\big|\\
&&+
\sum_{t=1}^n\big|f_{\s M}\big(Z_t(0;\hat\T_n^\circ(0))\big)
-f_{\s M}\big(\tilde Z_t(0;\hat\T_n^\circ(0))\big)\big|\\
&=:& 
\Gamma_{n,1}^M(s)+\Gamma_{n,2}^M(s)+\Gamma_{n,3}^M(s).
\end{eqnarray*}
Observe that \eqref{diff.ep} remains valid when $\TR$ is replaced by any estimator taking values in $\Theta$. Hence, using the Lipschitz continuity of $f_{\s M}$, we obtain
\begin{equation*}
\Gamma_{n,1}^M(s)+\Gamma_{n,3}^M(s)=O_{\uP}(1).
\end{equation*}
Following the same argument used for $II_t$ in \eqref{fm.split}, we also have
\begin{equation*}
\Gamma_{n,2}^M(s)
=
O_{\uP}\big(n\big\|\hat\T_n^\circ(s)-\hat\T_n^\circ(0)\big\|\big).
\end{equation*}
Consequently, we have
\begin{equation*}
B_n^M(s)
\ \le\
\Psi_n^M(s)+O_{\uP}(1)+O_{\uP}\big(n\big\|\hat\T_n^\circ(s)-\hat\T_n^\circ(0)\big\|\big).
\end{equation*}
\enlargethispage{2\baselineskip}

Thus, a set of sufficient conditions for $B_n^M(s_n)=o_{\uP}(\sqrt n)$ is
\begin{equation*}\label{eq:B_suff}
\Psi_n^M(s_n)=o_{\uP}(\sqrt n)
\qquad\text{and}\qquad
\sqrt n\,
\big\|\hat\T_n^\circ(s_n)-\hat\T_n^\circ(0)\big\|
=o_{\uP}(1).
\end{equation*}

The first condition requires the cumulative distortion of the truncated residuals to be small, more exactly, relative to $\sqrt{n}$. This may occur when only a small number of residuals are distorted, or when the individual distortions are sufficiently small. As mentioned above, outliers affect the fitted conditional variances at the subsequent times. However, this propagated effect is expected to decay rapidly in many conditionally heteroscedastic models because the fitted conditional variances are usually given by a recursion in which the influence of a past observation diminishes at each step. Moreover, since outliers are usually rare relative to the sample size in empirical applications, the number of distorted terms is likely to be small. Hence, the first condition seems plausible in practice. The second condition also seems reasonable, particularly when the degree of contamination is not too severe, because a robust estimator based on contaminated observations is expected to yield an estimate close to that obtained from the uncontaminated ones.  This also indicates that not only the truncation, but also the robustness of the employed estimator plays an important role in achieving the robustness of the proposed tests.

\begin{table}[!t]
\renewcommand\arraystretch{1.12}
\tabcolsep=5pt
  \centering
  \caption{Empirical sizes and powers of the naive tests and proposed robust tests without contamination}\label{tab:no.out}
  {\scriptsize
    \begin{tabular}{cccccccccccccccc}
    \hline
          & \multicolumn{3}{c}{size} &       & \multicolumn{11}{c}{power} \\
\cline{2-4}\cline{6-16}          & \multicolumn{3}{c}{$\T=(1,0.3,0.4)$} &       & \multicolumn{3}{c}{$\omega: 1 \rightarrow 1.5$} &       & \multicolumn{3}{c}{$\alpha: 0.3 \rightarrow 0.5$} &       & \multicolumn{3}{c}{$\beta: 0.4 \rightarrow 0.2$} \\
    n     & 500   & 1000  & 2000  &       & 500   & 1000  & 2000  &       & 500   & 1000  & 2000  &       & 500   & 1000  & 2000 \\
    \hline
    $T_n$ & 0.030 & 0.043 & 0.052 &       & 0.308 & 0.682 & 0.947 &       & 0.291 & 0.633 & 0.935 &       & 0.428 & 0.837 & 0.992 \\
    $T_n^9(\hat\T_n)$ & 0.031 & 0.043 & 0.050 &       & 0.308 & 0.677 & 0.947 &       & 0.295 & 0.633 & 0.936 &       & 0.436 & 0.838 & 0.990 \\
    $T_n^{16}(\hat\T_n)$ & 0.030 & 0.043 & 0.052 &       & 0.308 & 0.683 & 0.947 &       & 0.291 & 0.632 & 0.935 &       & 0.429 & 0.837 & 0.992 \\
    $T_n^9(\hat\T^R_n)$ & 0.031 & 0.040 & 0.048 &       & 0.309 & 0.674 & 0.946 &       & 0.296 & 0.637 & 0.938 &       & 0.430 & 0.842 & 0.990 \\
    $T_n^{16}(\hat\T^R_n)$ & 0.030 & 0.042 & 0.051 &       & 0.307 & 0.677 & 0.948 &       & 0.290 & 0.635 & 0.939 &       & 0.432 & 0.840 & 0.993 \\
    \hline
    $SN_n$ & 0.052 & 0.062 & 0.049 &       & 0.323 & 0.588 & 0.822 &       & 0.306 & 0.560 & 0.827 &       & 0.434 & 0.729 & 0.940 \\
    $SN_n^9(\hat\T_n)$ & 0.057 & 0.060 & 0.046 &       & 0.322 & 0.593 & 0.822 &       & 0.315 & 0.566 & 0.830 &       & 0.431 & 0.725 & 0.944 \\
    $SN_n^{16}(\hat\T_n)$ & 0.053 & 0.062 & 0.049 &       & 0.323 & 0.590 & 0.822 &       & 0.307 & 0.560 & 0.828 &       & 0.434 & 0.728 & 0.942 \\
    $SN_n^9(\hat\T^R_n)$ & 0.055 & 0.061 & 0.047 &       & 0.323 & 0.591 & 0.824 &       & 0.311 & 0.564 & 0.832 &       & 0.434 & 0.724 & 0.944 \\
    $SN_n^{16}(\hat\T^R_n)$ & 0.052 & 0.061 & 0.051 &       & 0.319 & 0.593 & 0.824 &       & 0.303 & 0.560 & 0.829 &       & 0.433 & 0.729 & 0.941 \\
    \hline
          & \multicolumn{3}{c}{size} &       & \multicolumn{11}{c}{power} \\
\cline{2-4}\cline{6-16}          & \multicolumn{3}{c}{$\T=(1,0.1,0.85)$} &       & \multicolumn{3}{c}{$\omega: 1 \rightarrow 2$} &       & \multicolumn{3}{c}{$\alpha: 0.1 \rightarrow 0.05$} &       & \multicolumn{3}{c}{$\beta: 0.85 \rightarrow 0.8$} \\
    n     & 500   & 1000  & 2000  &       & 500   & 1000  & 2000  &       & 500   & 1000  & 2000  &       & 500   & 1000  & 2000 \\
    \hline
    $T_n$ & 0.058 & 0.053 & 0.049 &       & 0.119 & 0.393 & 0.895 &       & 0.357 & 0.660 & 0.977 &       & 0.326 & 0.651 & 0.970 \\
    $T_n^9(\hat\T_n)$ & 0.060 & 0.051 & 0.048 &       & 0.138 & 0.405 & 0.896 &       & 0.365 & 0.662 & 0.975 &       & 0.324 & 0.646 & 0.968 \\
    $T_n^{16}(\hat\T_n)$ & 0.058 & 0.052 & 0.049 &       & 0.121 & 0.395 & 0.897 &       & 0.358 & 0.660 & 0.977 &       & 0.326 & 0.651 & 0.969 \\
    $T_n^9(\hat\T^R_n)$ & 0.057 & 0.050 & 0.047 &       & 0.115 & 0.384 & 0.882 &       & 0.364 & 0.649 & 0.973 &       & 0.310 & 0.643 & 0.967 \\
    $T_n^{16}(\hat\T^R_n)$ & 0.057 & 0.051 & 0.047 &       & 0.107 & 0.377 & 0.877 &       & 0.362 & 0.655 & 0.975 &       & 0.311 & 0.644 & 0.969 \\
    \hline
    $SN_n$ & 0.050 & 0.049 & 0.052 &       & 0.207 & 0.481 & 0.828 &       & 0.377 & 0.642 & 0.920 &       & 0.345 & 0.664 & 0.913 \\
    $SN_n^9(\hat\T_n)$ & 0.049 & 0.049 & 0.052 &       & 0.211 & 0.484 & 0.832 &       & 0.381 & 0.642 & 0.923 &       & 0.353 & 0.665 & 0.916 \\
    $SN_n^{16}(\hat\T_n)$ & 0.050 & 0.050 & 0.052 &       & 0.209 & 0.484 & 0.833 &       & 0.376 & 0.648 & 0.920 &       & 0.348 & 0.665 & 0.915 \\
    $SN_n^9(\hat\T^R_n)$ & 0.051 & 0.048 & 0.052 &       & 0.206 & 0.482 & 0.826 &       & 0.382 & 0.634 & 0.922 &       & 0.343 & 0.665 & 0.915 \\
    $SN_n^{16}(\hat\T^R_n)$ & 0.052 & 0.050 & 0.051 &       & 0.202 & 0.475 & 0.825 &       & 0.376 & 0.635 & 0.921 &       & 0.335 & 0.666 & 0.912 \\
    \hline
    \end{tabular}
    }
\end{table}%

\section{Simulation study}\label{sec:simul}
In this section, we evaluate the performance of $T_n^M(\hat\T_n^R)$ and $SN_n^M(\hat\T_n^R)$ and compare them with the naive tests $T_n$ and $SN_n$ within the following GARCH(1,1) model with the parameter $\T=(\omega, \alpha,\beta)$:
\[X_{0,t}=\sigma_t(\T)\ep_t,\quad \sigma_t^2(\T)=\omega+\alpha X_{0,t-1}^2+\beta \sigma_{t-1}^2(\T),\]
where $\{\ep_t\}$ is a sequence of i.i.d. 
random variables from $N(0,1)$. Additionally, we assess the performance of $T_n^M(\hat\T_n)$ and $SN_n^M(\hat\T_n)$, where only truncation is applied and $\hat\T_n$ is the QMLE.

We consider the AOs and IOs introduced in Section \ref{sec:outlier_behavior} as
contamination mechanisms. The sample $\{X_t\}$ contaminated by IO is generated by replacing $\ep_t$ with $\ep_t + s\cdot\text{sign} (\ep_t) P_t$, where $P_t$ are i.i.d. Bernoulli random variables with a success probability of $p$. The AO-contaminated sample is generated as $X_t=X_{0,t} + s\sqrt{\omega/(1-\alpha-\beta)}\cdot \text{sign} (X_{0,t}) P_t$. We consider $(p,s)=(0.5\%,5)$ and $(1\%,10)$ to evaluate performance under scenarios of mild and severe contamination.
\begin{table}[!t]
\renewcommand\arraystretch{1.12}
\tabcolsep=5pt
  \centering
  \caption{Empirical sizes and powers of the naive tests and proposed robust tests with IO contamination at $p=$0.5\% and $s=$5 }\label{tab:IO.mild}
{\scriptsize
    \begin{tabular}{cccccccccccccccc}
    \hline
          & \multicolumn{3}{c}{size} &       & \multicolumn{11}{c}{power} \\
\cline{2-4}\cline{6-16}           & \multicolumn{3}{c}{$\T=(1,0.3,0.4)$} &       & \multicolumn{3}{c}{$\omega: 1 \rightarrow 1.5$} &       & \multicolumn{3}{c}{$\alpha: 0.3 \rightarrow 0.5$} &       & \multicolumn{3}{c}{$\beta: 0.4 \rightarrow 0.2$} \\
    n     & 500   & 1000  & 2000  &       & 500   & 1000  & 2000  &       & 500   & 1000  & 2000  &       & 500   & 1000  & 2000 \\
    \hline
    $T_n$ & 0.021 & 0.030 & 0.030 &       & 0.087 & 0.212 & \tcr{0.467} &       & 0.095 & 0.230 & \tcr{0.487} &       & 0.173 & 0.373 & \tcr{0.702} \\
    $T_n^9(\hat\T_n)$ & 0.043 & 0.047 & 0.052 &       & 0.242 & 0.538 & 0.853 &       & 0.250 & 0.556 & 0.881 &       & 0.388 & 0.760 & 0.977 \\
    $T_n^{16}(\hat\T_n)$ & 0.037 & 0.039 & 0.042 &       & 0.161 & 0.381 & 0.684 &       & 0.184 & 0.393 & 0.743 &       & 0.278 & 0.576 & 0.889 \\
    $T_n^9(\hat\T^R_n)$ & 0.030 & 0.044 & 0.045 &       & 0.234 & 0.562 & \tcb{0.873} &       & 0.249 & 0.567 & \tcb{0.901} &       & 0.413 & 0.792 & \tcb{0.985} \\
    $T_n^{16}(\hat\T^R_n)$ & 0.025 & 0.044 & 0.037 &       & 0.168 & 0.416 & 0.727 &       & 0.182 & 0.416 & 0.768 &       & 0.307 & 0.633 & 0.923 \\
    \hline
    $SN_n$ & 0.056 & 0.058 & 0.047 &       & 0.145 & 0.248 & \tcr{0.420} &       & 0.149 & 0.251 & \tcr{0.415} &       & 0.209 & 0.375 & \tcr{0.577} \\
    $SN_n^9(\hat\T_n)$ & 0.048 & 0.063 & 0.049 &       & 0.266 & 0.460 & 0.728 &       & 0.261 & 0.455 & 0.769 &       & 0.388 & 0.668 & 0.889 \\
    $SN_n^{16}(\hat\T_n)$ & 0.050 & 0.066 & 0.048 &       & 0.201 & 0.350 & 0.576 &       & 0.197 & 0.348 & 0.636 &       & 0.295 & 0.523 & 0.779 \\
    $SN_n^9(\hat\T^R_n)$ & 0.049 & 0.064 & 0.048 &       & 0.285 & 0.489 & \tcb{0.744} &       & 0.283 & 0.474 & \tcb{0.790} &       & 0.407 & 0.689 & \tcb{0.900} \\
    $SN_n^{16}(\hat\T^R_n)$ & 0.050 & 0.063 & 0.046 &       & 0.216 & 0.370 & 0.633 &       & 0.214 & 0.369 & 0.671 &       & 0.314 & 0.581 & 0.808 \\
    \hline
          & \multicolumn{3}{c}{size} &       & \multicolumn{11}{c}{power} \\
\cline{2-4}\cline{6-16}           & \multicolumn{3}{c}{$\T=(1,0.1,0.85)$} &       & \multicolumn{3}{c}{$\omega: 1 \rightarrow 2$} &       & \multicolumn{3}{c}{$\alpha: 0.1 \rightarrow 0.05$} &       & \multicolumn{3}{c}{$\beta: 0.85 \rightarrow 0.8$} \\
    n     & 500   & 1000  & 2000  &       & 500   & 1000  & 2000  &       & 500   & 1000  & 2000  &       & 500   & 1000  & 2000 \\
    \hline
    $T_n$ & 0.029 & 0.038 & 0.027 &       & 0.051 & 0.059 & \tcr{0.236} &       & 0.141 & 0.224 & \tcr{0.592} &       & 0.132 & 0.209 & \tcr{0.523} \\
    $T_n^9(\hat\T_n)$ & 0.071 & 0.068 & 0.060 &       & 0.167 & 0.298 & 0.678 &       & 0.386 & 0.647 & 0.942 &       & 0.346 & 0.612 & 0.912 \\
    $T_n^{16}(\hat\T_n)$ & 0.057 & 0.052 & 0.047 &       & 0.097 & 0.178 & 0.493 &       & 0.254 & 0.468 & 0.858 &       & 0.221 & 0.409 & 0.788 \\
    $T_n^9(\hat\T^R_n)$ & 0.048 & 0.042 & 0.045 &       & 0.112 & 0.290 & \tcb{0.737} &       & 0.350 & 0.702 & \tcb{0.984} &       & 0.317 & 0.641 & \tcb{0.973} \\
    $T_n^{16}(\hat\T^R_n)$ & 0.049 & 0.042 & 0.034 &       & 0.064 & 0.159 & 0.527 &       & 0.240 & 0.515 & 0.910 &       & 0.216 & 0.450 & 0.867 \\
    \hline
    $SN_n$ & 0.058 & 0.063 & 0.047 &       & 0.112 & 0.165 & \tcr{0.339} &       & 0.199 & 0.318 & \tcr{0.607} &       & 0.193 & 0.304 & \tcr{0.547} \\
    $SN_n^9(\hat\T_n)$ & 0.061 & 0.058 & 0.051 &       & 0.167 & 0.342 & 0.656 &       & 0.362 & 0.571 & 0.913 &       & 0.340 & 0.564 & 0.885 \\
    $SN_n^{16}(\hat\T_n)$ & 0.059 & 0.059 & 0.035 &       & 0.131 & 0.252 & 0.496 &       & 0.278 & 0.453 & 0.791 &       & 0.265 & 0.436 & 0.741 \\
    $SN_n^9(\hat\T^R_n)$ & 0.064 & 0.058 & 0.048 &       & 0.154 & 0.380 & \tcb{0.687} &       & 0.360 & 0.626 & \tcb{0.928} &       & 0.358 & 0.600 & \tcb{0.898} \\
    $SN_n^{16}(\hat\T^R_n)$ & 0.058 & 0.055 & 0.040 &       & 0.128 & 0.281 & 0.550 &       & 0.286 & 0.494 & 0.843 &       & 0.282 & 0.483 & 0.772 \\
    \hline
    \end{tabular}
    }
  \label{tab:addlabel}%
\end{table}

As a robust estimator for $T_n^M(\hat\T_n^R)$ and $SN_n^M(\hat\T_n^R)$, we consider the minimum density power divergence estimator (MDPDE) introduced by \cite{lee:song:2009}. The MDPDE is defined as a minimizer of the empirical version of the density power divergence with a control parameter, say $\gamma$. This  estimator is $\sqrt{n}$-consistent and has a strong robust property with little loss in asymptotic efficiency relative to the MLE when $\gamma$ is close to zero.  In this simulation, we set $\gamma=0.1$. For more details on the MDPDE for GARCH models, we refer to \cite{lee:song:2009}. Meanwhile, since the error term in the GARCH models has unit variance, ideal residuals are expected to range usually between $-4$ and $4$. Hence,  we consider $M=3^2$ and $4^2$ for $f_{M,\delta}$. Further, since all results in Section \ref{sec:robust_tests} hold for any $\delta>0$ and since $f_{M,\delta}$ converges to $f^{tr}_{\s M}$ in (\ref{f.tr}) as $\delta\rightarrow 0$, we use $f^{tr}_{\s M}$ instead of $f_{M,\delta}$ for truncation.

To examine the empirical sizes, two parameters $\T=(1,0.3,0.4)$ and $(1,0.1,0.85)$ are considered. The latter is employed to assess performance in a more volatile situation, which often arises in real data analysis. For empirical powers, we change the parameter at the midpoint $t=n/2$. To remove initialization effects, 1000 initial observations are discarded, and empirical sizes and powers are calculated from 2000 repetitions at a significance level of 5\%. 

Table \ref{tab:no.out} presents the results under no contamination (i.e., $p=0$). It can be seen that all tests produce appropriate empirical sizes and reasonable powers. The naive test $T_n$ (resp. $SN_n$) and the robust tests $T_n^M(\hat\T_n)$, $T_n^M(\hat\T^R_n)$ (resp. $SN_n^M(\hat\T_n), SN_n^M(\hat\T_n^R))$ perform almost similarly. Notably, the residual-based CUSUM tests, both naive and robust, yield higher powers compared to the self-normalized tests when $n=2000$. However, when $n=500$, the self-normalized tests perform slightly better than the CUSUM tests.
\begin{table}[!t]
\renewcommand\arraystretch{1.12}
\tabcolsep=5pt
  \centering
  \caption{Empirical sizes and powers of the naive tests and proposed robust tests with IO contamination at $p=$1\% and $s=$10 }\label{tab:IO.sev}
  {\scriptsize
    \begin{tabular}{cccccccccccccccc}
    \hline
          & \multicolumn{3}{c}{size} &       & \multicolumn{11}{c}{power} \\
\cline{2-4}\cline{6-16}          & \multicolumn{3}{c}{$\T=(1,0.3,0.4)$} &       & \multicolumn{3}{c}{$\omega: 1 \rightarrow 1.5$} &       & \multicolumn{3}{c}{$\alpha: 0.3 \rightarrow 0.5$} &       & \multicolumn{3}{c}{$\beta: 0.4 \rightarrow 0.2$} \\
    n     & 500   & 1000  & 2000  &       & 500   & 1000  & 2000  &       & 500   & 1000  & 2000  &       & 500   & 1000  & 2000 \\
    \hline
    $T_n$ & 0.018 & 0.029 & 0.041 &       & 0.027 & 0.060 & \tcr{0.122} &       & 0.043 & 0.099 & \tcr{0.206} &       & 0.029 & 0.079 & \tcr{0.168} \\
    $T_n^9(\hat\T_n)$ & 0.080 & 0.077 & \tco{0.096} &       & 0.209 & 0.347 & 0.577 &       & 0.419 & 0.667 & 0.893 &       & 0.327 & 0.583 & 0.842 \\
    $T_n^{16}(\hat\T_n)$ & 0.044 & 0.060 & 0.068 &       & 0.115 & 0.190 & 0.285 &       & 0.232 & 0.374 & 0.592 &       & 0.165 & 0.313 & 0.494 \\
    $T_n^9(\hat\T^R_n)$ & 0.031 & 0.038 & 0.054 &       & 0.231 & 0.448 & \tcb{0.797} &       & 0.247 & 0.598 & \tcb{0.902} &       & 0.433 & 0.792 & \tcb{0.986} \\
    $T_n^{16}(\hat\T^R_n)$ & 0.024 & 0.032 & 0.051 &       & 0.164 & 0.305 & 0.558 &       & 0.142 & 0.365 & 0.707 &       & 0.257 & 0.555 & 0.884 \\
    \hline
    $SN_n$ & 0.070 & 0.051 & 0.066 &       & 0.073 & 0.104 & \tcr{0.141} &       & 0.070 & 0.098 & \tcr{0.181} &       & 0.100 & 0.132 & \tcr{0.202} \\
    $SN_n^9(\hat\T_n)$ & 0.048 & 0.048 & 0.062 &       & 0.153 & 0.260 & 0.452 &       & 0.222 & 0.381 & 0.602 &       & 0.255 & 0.429 & 0.691 \\
    $SN_n^{16}(\hat\T_n)$ & 0.055 & 0.054 & 0.067 &       & 0.111 & 0.165 & 0.239 &       & 0.148 & 0.218 & 0.364 &       & 0.158 & 0.256 & 0.401 \\
    $SN_n^9(\hat\T^R_n)$ & 0.049 & 0.048 & 0.062 &       & 0.242 & 0.408 & \tcb{0.692} &       & 0.296 & 0.506 & \tcb{0.768} &       & 0.406 & 0.645 &\tcb{0.914} \\
    $SN_n^{16}(\hat\T^R_n)$ & 0.046 & 0.049 & 0.060 &       & 0.191 & 0.299 & 0.526 &       & 0.191 & 0.348 & 0.565 &       & 0.287 & 0.471 & 0.758 \\
    \hline
          & \multicolumn{3}{c}{size} &       & \multicolumn{11}{c}{power} \\
\cline{2-4}\cline{6-16}         & \multicolumn{3}{c}{$\T=(1,0.1,0.85)$} &       & \multicolumn{3}{c}{$\omega: 1 \rightarrow 2$} &       & \multicolumn{3}{c}{$\alpha: 0.1 \rightarrow 0.05$} &       & \multicolumn{3}{c}{$\beta: 0.85 \rightarrow 0.8$} \\
    n     & 500   & 1000  & 2000  &       & 500   & 1000  & 2000  &       & 500   & 1000  & 2000  &       & 500   & 1000  & 2000 \\
    \hline
    $T_n$ & 0.053 & 0.078 & \tco{0.096} &       & 0.047 & 0.089 & \tcr{0.178} &       & 0.060 & 0.127 & \tcr{0.276} &       & 0.068 & 0.128 & \tcr{0.234} \\
    $T_n^9(\hat\T_n)$ & 0.396 & 0.436 & \tco{0.348} &       & 0.443 & 0.556 & 0.680 &       & 0.641 & 0.840 & 0.973 &       & 0.531 & 0.730 & 0.928 \\
    $T_n^{16}(\hat\T_n)$ & 0.213 & 0.252 & \tco{0.238} &       & 0.260 & 0.325 & 0.450 &       & 0.377 & 0.523 & 0.756 &       & 0.324 & 0.461 & 0.681 \\
    $T_n^9(\hat\T^R_n)$ & 0.044 & 0.051 & 0.037 &       & 0.052 & 0.142 & \tcb{0.337} &       & 0.508 & 0.903 & \tcb{0.997} &       & 0.320 & 0.683 & \tcb{0.977} \\
    $T_n^{16}(\hat\T^R_n)$ & 0.039 & 0.046 & 0.046 &       & 0.029 & 0.079 & 0.191 &       & 0.307 & 0.669 & 0.970 &       & 0.191 & 0.447 & 0.838 \\
    \hline
    $SN_n$ & 0.083 & 0.084 & 0.062 &       & 0.076 & 0.091 & \tcr{0.109} &       & 0.122 & 0.143 & \tcr{0.247} &       & 0.124 & 0.134 & \tcr{0.220} \\
    $SN_n^9(\hat\T_n)$ & 0.101 & 0.098 & 0.068 &       & 0.145 & 0.166 & 0.241 &       & 0.364 & 0.552 & 0.799 &       & 0.259 & 0.427 & 0.671 \\
    $SN_n^{16}(\hat\T_n)$ & 0.090 & 0.088 & 0.060 &       & 0.106 & 0.129 & 0.176 &       & 0.224 & 0.331 & 0.535 &       & 0.188 & 0.270 & 0.437 \\
    $SN_n^9(\hat\T^R_n)$ & 0.066 & 0.071 & 0.050 &       & 0.083 & 0.180 & \tcb{0.338} &       & 0.461 & 0.784 & \tcb{0.959} &       & 0.365 & 0.631 & \tcb{0.911} \\
    $SN_n^{16}(\hat\T^R_n)$ & 0.059 & 0.071 & 0.051 &       & 0.065 & 0.125 & 0.247 &       & 0.335 & 0.603 & 0.874 &       & 0.239 & 0.460 & 0.778 \\
    \hline
    \end{tabular}
    }
  \label{tab:addlabel}%
\end{table}%
Table \ref{tab:IO.mild} presents the results for IO contamination cases with $ p=0.5\%$ and $s=5$. First, we observe that $T_n$ tends to produce undersized values, while the remaining tests yield sizes close to the 5\% significance level. This size deflation of $T_n$ is discussed in Subsection \ref{subsec:naive_breakdown}. Overall, except for the conservative behavior of $T_n$, no serious size distortion is observed for any of the tests. Compared to the results in Table \ref{tab:no.out}, it is evident that $T_n$ and $SN_n$ experience substantial power losses. However, the power losses for the robust tests are comparatively minor, and in some cases, the power of the robust tests is even slightly higher than in the uncontaminated case. It is also noteworthy that the tests with only truncation applied, $T_n^M(\hat\T_n)$ and $SN_n^M(\hat\T_n)$, are quite robust. However, these tests show some power losses compared to the fully robustified tests $T_n^M(\hat\T^R_n)$ and $SN_n^M(\hat\T^R_n)$ when the degree of contamination is severe (see the results in Table \ref{tab:IO.sev} below).
 Additionally, the robust CUSUM tests $T_n^M(\hat\T_n)$ and $T_n^M(\hat\T_n^R)$ still outperform the robust self-normalized tests $SN_n^M(\hat\T_n)$ and $SN_n^M(\hat\T_n^R)$, respectively, when $n$ is large. It should be noted that the robust tests with $M=9$ show higher powers than those with $M=16$, and the fully robustified tests using the robust estimator $\hat\T_n^R$ tend to exhibit more power than those applying only truncation. Interestingly, $T_n^9(\hat\T_n)$ (resp. $SN_n^9(\hat\T_n)$) generally outperforms $T_n^{16}(\hat\T_n^R)$  (resp. $SN_n^{16}(\hat\T^R_n)$). These trends are similarly observed in the following contamination case. In sum, for this mild IO contamination case, $T_n^9(\hat\T_n^R)$ shows the strongest overall performance in most cases, particularly for large $n$.

\begin{table}[!t]
\renewcommand\arraystretch{1.12}
\tabcolsep=5pt
  \centering
  \caption{Empirical sizes and powers of the naive tests and proposed robust tests with AO contamination at $p=$0.5\% and $s=$5 }\label{tab:AO.mild}
{\scriptsize
    \begin{tabular}{cccccccccccccccc}
    \hline
          & \multicolumn{3}{c}{size} &       & \multicolumn{11}{c}{power} \\
\cline{2-4}\cline{6-16}          & \multicolumn{3}{c}{$\T=(1,0.3,0.4)$} &       & \multicolumn{3}{c}{$\omega: 1 \rightarrow 1.5$} &       & \multicolumn{3}{c}{$\alpha: 0.3 \rightarrow 0.5$} &       & \multicolumn{3}{c}{$\beta: 0.4 \rightarrow 0.2$} \\
    n     & 500   & 1000  & 2000  &       & 500   & 1000  & 2000  &       & 500   & 1000  & 2000  &       & 500   & 1000  & 2000 \\
\hline
    $T_n$                  & 0.012 & 0.031 & 0.022 & &0.090 & 0.212 & \tcr{0.414} & &0.097 & 0.173 & \tcr{0.388} & &0.116 & 0.221 & \tcr{0.355} \\
    $T_n^9(\hat\T_n)$      & 0.061 & 0.080 & 0.072 & &0.284 & 0.624 & \tcb{0.919} & &0.320 & 0.673 & \tcb{0.945}& & 0.462 & 0.811 & 0.987 \\
    $T_n^{16}(\hat\T_n)$   & 0.037 & 0.054 & 0.051& & 0.182 & 0.466 & 0.797& & 0.211 & 0.509 & 0.841& & 0.324 & 0.657 & 0.913 \\
    $T_n^9(\hat\T^R_n)$    & 0.033 & 0.056 & 0.050 & &0.269 & 0.619 & 0.916 & &0.255 & 0.574 & 0.913 & &0.480 & 0.826 & \tcb{0.990} \\
    $T_n^{16}(\hat\T^R_n)$ & 0.024 & 0.044 & 0.036 & &0.173 & 0.475 & 0.811& & 0.165 & 0.414 & 0.803 & &0.338 & 0.707 & 0.941 \\
\hline
    $SN_n$                 & 0.063 & 0.056 & 0.044 && 0.159 & 0.248 & \tcr{0.428} && 0.142 & 0.253 & \tcr{0.386} &       & 0.155 & 0.228 & \tcr{0.334} \\
    $SN_n^9(\hat\T_n)$     & 0.056 & 0.053 & 0.051 &       & 0.289 & 0.537 & 0.778 &       & 0.305 & 0.507 & \tcb{0.779} &       & 0.398 & 0.651 & 0.891 \\
    $SN_n^{16}(\hat\T_n)$  & 0.050 & 0.055 & 0.053 &       & 0.239 & 0.459 & 0.697 &       & 0.246 & 0.425 & 0.681 &       & 0.317 & 0.541 & 0.803 \\
    $SN_n^9(\hat\T^R_n)$   & 0.052 & 0.055 & 0.052 &       & 0.310 & 0.562 & \tcb{0.799} &       & 0.301 & 0.500 & 0.778 &       & 0.419 & 0.692 & \tcb{0.906} \\
    $SN_n^{16}(\hat\T^R_n)$ & 0.045 & 0.053 & 0.053 &       & 0.257 & 0.483 & 0.721 &       & 0.242 & 0.420 & 0.673 &       & 0.345 & 0.590 & 0.844 \\
    \hline
          & \multicolumn{3}{c}{size} &       & \multicolumn{11}{c}{power} \\
\cline{2-4}\cline{6-16}          & \multicolumn{3}{c}{$\T=(1,0.1,0.85)$} &       & \multicolumn{3}{c}{$\omega: 1 \rightarrow 2$} &       & \multicolumn{3}{c}{$\alpha: 0.1 \rightarrow 0.05$} &       & \multicolumn{3}{c}{$\beta: 0.85 \rightarrow 0.8$} \\
    n     & 500   & 1000  & 2000  &       & 500   & 1000  & 2000  &       & 500   & 1000  & 2000  &       & 500   & 1000  & 2000 \\
    \hline
    $T_n$                  & 0.022 & 0.020 & 0.020 &       & 0.052 & 0.117 & \tcr{0.401} &       & 0.101 & 0.151 & \tcr{0.329} &       & 0.094 & 0.141 & \tcr{0.285} \\
    $T_n^9(\hat\T_n)$      & 0.172 & 0.182 & \tco{0.237} &       & 0.368 & 0.697 & 0.948 &       & 0.598 & 0.851 & 0.985 &       & 0.579 & 0.850 & 0.980 \\
    $T_n^{16}(\hat\T_n)$   & 0.081 & 0.103 & \tco{0.119} &       & 0.208 & 0.505 & 0.893 &       & 0.441 & 0.769 & 0.973 &       & 0.415 & 0.742 & 0.960 \\
    $T_n^9(\hat\T^R_n)$    & 0.111 & 0.133 & \tco{0.140} &       & 0.271 & 0.628 & 0.938 &       & 0.559 & 0.865 & 0.993 &       & 0.526 & 0.857 & 0.988 \\
    $T_n^{16}(\hat\T^R_n)$ & 0.051 & 0.067 & 0.070 &       & 0.101 & 0.358 & \tcb{0.847} &       & 0.403 & 0.764 & \tcb{0.987} &       & 0.361 & 0.741 & \tcb{0.980} \\
    \hline
    $SN_n$                  & 0.044 & 0.052 & 0.054 &       & 0.118 & 0.272 & \tcr{0.564} &       & 0.173 & 0.235 & \tcr{0.428} &       & 0.165 & 0.243 & \tcr{0.415} \\
    $SN_n^9(\hat\T_n)$      & 0.065 & 0.068 & 0.056 &       & 0.269 & 0.516 & 0.826 &       & 0.421 & 0.639 & 0.882 &       & 0.386 & 0.637 & 0.873 \\
    $SN_n^{16}(\hat\T_n)$   & 0.059 & 0.062 & 0.059 &       & 0.217 & 0.461 & 0.789 &       & 0.341 & 0.577 & 0.852 &       & 0.334 & 0.587 & 0.847 \\
    $SN_n^9(\hat\T^R_n)$    & 0.067 & 0.070 & 0.056 &       & 0.250 & 0.508 & \tcb{0.827} &       & 0.397 & 0.636 & \tcb{0.889} &       & 0.383 & 0.644 & \tcb{0.884} \\
    $SN_n^{16}(\hat\T^R_n)$ & 0.062 & 0.064 & 0.056 &       & 0.207 & 0.454 & 0.794 &       & 0.347 & 0.596 & 0.878 &       & 0.352 & 0.616 & 0.881 \\
    \hline
    \end{tabular}
    }
\end{table}%

In Table \ref{tab:IO.sev}, which presents the results for a more severe contamination case, we can observe that the power losses of the naive tests become substantial, while $T_n^9(\hat\T_n^R)$ and $SN_n^9(\hat\T_n^R)$ exhibit comparatively higher powers in most cases while maintaining stable sizes, demonstrating their robustness even under significant contamination. It is important to note that, unlike the mild contamination case above,  the robust tests $T_n^9(\hat\T_n)$ and $T_n^{16}(\hat\T_n)$ show severe size distortions in the highly volatile case of $\theta = (1, 0.1, 0.85)$,  highlighting the limitations of applying truncation alone without a robust estimator.  Notably, the robust and naive self-normalized tests, as well as the fully robust CUSUM tests $T_n^9(\hat\T_n^R)$ and $T_n^{16}(\hat\T_n^R)$, exhibit no size distortions when $n$ is large. As in the case above, $T_n^9(\hat\T_n^R)$ generally performs best among the tests yielding no size distortions.

While, in the IO contamination cases above, $T_n^M(\hat\T_n^R)$ maintains stable sizes and performs better than $SN_n^M(\hat\T_n^R)$, this test is observed to yield unstable sizes in the following AO contamination cases, particularly when the process is highly volatile and the degree of contamination is severe. We can clearly see this in the size columns in Tables \ref{tab:AO.mild} and \ref{tab:AO.sev}, which report the empirical sizes and powers for the mild and severe AO contamination cases, respectively. Although, for the moderately volatile case of $\T = (1, 0.3, 0.4)$, $T_n^M(\hat\T_n)$ and $T_n^M(\hat\T_n^R)$ exhibit stable sizes under mild contamination, as shown in Table \ref{tab:AO.mild}, both tests begin to exhibit some size distortions under severe contamination, as seen in Table \ref{tab:AO.sev}. In contrast, $SN_n^M(\hat\T_n)$ and $SN_n^M(\hat\T_n^R)$ provide substantially more stable size than the robust CUSUM tests, particularly as the sample size increases. Notably, $SN_n^9(\hat\T_n^R)$ demonstrates good powers across all cases presented in both Tables \ref{tab:AO.mild} and \ref{tab:AO.sev}. Meanwhile, compared to the results in Table \ref{tab:IO.mild}, $T_n$ and $SN_n$ show more significant power losses in most cases, indicating that the naive tests are more sensitive to AO contamination.
\begin{table}[!t]
\renewcommand\arraystretch{1.12}
\tabcolsep=5pt
  \centering
  \caption{Empirical sizes and powers of the naive tests and proposed robust tests with AO contamination at $p=1\%$ and $s=10$}\label{tab:AO.sev}
  {\scriptsize
    \begin{tabular}{cccccccccccccccc}
    \hline
          & \multicolumn{3}{c}{size} &       & \multicolumn{11}{c}{power} \\
\cline{2-4}\cline{6-16}          & \multicolumn{3}{c}{$\T=(1,0.3,0.4)$} &       & \multicolumn{3}{c}{$\omega: 1 \rightarrow 1.5$} &       & \multicolumn{3}{c}{$\alpha: 0.3 \rightarrow 0.5$} &       & \multicolumn{3}{c}{$\beta: 0.4 \rightarrow 0.2$} \\
    n & 500 & 1000 & 2000 &  & 500 & 1000 & 2000 &  & 500 & 1000 & 2000 &  & 500 & 1000 & 2000 \\
    \hline
    $T_n$ & 0.006 & 0.015 & 0.021 &  & 0.005 & 0.019 & \tcr{0.053} &  & 0.009 & 0.022 & \tcr{0.051} &  & 0.012 & 0.027 & \tcr{0.059} \\
    $T_n^9(\hat\T_n)$ & 0.134 & 0.134 & \tco{0.157} &  & 0.234 & 0.500 & 0.790 &  & 0.404 & 0.731 & 0.954 &  & 0.411 & 0.676 & 0.910 \\
    $T_n^{16}(\hat\T_n)$ & 0.047 & 0.069 & 0.069 &  & 0.097 & 0.244 & 0.502 &  & 0.183 & 0.439 & 0.738 &  & 0.193 & 0.358 & 0.613 \\
    $T_n^9(\hat\T^R_n)$ & 0.142 & 0.128 & \tco{0.142} &  & 0.303 & 0.592 & \tcb{0.863} &  & 0.351 & 0.680 & \tcb{0.932} &  & 0.504 & 0.767 & \tcb{0.951} \\
    $T_n^{16}(\hat\T^R_n)$ & 0.064 & 0.074 & 0.074 &  & 0.161 & 0.361 & 0.660 &  & 0.166 & 0.433 & 0.766 &  & 0.290 & 0.510 & 0.764 \\
    \hline
    $SN_n$ & 0.060 & 0.058 & 0.048 &  & 0.045 & 0.057 & \tcr{0.088} &  & 0.054 & 0.068 & \tcr{0.097} &  & 0.062 & 0.062 & \tcr{0.096} \\
    $SN_n^9(\hat\T_n)$ & 0.072 & 0.057 & 0.057 &  & 0.149 & 0.300 & 0.515 &  & 0.214 & 0.432 & 0.713 &  & 0.298 & 0.455 & 0.708 \\
    $SN_n^{16}(\hat\T_n)$ & 0.056 & 0.065 & 0.048 &  & 0.095 & 0.190 & 0.353 &  & 0.141 & 0.307 & 0.528 &  & 0.178 & 0.281 & 0.471 \\
    $SN_n^9(\hat\T^R_n)$ & 0.071 & 0.060 & 0.051 &  & 0.197 & 0.374 & \tcb{0.640} &  & 0.258 & 0.489 & \tcb{0.751} &  & 0.342 & 0.527 & \tcb{0.779} \\
    $SN_n^{16}(\hat\T^R_n)$ & 0.059 & 0.063 & 0.048 &  & 0.146 & 0.287 & 0.498 &  & 0.180 & 0.377 & 0.629 &  & 0.235 & 0.377 & 0.597 \\
    \hline
          & \multicolumn{3}{c}{size} &       & \multicolumn{11}{c}{power} \\
\cline{2-4}\cline{6-16}          & \multicolumn{3}{c}{$\T=(1,0.1,0.85)$} &       & \multicolumn{3}{c}{$\omega: 1 \rightarrow 2$} &       & \multicolumn{3}{c}{$\alpha: 0.1 \rightarrow 0.05$} &       & \multicolumn{3}{c}{$\beta: 0.85 \rightarrow 0.8$} \\
    n & 500 & 1000 & 2000 &  & 500 & 1000 & 2000 &  & 500 & 1000 & 2000 &  & 500 & 1000 & 2000 \\
    \hline
    $T_n$ & 0.009 & 0.012 & 0.014 &  & 0.006 & 0.017 & \tcr{0.060} &  & 0.010 & 0.029 & \tcr{0.051} &  & 0.014 & 0.021 & \tcr{0.060} \\
    $T_n^9(\hat\T_n)$ & 0.292 & 0.378 & \tco{0.453} &  & 0.505 & 0.794 & 0.967 &  & 0.609 & 0.854 & 0.983 &  & 0.618 & 0.855 & 0.979 \\
    $T_n^{16}(\hat\T_n)$ & 0.120 & 0.169 & \tco{0.192} &  & 0.258 & 0.553 & 0.859 &  & 0.336 & 0.574 & 0.829 &  & 0.340 & 0.571 & 0.834 \\
    $T_n^9(\hat\T^R_n)$ & 0.334 & 0.424 & \tco{0.528} &  & 0.572 & 0.855 & \tcb{0.982} &  & 0.702 & 0.905 & \tcb{0.988} &  & 0.708 & 0.895 & \tcb{0.988} \\
    $T_n^{16}(\hat\T^R_n)$ & 0.158 & 0.221 & \tco{0.304} &  & 0.322 & 0.680 & 0.925 &  & 0.459 & 0.714 & 0.895 &  & 0.468 & 0.684 & 0.899 \\
    \hline
    $SN_n$ & 0.043 & 0.038 & 0.037 &  & 0.033 & 0.053 & \tcr{0.119} &  & 0.054 & 0.087 & \tcr{0.117} &  & 0.077 & 0.069 & \tcr{0.128} \\
    $SN_n^9(\hat\T_n)$ & 0.108 & 0.079 & 0.067 &  & 0.220 & 0.363 & 0.625 &  & 0.368 & 0.553 & 0.771 &  & 0.383 & 0.535 & 0.761 \\
    $SN_n^{16}(\hat\T_n)$ & 0.072 & 0.060 & 0.057 &  & 0.147 & 0.284 & 0.519 &  & 0.234 & 0.413 & 0.597 &  & 0.263 & 0.377 & 0.593 \\
    $SN_n^9(\hat\T^R_n)$ & 0.109 & 0.080 & 0.064 &  & 0.249 & 0.447 & \tcb{0.710} &  & 0.401 & 0.580 & \tcb{0.799} &  & 0.413 & 0.556 & \tcb{0.784} \\
    $SN_n^{16}(\hat\T^R_n)$ & 0.077 & 0.067 & 0.054 &  & 0.197 & 0.398 & 0.664 &  & 0.281 & 0.469 & 0.683 &  & 0.322 & 0.435 & 0.663 \\
    \hline
    \end{tabular}
    }
\end{table}%

Our empirical findings are summarized as follows: (i) In the absence of
contamination, the naive and proposed tests perform similarly. (ii) The naive tests $T_n$ and $SN_n$ tend to become conservative and exhibit significant power losses in the presence of outliers, which is consistent with the breakdown mechanism described in Subsection \ref{subsec:naive_breakdown}, whereas our proposed tests
$T_n^M(\hat\T_n^R)$ and $SN_n^M(\hat\T_n^R)$ demonstrate strong robustness.
(iii) Truncation alone is often effective under mild contamination, but
robust estimation becomes increasingly important under severe contamination and
high persistence. (iv) The fully robust CUSUM test $T_n^M(\hat\T_n^R)$ is highly effective under IO contamination, outperforming the robust self-normalized test $SN_n^M(\hat\T_n^R)$. (v) $SN_n^M(\hat\T_n^R)$ generally performs well across all cases and is particularly reliable under severe AO contamination and highly volatile scenarios, where $T_n^M(\hat\T_n^R)$ exhibits size distortions. Overall, our simulation results confirm the validity and effectiveness of the proposed tests in the presence of outliers.


\section{Real data analysis}\label{sec:real_data}
\begin{figure}[!t]
    \centering
    \includegraphics[height=0.45\textwidth,width=1\textwidth]{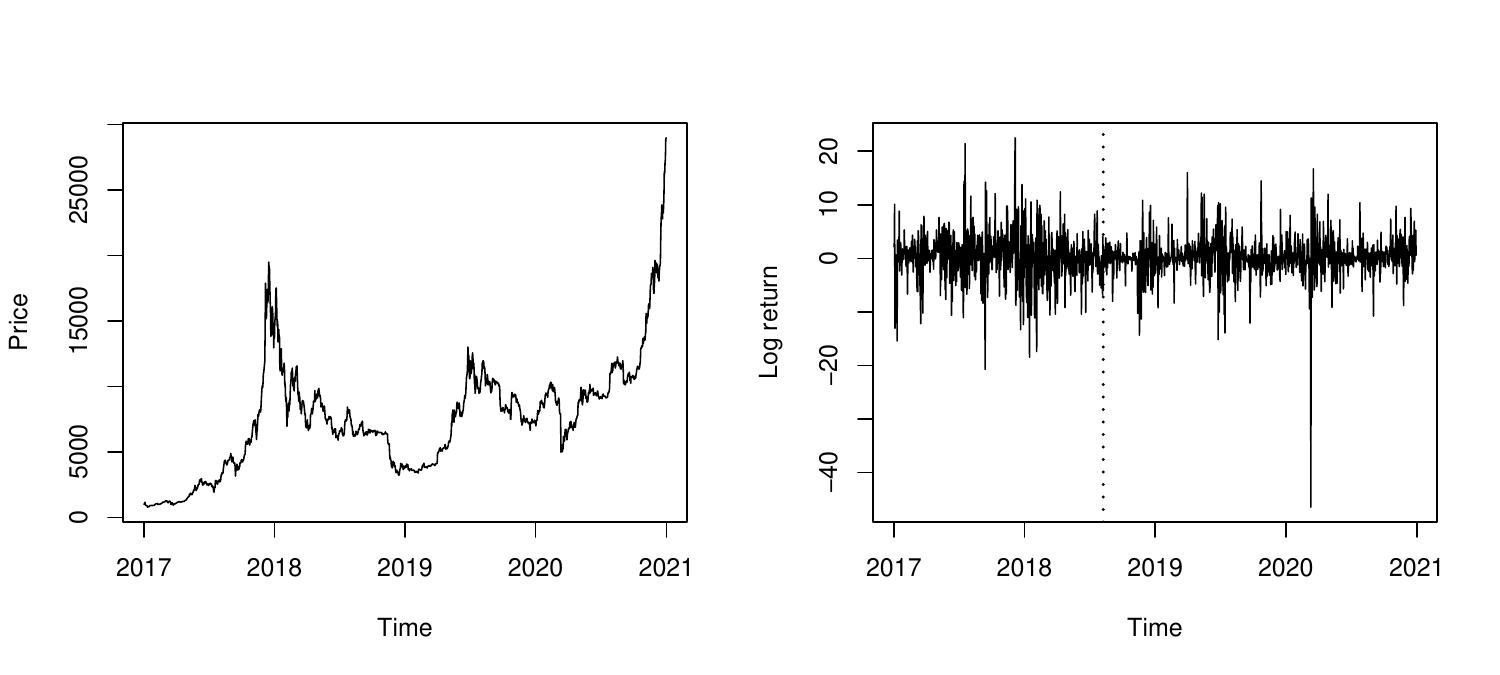}
    \vspace{-0.8cm}
    \caption{Plots of Bitcoin price series (L) and log-return series (R) from
    January 2017 to December 2020. The dotted vertical line in the right
    panel indicates the estimated change point at $t=586$ (August 10, 2018).}
    \label{Fig:btc}
\end{figure}

In this section, we present a real data application to Bitcoin. The dataset consists of daily closing prices from January 1, 2017 to December 31, 2020, totally 1,460 observations. The price series $\{S_t\}$ and its log return series $\{r_t\}$, where $S_t$ represents the Bitcoin price at time $t$ and $r_t = 100 \log(S_t/S_{t-1})$, are shown in the left and right panels of Figure \ref{Fig:btc}, respectively. We can observe that the return series exhibits typical volatility clustering. Since the Ljung–Box and LM-ARCH tests strongly suggest the presence of an ARCH effect, we fit a GARCH(1,1) model with the parameter $(\omega,\alpha,\beta)$ to the return series, as is commonly done in empirical studies for simplicity.

The QML estimates for the entire series are obtained as $\hat\omega=1.39$, $\hat\alpha=0.16$, and $\hat\beta=0.78$. On the other hand, the estimates from the MDPDE with $\gamma=0.1$ are $\hat\omega=0.33$, $\hat\alpha=0.10$, and $\hat\beta=0.86$. When data does not contain outliers, the QMLE and MDPDE typically yield similar estimates. The observed discrepancy between the two sets of estimates suggests the potential presence of outliers that may be affecting the QML estimates. Actually, in the return series shown in the right panel of Figure \ref{Fig:btc}, we can see some deviating observations and one large outlier, which could interfere with accurate statistical inference.

To examine whether parameter changes occurred during the period, we conduct the robust tests $T_n^M(\hat\T_n^R)$ and $SN_n^M(\hat\T_n^R)$, where $\hat\T_n^R$ represents the MDPDE with $\gamma=0.1$, using $M=9$ and $M=16$, as well as the naive tests $T_n$ and $SN_n$. Results of the naive and robust tests for parameter changes are presented in Table \ref{tab:real}. While the two naive tests do not reject the null hypothesis of no parameter change, the robust tests $T_n^9(\hat\T_n^R)$, $SN_n^9(\hat\T_n^R)$, and $SN_n^{16}(\hat\T_n^R)$ reject the null hypothesis.  Based on the findings from the simulation study, we infer that the naive tests fail to reject the null hypothesis due to the impact of outlying observations.

Using the estimator  mentioned in Remark \ref{rmk:multi.chg}, we locate the change point. The estimated change points are  $\hat k^*=586$ (August 10, 2018) based on $T_n^9(\hat\T_n^R)$ and  $SN_n^9(\hat\T_n^R)$, and $\hat k^*=569$ (July 24, 2018) based on $SN_n^{16}(\hat\T_n^R)$. We implement the binary segmentation procedure to detect additional changes, but no further parameter changes are identified.  Referring to the simulation results, where $SN_n^9(\hat\T_n^R)$ generally produces higher power than $SN_n^{16}(\hat\T_n^R)$, and considering the smaller p-value of $SN_n^9(\hat\T_n^R)$, we decide to locate the change point at $t = 586$ (dashed vertical line in the right panel of Figure \ref{Fig:btc}).  Consequently, the data is divided into two sub-periods.

\begin{table}[!t]
\renewcommand{\arraystretch}{1.4}
\tabcolsep=8pt
\centering
\caption{Results of the naive tests and robust tests for Bitcoin returns}
\label{tab:real}
\vspace{0.2cm}

\begin{tabular}{cccccccc}
\hline
& \multicolumn{2}{c}{naive tests}
& & \multicolumn{4}{c}{robust tests} \\
\cline{2-3}\cline{5-8}
test
& $T_n$ & $SN_n$
& & $T_n^9(\hat\T^R_n)$ & $T_n^{16}(\hat\T^R_n)$
& $SN_n^9(\hat\T^R_n)$ & $SN_n^{16}(\hat\T^R_n)$ \\
\hline
statistics
& 0.51 & 3.18
& & 1.43$^{*}$ & 1.01 & 105.1$^{**}$ & 76.2$^{**}$ \\
\hline
$\hat k^*$
& $\cdot$ & $\cdot$
& & 586 & $\cdot$ & 586 & 569 \\
\hline
\end{tabular}

\begin{tablenotes}
\footnotesize
\item \hspace{1.5cm}$^{*}$ and $^{**}$ denote significance at the 5\% and
1\% level, respectively.
\end{tablenotes}
\end{table}
\noindent

The estimation results are as follows: For the first sub-period, i.e., the data before August 10, 2018, the conditional variance, estimated using the MDPDE with $\gamma = 0.1$, is given by:
\[\hat\sigma_t^2=1.37+0.13\,r_{t-1}^2+0.80\,\hat\sigma_{t-1}^2,\]
and for the second sub-period:
\[\hat\sigma_t^2=0.23+0.06\,r_{t-1}^2+0.89\,\hat\sigma_{t-1}^2.\]
We can observe that the estimates differ significantly between the two periods. This result indicates that the parameters experience a substantial change.

\enlargethispage{2\baselineskip}

\section{Concluding remark}\label{sec:conclude}
In this study, we addressed the development of robust tests for parameter
changes in conditionally heteroscedastic time series models, particularly in
the presence of outliers. To mitigate the impact of outliers, we introduced a
two-step procedure comprising robust estimation and residual truncation. Based
on this procedure, we proposed the robust CUSUM of squares test and its
self-normalized counterpart. We derived their limiting null distributions and
established their consistency under an appropriate condition. We also investigated the effect of outliers on
the naive and proposed tests. In particular, we showed that large outliers can
seriously affect the behavior of the naive tests, whereas the proposed tests
can remain close to their uncontaminated counterparts under suitable
conditions. Simulation studies
demonstrated the strong robustness of the proposed tests against outliers, and
the results from real data analysis further illustrated their practical
usefulness.

Moving forward, extending the proposed methodology to other models, including
multivariate GARCH models, is a natural and interesting direction. While this
study focuses on retrospective parameter change tests, exploring a sequential
framework also holds significant promise, as monitoring tests for parameter
changes based on truncated residuals are expected to perform well in the
presence of outliers. Another important direction is to investigate parameter change tests under model misspecification. This is a practical issue because
parametric models are rarely perfectly specified in real data analysis.
Recent work, such as \cite{francq2026breaks}, has  considered change point
testing under model misspecification. Since model misspecification can often
occur in practice, developing robust change point tests that remain reliable
under such settings would be meaningful.  We leave these problems as promising topics for future research.

\section{Appendix}
In this appendix, we provide several lemmas needed for the proofs of
Theorem  \ref{Theorem.main.SN} and Corollary \ref{Cor:SN}. Lemmas \ref{lem:cmt} and \ref{lem:sn} play a central role in the
proof of Theorem \ref{Theorem.main.SN}, while Lemma \ref{lem:Vunif} is used to establish Lemma \ref{lem:sn}. To verify Corollary \ref{Cor:SN}, we establish Lemma \ref{lem:diff_G} and then apply Theorem 3.2 of \cite{billingsley:1999}.  In what follows, $\|\cdot\|_\infty$ denotes the supremum norm.

\begin{lemma}\label{lem:cmt}
Let $g\in C[0,1]\cap D_0^+$. For a sequence $\{g_n\} \subset D_0$, if $g_n \to g$ in $D[0,1]$ under the Skorokhod $J_1$ topology, then we have $\Phi(g_n)\to\Phi(g)$. In particular, $\Phi$ is continuous at $\tau_{\s M} B^o$ almost surely.
\end{lemma}
\begin{proof}
Since $g$ is continuous, we have $\|g\|_\infty <\infty$. Furthermore, note that  the convergence $g_n\to g$ in the $J_1$ topology implies $\|g_n-g\|_\infty\to0$, which also asserts  $\sup_n \|g_n\|_\infty <\infty.$ Hence, we can see that 
\begin{eqnarray}\label{eq:Vcont}
\big\|V_{g_n}-V_g\big\|_\infty
\ \les\
\|g_n-g\|_\infty (\|g_n\|_\infty+\|g\|_\infty) = o(1).
\end{eqnarray}
Note that $|\inf_{0\le t\le1}V_{g_n}(t)-\inf_{0\le t\le1}V_g(t)| \le \|V_{g_n} -V_g\|_\infty$. Thus, since $\inf_{0\le t\le1}V_g(t)>0$, it follows from \eqref{eq:Vcont} that $\inf_{0\le t\le 1} V_{g_n}(t)>0$ for all $n$ large enough. Thus, for all such $n$, we have 
\begin{eqnarray*}
    \sup_{0\leq t\leq1}  \Big|  \frac{g_n^2(t)}{V_{g_n}(t)}  - \frac{g^2(t)}{V_g(t)}\Big|
    &\leq&
    \sup_{0\leq t\leq1}
    \frac{|g_n^2(t)-g^2(t)|}{V_{g_n}(t)}
    +
    \sup_{0\leq t\leq1}
    \frac{g^2(t)|V_{g_n}(t)-V_g(t)|}
    {V_{g_n}(t)V_g(t)}  \\
    &\les&
    \|g_n-g\|_\infty\big(\|g_n\|_\infty+\|g\|_\infty\big)
    +
    \|g\|_\infty^2 \big\|V_{g_n}-V_g\big\|_\infty
    =o(1),
\end{eqnarray*}
which yields the first part of the lemma. 

To prove  the second assertion, it suffices to show that 
\begin{equation}\label{eq:inf_BB}
  \inf_{0\le t\le1}V_{\tau_{\s M} B^o}(t)\ >\ 0\quad a.s.  
\end{equation}
Let $g$ be a sample path of $\tau_{\s M} B^o$ and suppose that $\inf_{0\le t\le1}V_g(t)=0$.
Since $g$ is continuous, $V_g$ is also continuous on $[0,1]$. Hence, the infimum of $V_g$ is attained at some $t^*\in[0,1]$. If $t^*\in(0,1)$, then $g(u)=\tfrac{u}{t^*}g(t^*)$ for $u\in[0,t^*]$ and $g(u)=\tfrac{1-u}{1-t^*}g(t^*)$ for $u\in[t^*,1]$, and thus $g$ is linear on $[0,t^*]$ and on $[t^*,1]$. If $t^*\in\{0,1\}$, then $\int_0^1 g^2(u)\,du=0$, implying that $g(u)=0$ on $[0,1]$. In either case, $g$ is linear on some nondegenerate subinterval of $[0,1]$. However, since the sample paths of $\tau_{\s M} B^o$ are almost surely nonlinear on every nondegenerate subinterval of $[0,1]$, the event $\big\{\inf_{0\le t\le1}V_{\tau_{\s M} B^o}(t)=0\big\}$ has probability zero. We therefore obtain (\ref{eq:inf_BB}). This completes the proof.
\end{proof}
From now on, we denote $\tilde f_t=f_{\s M}(\tilde e^2_t(\TR))$ and $\bar f=\frac1n\sum_{t=1}^n\tilde f_t$.
Note that 
\[
\GtM(u)=\frac{1}{\sqrt{n}} \sum_{t=1}^{\lfloor nu\rfloor} (\tilde f_t -\bar f).
\]

\begin{lemma}\label{lem:Vunif}
Suppose that assumptions {\bf A1}--{\bf A6} are satisfied. Then, under $H_0$, we have
\begin{eqnarray*}
\max_{1\le k\le n-1}\Big|\,\frac{1}{n^2}V_{n,k}\big(f_{\s M}(\tilde e^2(\hat\T^R_n))\big)-V_{\GtM}\!\big(\tfrac{k}{n}\big)\,\Big|=o_{P}(1).
\end{eqnarray*}
\end{lemma}

\begin{proof}
For $1\le k \le n-1$, letting 
\begin{eqnarray}\label{eq:phi_psi}
    \phi_{n,k}(u):=\GtM(u)-\frac{u}{k/n}\GtM\big(\tfrac kn\big)\quad  \text{and}\quad  \psi_{n,k}(u):=\GtM(u)-\frac{1-u}{1-k/n}\GtM\big(\tfrac kn\big),
\end{eqnarray}
by an elementary calculation, we can express that 
\begin{eqnarray*}
\frac{1}{n^2}V_{n,k}\big(f_{\s M}\big(\tilde e^2(\TR)\big)\big)
=\frac1n\sum_{t=1}^{k}\phi^2_{n,k}\big(\tfrac{t}{n}\big)
+\frac1n\sum_{t=k+1}^{n}\psi^2_{n,k}\big(\tfrac{t-1}{n}\big).
\end{eqnarray*}
Set $M_n:=\sup_{0\le u\le1}|\GtM(u)|$. By Theorem \ref{Theorem.main}, we have $M_n=O_{\uP}(1)$.  Recall also that $0\le \tilde f_t\le M$ by the definition of $f_{\s M}$. Then, for $\tfrac {t-1}{n} < u \le \tfrac tn$, we have
\begin{eqnarray*}
\big|\GtM(u)-\GtM\big(\tfrac{t}{n}\big)\big|
\ \le\ \frac{1}{\sqrt n} \big|\tilde f_t - \bar f\big|
\ \le\ \frac{M}{\sqrt n},
\end{eqnarray*}
and thus
\begin{eqnarray*}
\big|\phi_{n,k}(u)-\phi_{n,k}\big(\tfrac tn\big)\big| 
\ \le\ 
\big|\GtM(u)- \GtM\big(\tfrac{t}{n}\big)\big|
+\frac{|u-\frac{t}{n}|}{k/n}\,\big|\GtM\big(\tfrac kn\big)\big|
\ \le\ 
\frac{M}{\sqrt n}+\frac{M_n}{k}.
\end{eqnarray*}
We can also readily see that 
\[ \max_{1\le k\le n-1}\, \sup_{0\le u \le k/n} |\phi_{n,k}(u)|\ \le\ 2 M_n.\]
Hence, using the above, we obtain that 
\begin{eqnarray}\label{numer_conv}
   \Big|  \int_0^{\tfrac kn} \phi_{n,k}^2(u)du -\frac1n\sum_{t=1}^{k}\phi^2_{n,k}\big(\tfrac{t}{n}\big) \Big| 
   &\le&
   \frac kn\max_{1\le t \le k}\,\, \sup_{ u \in \big(\tfrac{t-1}{n},\tfrac tn\big]} \big|\phi_{n,k}^2(u)-\phi_{n,k}^2 \big(\tfrac tn\big)\big| \nonumber\\
   &\les&
   M_n\Big( \frac{M}{\sqrt n} +\frac{M_n}{n}\Big).  
\end{eqnarray}
Next, one can see that for $\tfrac{t-1}{n}\le u<\tfrac tn$,
\begin{eqnarray*}
\big|\psi_{n,k}(u)-\psi_{n,k}\big(\tfrac{t-1}n\big)\big|
=\frac{|u-\frac{t-1}{n}|}{1-k/n}\,\Big|\GtM\big(\tfrac kn\big)\Big|
\ \le\ \frac{M_n}{n-k},
\end{eqnarray*}
and thus, in a similar manner as above,
\begin{eqnarray*}\label{denom_conv}
   \Big|  \int^1_{\tfrac kn} \psi_{n,k}^2(u)du -\frac1n\sum_{t=k+1}^{n}\psi^2_{n,k}\big(\tfrac{t-1}{n}\big) \Big|
   \  \les\
   \frac{M_n^2}{n},
\end{eqnarray*}
which together with (\ref{numer_conv}) yields
\begin{eqnarray*}
\max_{1\le k\le n-1}\Big|\,\frac{1}{n^2}V_{n,k}\big(f_{\s M}(\tilde e^2(\TR))\big)-V_{\GtM}\!\big(\tfrac{k}{n}\big)\,\Big|
\ \les\ 
M_n\Big( \frac{M}{\sqrt n} +\frac{M_n}{n}\Big) =O_{\uP}\Big(\frac{1}{\sqrt{n}}\Big).
\end{eqnarray*}
This completes the proof.
\end{proof}

\begin{lemma}\label{lem:sn}
Suppose that assumptions {\bf A1}--{\bf A6} are satisfied. Then, under $H_0$, we have
\[
SN_n^M(\TR)=\Phi\big(\GtM\big)+o_{P}(1).
\]
\end{lemma}

\begin{proof}
For notational simplicity, we write $\Gt$ for $\GtM\!.$
Let $m_n:=\inf_{0\le t\le1} V_{\Gt}(t)$ and $\rho_n:=\max_{1\le k\le n-1}|R_{n,k}|,$  where
\begin{eqnarray*}
R_{n,k}=\frac{1}{n^2}V_{n,k}\big(f_{\s M}\big(\tilde e^2(\hat\T^R_n)\big)\big)-V_{\Gt}\big(\tfrac kn\big),\qquad 1\le k \le n-1.
\end{eqnarray*}
Recall from \eqref{Gn.conv} that $\Gt\stackrel{w}{\longrightarrow}\tau_{\s M}B^o$ in $D[0,1]$,  and note that the mapping $g\mapsto\inf_{0\le t\le1}V_g(t)$ is continuous at every $g\in C[0,1]\cap D_0$ by \eqref{eq:Vcont}. Since the sample paths of $\tau_{\s M}B^o$ are almost surely continuous, we have by the continuous mapping theorem that 
\begin{equation}\label{eq:m_n}
   m_n\stackrel{d}{\longrightarrow}m:=\inf_{0\le t\le1}V_{\tau_{\s M}B^o}(t).
 \end{equation}

Next, let $\cA_n:=\{\rho_n< m_n/2\}$. Since $\rho_n \ge 0$,  we can see that   $m_n>0$ on $\cA_n$. Thus, on this event, we have that $\Gt \in D_0^+$ and that for each $1\le k\le n-1$,
\begin{eqnarray*}
\frac{1}{n^2}V_{n,k}\big(f_{\s M}\big(\tilde e^2(\hat\T^R_n)\big)\big)=V_{\Gt}\big(\tfrac kn\big)+R_{n,k}\ \ge\ \frac{m_n}{2}\ >\ 0.
\end{eqnarray*}
Consequently, we have that
\begin{eqnarray}\label{eq:Phi_SN_onA}
\Phi\big(\Gt\big)=\sup_{0\le t \le1}\frac{\Gt^2(t)}{V_{\Gt}(t)}
\qquad\text{and}\qquad
SN_n^M(\TR)=\max_{1\le k\le n-1}\frac{\Gt^2\big(\tfrac kn\big)}{V_{\Gt}\big(\tfrac kn\big)+R_{n,k}}
\quad\text{on }\cA_n.
\end{eqnarray}

Now, for $g\in D_0$ and $0< t<1$, let $\ell_{g,t}$ be the function on $[0,1]$ defined by $\ell_{g,t}(u)=\frac ut g(t)$ for $0\le u\le t$ and $\ell_{g,t}(u)=\frac{1-u}{1-t}g(t)$ for $t<u\le1$. We set $\ell_{g,0}= \ell_{g,1}\equiv0$. Then, we can express that $V_g(t)=\int_0^1\big(g(u)-\ell_{g,t}(u)\big)^2du$ for all $0\le t\le1$. Since $\|\ell_{g,t}\|_\infty=|g(t)|$, we have that for all $s,t\in[0,1]$,
\begin{eqnarray}\label{eq:diff_V}
\big|V_g(t)-V_g(s)\big|
\ \le\ 
\big\|\ell_{g,t}-\ell_{g,s}\big\|_\infty\,\big\|2g-\ell_{g,t}-\ell_{g,s}\big\|_\infty
\ \les\  \|g\|_\infty\big\|\ell_{g,t}-\ell_{g,s}\big\|_\infty.
\end{eqnarray}
Let $c_k:=\Gt\big(\tfrac kn\big)$ for $0\le k\le n-1$. 
Since $\big|\sum_{t=1}^{k}\big(\tilde f_t-\bar f\big)\big|
=\big|\sum_{t=k+1}^{n}\big(\tilde f_t-\bar f\big)\big|$ and $0\le\tilde f_t\le M$, we have
\begin{eqnarray*}
|c_k|=\frac{1}{\sqrt n}\Big|\sum_{t=1}^{k}\big(\tilde f_t-\bar f\big)\Big|
\ \le\ \frac{M}{\sqrt n}\big(k\wedge(n-k)\big).
\end{eqnarray*}
Note that $\Gt(t)=c_k$ for $\tfrac kn \le t < \tfrac{k+1}{n}$. Using the above, we obtain by an elementary calculation that for $\tfrac kn \le t < \tfrac{k+1}{n},$
\begin{eqnarray*}
\big\|\ell_{\Gt,t}-\ell_{\Gt,k/n}\big\|_\infty
\ \le\ |c_k|\bigg(\frac{t-\frac kn}{t}\vee\frac{t-\frac kn}{1-\frac kn}\bigg)
\ \le\ |c_k|\Big(\frac1k\vee\frac{1}{n-k}\Big)
\ \le\ \frac{M}{\sqrt n}.
\end{eqnarray*}
Therefore, combining this with \eqref{eq:diff_V}, we have
\begin{eqnarray}\label{V.disc}
\max_{0\le k\le n-1}\ \sup_{\tfrac kn \le t < \tfrac{k+1}{n}}\big|V_{\Gt}(t)-V_{\Gt}\big(\tfrac kn\big)\big|
\ \les\ \frac{MM_n}{\sqrt n},
\end{eqnarray}
where $M_n$ is the one defined in Lemma \ref{lem:Vunif}.
Recall \eqref{eq:Phi_SN_onA}. Using \eqref{V.disc}, we obtain that on $\cA_n$,
\begin{eqnarray}\label{eq:err1}
\Big|\Phi\big(\Gt\big)-\max_{1\le k\le n-1}\frac{\Gt^2(k/n)}{V_{\Gt}(k/n)}\Big|
\ \le\ 
\max_{0\le k\le n-1}\ \sup_{\tfrac kn \le t < \tfrac{k+1}{n}}\Big|\frac{\Gt^2(t)}{V_{\Gt}(t)}-\frac{\Gt^2(k/n)}{V_{\Gt}(k/n)}\Big|
\ \les\
\frac{MM_n^3}{m_n^2} \,\frac{1}{\sqrt n}
\end{eqnarray}
 and
\begin{eqnarray}\label{eq:err2}
\Big|SN_n^M(\TR)
-\max_{1\le k\le n-1}\frac{\Gt^2(k/n)}{V_{\Gt}(k/n)}\Big|
\ \le\
\max_{1\le k\le n-1}\frac{\Gt^2(k/n)|R_{n,k}|}{\big(V_{\Gt}(k/n)+R_{n,k}\big)V_{\Gt}(k/n)}
\ \les\
\frac{M_n^2}{m_n^2} \rho_n.
\end{eqnarray}
Letting $\mathcal{B}_n(\kappa):=\{m_n>\kappa, \rho_n<\kappa/2\}$, since $\mathcal{B}_n(\kappa)\subset\cA_n$ for all $\kappa>0$, we have from \eqref{eq:err1} and \eqref{eq:err2} that
\begin{eqnarray*}
\big|SN_n^M(\hat\theta_n^{\revR})-\Phi(\widetilde{G}_n)\big|
\ \le\ \frac{MM_n^3}{m_n^2}\frac{1}{\sqrt{n}}+\frac{M_n^2}{m_n^2}\rho_n
\ \le\ \frac{1}{\kappa^2}\Big(\frac{MM_n^3}{\sqrt{n}}+M_n^2\rho_n\Big)
\quad\text{on }\mathcal{B}_n(\kappa).
\end{eqnarray*}
Hence, we obtain that for any $\eta>0$,
\begin{eqnarray*}
\uP\big(|SN_n^M(\hat\theta_n^{\revR})-\Phi(\widetilde{G}_n)|>\eta\big)
&\le& \uP(m_n\le\kappa)+\uP(\rho_n\ge\kappa/2)
+\uP\Big(\frac{MM_n^3}{\sqrt{n}}+M_n^2\rho_n\ge\kappa^2\eta\Big).
\end{eqnarray*}
Since $m_n\stackrel{d}{\longrightarrow}m$ by \eqref{eq:m_n}, $\rho_n=o_{\uP}(1)$ by Lemma \ref{lem:Vunif}, and $M_n=O_{\uP}(1)$, 
we have that for any $\eta>0$ and $\kappa>0$,
\begin{eqnarray*}
\limsup_{n\to\infty}
\uP\big(\,|SN_n^M(\TR)-\Phi(\Gt)|>\eta\big)
\ \le\ \uP(m\le\kappa).
\end{eqnarray*}
Letting $\kappa\downarrow0$ and recalling that $\uP(m>0)=1$ by \eqref{eq:inf_BB},
we obtain the lemma.
\end{proof}
 To prove Corollary \ref{Cor:SN}, we redefine the process $\GtM= \{ \GtM(u)\mid 0\le u \le 1\}$ by 
\begin{eqnarray*}
    \GtM(u):=\frac{1}{\sqrt{n}}D_{n,\lfloor nu\rfloor} \big(f_{\s M}(\tilde e^2)\big),
\end{eqnarray*}
where $f_{\s M}(\tilde e^2)=(f_{\s M}(\tilde e_1^2(\hat\T_n)),\cdots, f_{\s M}(\tilde e_n^2(\hat\T_n))). $
Here, $\hat\T_n$ is the estimator used in $SN_n$ and satisfies $\hat\theta_n\to\theta_0$ almost surely and
$\sqrt n\|\hat\theta_n-\theta_0\|=O_{\uP}(1)$. For the non-truncated case, we also define
\begin{eqnarray*}
    \Gto(u):=\frac{1}{\sqrt{n}}D_{n,\lfloor nu\rfloor} \big(\tilde e^2 \big),
\end{eqnarray*}
where $\tilde e^2=(\tilde e_1^2(\hat\T_n),\cdots, \tilde e_n^2(\hat\T_n)). $

\begin{lemma}\label{lem:diff_G}
Suppose that assumptions {\bf A1}--{\bf A5} are satisfied. Under $H_0$, we have that for any $\eta>0$,
\[\lim_{M\rightarrow\infty}\limsup_{n\rightarrow\infty} 
\uP\big(\,\big\|\Gto-\GtM\big\|_\infty\geq\eta\,\big)=0.\]
\end{lemma}
\begin{proof}
Let $G_n^\infty$ and $G_n^{M}$ be the counterparts of $\Gto$ and $\GtM$, respectively, obtained by replacing $\tilde e^2_t(\hat\T_n)$ with $\ep_t^2$, and
set $h_{\s M}(x):=x-f_{\s M}(x)$ for $x \ge 0$. Since $0\leq h_{\s M}(\ep_1^2)\leq\ep_1^2$ and $\E\ep_1^4<\infty$, we have $\varsigma_{\s M}^2:=\Var(h_{\s M}(\ep_1^2))<\infty$, and thus, by the invariance principle and the continuous mapping theorem, we have
\begin{eqnarray*}
    \big\|G^\infty_n-G_n^{M}\big\|_\infty
    =\frac{1}{\sqrt{n}} \max_{1\le k \le n}\Big|\sum_{t=1}^k h_{\s M}(\ep_t^2) -\frac{k}{n}\sum_{t=1}^nh_{\s M}(\ep_t^2) \Big|
    \stackrel{d}{\longrightarrow}\varsigma_{\s M}\|B^o\|_\infty\quad\text{as}\ \ n\rightarrow\infty.
\end{eqnarray*}
Here, we note that this convergence trivially holds in the case of $\varsigma_{\s M}^2=0$. Using the dominated convergence theorem,
one can also see that  $\varsigma_{\s M}^2\rightarrow 0$ as $M\to\infty$,  and hence $\varsigma_{\s M}\|B^o\|_\infty=o_{\uP}(1)$. Therefore, it follows from the portmanteau theorem that, for any $\eta>0$, 
\begin{eqnarray}\label{eq:lim_iii}
    \limsup_{n\rightarrow\infty}\uP\big(\big\|G^\infty_n-G_n^{M}\big\|_\infty\geq\eta\big)
    \ \leq\ 
    \uP\big(\varsigma_{\s M}\|B^o\|_\infty\geq\eta\big)
    \ \rightarrow\ 0\quad\text{as }M\to\infty.
\end{eqnarray}
Since $h_{\s M}$ is continuously differentiable and Lipschitz continuous with a Lipschitz constant of one, \eqref{main.conv} also holds for $h_{\s M}$ in place of $f_{\s M}$. Thus, we have
\begin{eqnarray*}
    &&\big\|\big(\Gto-\GtM\big)-\big(G_n^\infty-G_n^{M}\big)\big\|_\infty\\
    &&=
    \frac{1}{\sqrt{n}}\max_{1\leq k\leq n } \Big|\sum_{t=1}^k \big(h_{\s M}(\te^2(\hat\T_n))-h_{\s M}(\epsilon_t^2)\big) 
-\frac{k}{n}\sum_{t=1}^n \big(h_{\s M}(\te^2(\hat\T_n))-h_{\s M}(\epsilon_t^2)\big) \Big|\\
    &&=o_{\uP}(1),
\end{eqnarray*}
which together with \eqref{eq:lim_iii} asserts the lemma.
\end{proof}

\noindent{\bf Proof of Corollary \ref{Cor:SN}}\\
As stated in Remark \ref{rmk:A5}, Theorem \ref{Theorem.main.SN} remains valid for any estimator that is strongly consistent and $\sqrt{n}$-consistent.  Hence, for the estimator $\hat\T_n$ used in $SN_n$ and each fixed $M$ such that $\tau_{\s M}^2>0$, it holds that
\[
SN_n^M(\hat\T_n) \stackrel{d}{\longrightarrow}\Phi(B^o).
\]
Since the limiting distribution does not depend on $M$, the second condition of Theorem 3.2 of \cite{billingsley:1999} is immediate. Therefore, it suffices to show that, for any $\eta>0$,
\begin{equation}\label{eq:SN_Bill}
\lim_{M\to\infty}\limsup_{n\to\infty}\uP\big(|SN_n-SN_n^M(\hat\T_n)|>\eta\big)=0.
\end{equation}

For $g\in D[0,1]$, define
\begin{eqnarray*}
\phi_{n,k}(t \mid g):=
\begin{cases}
g\big(\tfrac tn\big)-\tfrac tk\,g\big(\frac kn\big)& \text{if }1\le t\le k,\\[8pt]
g\big(\tfrac{t-1}{n}\big)-\frac{n-t+1}{n-k}\,g\big(\tfrac kn\big)&\text{if } k+1\le t\le n.
\end{cases}
\end{eqnarray*}
One can easily see that for all $1\le k \le n-1$ and $1\le t \le n$,
\begin{eqnarray}\label{eq:phi_psi_bdd}
    |\phi_{n,k}(t\mid g)|\le 2\|g\|_\infty.
\end{eqnarray}
By a direct calculation, we also have that
\begin{eqnarray*}
\frac{1}{n^2}V_{n,k}(\tilde e^2)
=\frac1n\sum_{t=1}^{n}\phi^2_{n,k}(t\mid \Gto)
\quad\text{and}\quad
\frac{1}{n^2}V_{n,k}\big(f_{\s M}(\tilde e^2 )\big)
=\frac1n\sum_{t=1}^{n}\phi^2_{n,k}(t \mid \GtM).
\end{eqnarray*}
Since $\phi_{n,k}(t\mid \Gto)\pm\phi_{n,k}(t\mid\GtM)=\phi_{n,k}(t\mid\Gto\pm\GtM)$, it follows from \eqref{eq:phi_psi_bdd} that
\begin{eqnarray*}
\frac{1}{n^2}\big|V_{n,k}(\tilde e^2)-V_{n,k}(f_{\s M}(\tilde e^2))\big|
&\le&\frac1n\sum_{t=1}^{n}\big|\phi_{n,k}(t\mid\Gto-\GtM)\big|\,\big|\phi_{n,k}(t\mid\Gto+\GtM)\big|\\
&\le& 4\|\Gto-\GtM\|_\infty \|\Gto+\GtM\|_\infty
\end{eqnarray*}
Now, let $\Delta_n^M:=\|\Gto-\GtM\|_\infty$, $K_n:=\|\Gto\|_\infty,$ and
$\underline V_n^M:=\min_{1\le k\le n-1}\frac1{n^2}V_{n,k}\big(f_{\s M}(\tilde e^2)\big)$, 
and set $\cA_n:=\{\Delta_n^M\le\delta',\ K_{n}\le K,\ \underline V_n^M>\kappa\}$, where $\delta', K,$ and $\kappa$ are positive constants satisfying $4\delta'(2K+\delta')<\kappa/2$.
Since $\|\Gto+\GtM\|_\infty\le 2K_{n}+\Delta_n^M$,  we have by the above that 
\[
\max_{1\le k \le  n-1} \frac{1}{n^2}\big|V_{n,k}(\tilde e^2)-V_{n,k}(f_{\s M}(\tilde e^2))\big|
\ \le\  4\Delta_n^M\big(2K_n+\Delta_n^M\big),
\]
and thus
\[
\min_{1\le k\le n-1}\frac1{n^2}V_{n,k}(\tilde e^2)>\frac{\kappa}{2}\quad\text{on }\cA_n.
\]
Note also that  $\big|\big(\Gto\big(\tfrac kn\big)\big)^2-\big(\GtM\big(\tfrac kn\big)\big)^2\big|
\le\Delta_n^M\big(2K_n+\Delta_n^M\big)$. Then, we have that on $\cA_n$, 
\begin{eqnarray*}
 |SN_n-SN_n^M(\hat\T_n)| 
&\le&
\max_{1\le k \le n-1} \Big| \frac{\big(\Gto\big(\tfrac kn\big)\big)^2}{\frac1{n^2}V_{n,k}(\tilde e^2)}
\ -\ \frac{\big(\GtM\big(\tfrac kn\big)\big)^2}{\frac1{n^2}V_{n,k}\big(f_{\s M}(\tilde e^2)\big)}\Big|\\[5pt]
&\le& 
\frac{8K^2\delta'(2K+\delta')}{\kappa^2}+\frac{\delta'(2K+\delta')}{\kappa}
=:q(\delta'; K,\kappa).
\end{eqnarray*}
Consequently, we have that for any $\eta>0$,
\begin{eqnarray}\label{eq:SN_three}
\uP\big(|SN_n-SN_n^M|>\eta\big)
\!\!&\le&\!\! \uP(\cA_n^c )
+\uP\big(|SN_n-SN_n^M|>\eta , \cA_n\big)\nonumber\\
\!\!&\le&\!\! \uP\big(\Delta_n^M>\delta'\big)+\uP\big(K_{n}>K\big)+\uP\big(\underline V_n^M\le\kappa\big)
+\mathbf 1\big(q(\delta';K,\kappa)>\eta\big)
\end{eqnarray}
By Lemma \ref{lem:diff_G}, we have that
\[
\lim_{M\to\infty}\limsup_{n\to\infty}P(\Delta_n^M>\delta')=0.
\]
One can also see from Corollary \ref{Cor:Tn} that $ K_n=\|\Gto\|_\infty\stackrel{d}{\longrightarrow}\tau\|B^o\|_\infty$, thus we have
\begin{equation}\label{eq:Kn_tight}
\limsup_{n\to\infty}\uP(K_n \ge K)
\ \le\
\uP(\tau\|B^o\|_\infty \ge K )\rightarrow 0\quad\text{as }K\to\infty.
\end{equation}

Next, note that Lemma \ref{lem:Vunif} and the intermediate results
obtained in the proof of Lemma \ref{lem:sn} remain valid when $\TR$ is replaced by $\hat\T_n$.
Since $V_{\GtM}(1)=V_{\GtM}(0)$, we have $\inf_{0\le t\le1}V_{\GtM}(t)= \inf_{0\le t<1}V_{\GtM}(t)$. Then, it follows from Lemma \ref{lem:Vunif} and \eqref{V.disc} that
\begin{eqnarray}
&&\Big|\underline V_n^M-\inf_{0\le t\le1}V_{\GtM}(t) \Big| \notag\\
&&\le\
\Big|\min_{1\le k\le n-1}
\frac1{n^2}V_{n,k}\big(f_{\s M}(\tilde e^2)\big)
-\min_{1\le k\le n-1}V_{\GtM}\big(\tfrac kn\big)\Big|
+
\Big|\min_{1\le k\le n-1}V_{\GtM}\big(\tfrac kn\big)
-\inf_{0\le t <1}V_{\GtM}(t)\Big| \notag\\[5pt]
&&\le\
\max_{1\le k\le n-1}
\Big|
\frac1{n^2}V_{n,k}\big(f_{\s M}(\tilde e^2)\big)
-V_{\GtM}\big(\tfrac kn\big)\Big|
+
\max_{1\le k\le n-1} \sup_{\tfrac kn \le t < \tfrac{k+1}{n}} \big|V_{\GtM}(t)-V_{\GtM}\big(\tfrac kn\big)\big| \notag\\
&&=\ o_{\uP}(1).\label{eq:diff_V_v}
\end{eqnarray}
Moreover, we have from \eqref{eq:m_n} that
\[
\inf_{0\le t\le1}V_{\GtM}(t)
\stackrel{d}{\longrightarrow}
\inf_{0\le t\le1}V_{\tau_{\s M}B^o}(t)
=
\tau_{\s M}^2
\inf_{0\le t\le1}V_{B^o}(t),
\]
which together with \eqref{eq:diff_V_v} yields
\begin{eqnarray}\label{eq:d_V_nM}
 \underline V_n^M
\stackrel{d}{\longrightarrow}
\tau_{\s M}^2 \inf_{0\le t\le1}V_{B^o}(t). 
\end{eqnarray}
Recall from \eqref{eq:inf_BB} that $\inf_{0\le t\le1}V_{B^o}(t)>0$ almost surely.  Since $\tau_{\s M}^2\to\tau^2>0$ as $M\to\infty$, it follows from the portmanteau theorem  that
\begin{eqnarray}\label{eq:V_nM_tight}
\limsup_{M\to\infty}\limsup_{n\to\infty}
\uP(\underline V_n^M\le\kappa)
&\le&
\limsup_{M\to\infty}
\uP\big(\tau_{\s M}^2 \inf_{0\le t\le1}V_{B^o}(t) \le\kappa\big)\notag\\
&\le&
\uP\Big(\inf_{0\le t\le1}V_{B^o}(t)\le\frac{2\kappa}{\tau^2}\Big)
\rightarrow 0\quad\text{as } \kappa\to 0.
\end{eqnarray}
Therefore, first choose $K$ sufficiently large and $\kappa$ sufficiently small such that the probabilities on the right sides of \eqref{eq:Kn_tight} and \eqref{eq:V_nM_tight}  are arbitrarily small, respectively. Then, taking $\delta'>0$ sufficiently small such that $q(\delta';K,\kappa)<\eta$, we obtain from \eqref{eq:SN_three} that for any $\varepsilon>0$,
\begin{equation*}
\lim_{M\to\infty}\limsup_{n\to\infty}\uP\big(|SN_n-SN_n^M(\hat\T_n)|>\eta\big) \le \varepsilon,
\end{equation*}
\enlargethispage{2\baselineskip}
which establishes \eqref{eq:SN_Bill}. This completes the proof.
\hfill{$\Box$}\vspace{0.2cm}\\

\noindent{\bf Acknowledgments}\\
The author sincerely thanks the associate editor and the referees for their careful reviews and valuable suggestions, which substantially improved the quality of the paper.
This work was supported by the National Research Foundation of Korea(NRF) grant funded by the Ministry of Education (NRF-2019R1I1A3A01056924) and the Korean government (MSIT) (RS-2025-20122969).
\normalem

\putbib[jun]

\end{bibunit}

\clearpage

\pagenumbering{arabic}

\setcounter{section}{0}
\renewcommand{\thesection}{S\arabic{section}}

\setcounter{subsection}{0}
\renewcommand{\thesubsection}{\thesection.\arabic{subsection}}

\setcounter{equation}{0}
\renewcommand{\theequation}{S\arabic{equation}}

\setcounter{figure}{0}
\renewcommand{\thefigure}{S\arabic{figure}}

\setcounter{table}{0}
\renewcommand{\thetable}{S\arabic{table}}

\begin{bibunit}[apalike_noissue]

\vspace*{2em}

\begin{center}

{\normalfont\LARGE
Supplementary Material for\\[10pt]
``Robust tests for parameter change in conditionally heteroscedastic time series models''
\par}

\vskip 1.5em

{\normalfont\large Junmo Song\par}

\vskip 0.5em

{\normalfont Department of Statistics, Kyungpook National University\par}

\end{center}

\vskip 2em


\noindent 
In Sections \ref{supp:Sec1} and \ref{supp:Sec2}, we denote $x=(x_1,\cdots,x_n)$ and $y=(y_1,\cdots,y_n)$.
We redefine the function $\phi_{n,k}$ used in the proof of Corollary \ref{Cor:SN} as follows:
\[
\phi_{n,k}(t \mid x):=
\begin{cases}
\displaystyle
\sum_{j=1}^t x_j-\frac{t}{k}\sum_{j=1}^k x_j
&\text{if } 1\le t\le k,\\[15pt]
\displaystyle
\sum_{j=t}^n x_j-\frac{n-t+1}{n-k}\sum_{j=k+1}^n x_j
&\text{if } k+1\le t\le n.
\end{cases}
\]
We note that
\begin{eqnarray}\label{supp:eq:phi_bdd}
 \big| \phi_{n,k}(t \mid x)\big|
&\le&
\bigg\{\mathbf 1(t\le k)\sum_{j=1}^k \Big|\mathbf 1( j \le t) -\frac tk \Big| |x_j|\bigg\}
\vee
 \bigg\{\mathbf 1(t> k)\sum_{j=k+1}^n \Big| \mathbf 1( j \ge t ) -\frac{n-t+1}{n-k} \Big||x_j|\bigg\}\notag\\
&\le&
\sum_{j=1}^n |x_j|,   
\end{eqnarray}
and thus
\begin{equation*}
 V_{n,k}(x)=\sum_{t=1}^n\phi_{n,k}^2(t \mid x)=\|\phi_{n,k}(x)\|^2
 \ \le\
 n \Big(\sum_{j=1}^n |x_j| \Big)^2,
\end{equation*}
where $\phi_{n,k}(x)=(\phi_{n,k}(1 | x),\cdots, \phi_{n,k}(n | x)).$
Using the above, one can see that
\begin{equation}\label{supp:eq:V_norm}
\Big| \sqrt{V_{n,k}(x)} -\sqrt{V_{n,k}(y)} \Big|
\ \le\
\big\| \phi_{n,k}(x) -\phi_{n,k}(y)\big\|
=
\big\| \phi_{n,k}(x-y)\big\|
\ \le\
\sqrt{n}\sum_{j=1}^n |x_j-y_j|.
\end{equation}
\section{Proof of Proposition 1}\label{supp:Sec1}
Since
\begin{equation}\label{supp:eq:D_nk}
  |D_{n,k}(x)| =\Big|\sum_{t=1}^n \Big(\mathbf1(t\le k) -\frac kn\Big)x_t\Big|
  \ \le\ 
  \sum_{t=1}^n |x_t|
\end{equation}
and $n$ is fixed, we have 
\[
\big|D_{n,k}(r(s))\big| 
\ \le\
\sum_{t=1}^n |r_t(s)| = o(s^2),
\]
where $r(s)=(r_1(s),\cdots,r_n(s))$.
Recall that the sample size $n$, the uncontaminated sample $\{X_t(0)\}_{t=1}^n$, and the outlier
set $O_n$ are fixed. Therefore, it follows from (\ref{eq:ut}) that
\begin{eqnarray}\label{supp:eq:D_nk_Z}
D_{n,k}(\tilde Z(s))=D_{n,k}(\tilde Z(0))+s^2D_{n,k}(\xi)+o(s^2)
\end{eqnarray}
and
\begin{eqnarray}\label{supp:eq:hat_tau}
\hat\tau_n^2(s)
&=&\frac1n\sum_{t=1}^n
\big(\tilde Z_t(0)-{\bar {\tilde Z}}(0)+s^2(\xi_t-\bar\xi)+r_t(s)-\bar r(s)\big)^2 \notag\\
&=&\frac{s^4}{n}\sum_{t=1}^n(\xi_t-\bar\xi)^2+o(s^4),
\end{eqnarray}
where ${\bar {\tilde Z}}(0)$, $\bar \xi$, and $\bar r(s)$ denote the sample means of $\tilde Z(0)$, $\xi$, and $r(s)$, respectively.
Thus, the first result follows from \eqref{supp:eq:D_nk_Z} and \eqref{supp:eq:hat_tau}. 

Next, note that
\[
\phi_{n,k}(t\mid \tilde Z(s))
=
\phi_{n,k}(t\mid \tilde Z(0))
+s^2\phi_{n,k}(t \mid \xi)
+\phi_{n,k}(t \mid r(s)).
\]
Since $n$ is fixed, we have by \eqref{supp:eq:phi_bdd} that $\big|\phi_{n,k}(t\mid r(s))\big|=o(s^2)$.
Hence, one can see that
\[
V_{n,k}(\tilde Z(s))
=
\sum_{t=1}^n \phi^2_{n,k}(t\mid \tilde Z(s))
=
s^4V_{n,k}(\xi)+o(s^4),
\]
which together with \eqref{supp:eq:D_nk_Z} yields the second result. 
This completes the proof.

\section{Proof of Proposition 2}\label{supp:Sec2}
Put $Y_s:=f_{\s M}(\tilde Z(s))$. Recall that $D_{n,k}(x)-D_{n,k}(y)=D_{n,k}(x-y)$. Then, we have by \eqref{supp:eq:D_nk}  that
\begin{eqnarray}\label{supp:eq:D_diff}
\big|\,|D_{n,k}(Y_s)|-|D_{n,k}(Y_0)|\,\big|
\ \le\ B^M_n(s).
\end{eqnarray}
Let $\bar f_{\s M}(\tilde Z(s)):=\frac{1}{n}\sum_{t=1}^n f_{\s M}(\tilde Z_t(s))$. Since  
$|f_{\s M}(\tilde Z_t(s))-\bar f_{\s M}(\tilde Z(s))|\vee|f_{\s M}(\tilde Z_t(0))- \bar f_{\s M}(\tilde Z(0))|\le M$, we have 
\begin{eqnarray*}
&&\big|\hat\tau_{\s M}^2(s)-\hat\tau_{\s M}^2(0)\big|\\
&&=\ \frac 1n\Big|\sum_{t=1}^n\big(f_{\s M}(\tilde Z_t(s))-\bar f_{\s M}(\tilde Z(s))+f_{\s M}(\tilde Z_t(0))-\bar f_{\s M}(\tilde Z(0))\big)
\big(f_{\s M}(\tilde Z_t(s))-f_{\s M}(\tilde Z_t(0))\big)\Big|\\
&&\le\ \frac{2M B^M_n(s)}{n},
\end{eqnarray*}
and thus 
\begin{eqnarray}\label{supp:eq:tau_diff}
\big|\hat\tau_{\s M}(s)-\hat\tau_{\s M}(0)\big|
=\frac{\big|\hat\tau^2_{\s M}(s)-\hat\tau_{\s M}^2(0)\big|}
{\hat\tau_{\s M}(s)+\hat\tau_{\s M}(0)}
\ \le\ \frac{2M B^M_n(s)}{n\hat\tau_{\s M}(0)}.
\end{eqnarray}
Observe also that $D_{n,k}(x) =\sum_{t=1}^k (x_t -\bar x)=-\sum_{t=k+1}^n (x_t -\bar x)$, from which it follows
\begin{eqnarray*}
|D_{n,k}(x)|\ \le\ \frac{1}{2}\sum_{t=1}^{n}|x_t-\bar x|
\ \le\ \frac{\sqrt n}{2}\Big(\sum_{t=1}^{n}(x_t-\bar x)^2\Big)^{1/2}.
\end{eqnarray*}
Using this, we can see that for all $s\geq 0$,
\begin{eqnarray}\label{supp:eq:Tbdd}
\frac{|D_{n,k}(Y_s)|}{\hat\tau_{\s M}(s)}\ \le\ \frac{n}{2}.
\end{eqnarray}
Combining \eqref{supp:eq:D_diff} - \eqref{supp:eq:Tbdd}, we have
\begin{eqnarray*}
\bigg|\frac{|D_{n,k}(Y_s)|}{\hat\tau_{\s M}(s)}-\frac{|D_{n,k}(Y_0)|}{\hat\tau_{\s M}(0)}\bigg|
&\le&
\frac{\big|\,|D_{n,k}(Y_s)|-|D_{n,k}(Y_0)|\,\big|}{\hat\tau_{\s M}(0)}
+
\frac{|D_{n,k}(Y_s)|}{\hat\tau_{\s M}(s)}\,\frac{\big|\hat\tau_{\s M}(0)-\hat\tau_{\s M}(s)\big|}{\hat\tau_{\s M}(0)}\\[5pt]
&\le&
\frac{1}{\hat\tau_{\s M}(0)}\Big(1+\frac{M}{\hat\tau_{\s M}(0)}\Big)
B_n^M(s),
\end{eqnarray*}
and thus 
\begin{eqnarray*}
\big|T^M_n(s)-T^M_n(0)\big|
&\le&
\max_{1\le k \le n} \frac{1}{\sqrt{n}}\bigg|\frac{|D_{n,k}(Y_s)|}{\hat\tau_{\s M}(s)}-\frac{|D_{n,k}(Y_0)|}{\hat\tau_{\s M}(0)}\bigg|\\[5pt]
&\le&
\frac{1}{\hat\tau_{\s M}(0)}\Big(1+\frac{M}{\hat\tau_{\s M}(0)}\Big)
\frac{B_n^M(s)}{\sqrt{n}}.
\end{eqnarray*}

Next, we deal with the second inequality. 
By \eqref{supp:eq:V_norm}, we obtain
\begin{equation}\label{supp:eq:V_diff}
\Big|
\sqrt{V_{n,k}(Y_s)}
-\sqrt{V_{n,k}(Y_0)}
\Big|
\ \le\
\sqrt{n}B_n^M(s),
\end{equation}
and thus
\begin{equation}\label{supp:eq:V_Ys}
\sqrt{V_{n,k}(Y_s)}
\ \ge\
n \sqrt{\underline V_n^M(0)}
-\sqrt{n} B_n^M(s).
\end{equation}
Therefore, it follows from \eqref{supp:eq:D_diff}, \eqref{supp:eq:V_diff}, and \eqref{supp:eq:V_Ys} that
\begin{eqnarray*}
\bigg|
\frac{|D_{n,k}(Y_s)|}{\sqrt{V_{n,k}(Y_s)}} 
- 
\frac{|D_{n,k}(Y_0)|}{\sqrt{V_{n,k}(Y_0)}} 
\bigg|
&\le&
\frac{
\big|\,|D_{n,k}(Y_s)|-|D_{n,k}(Y_0)|\, \big|
}{\sqrt{V_{n,k}(Y_s)}} 
+
\frac{|D_{n,k}(Y_0)|}{\sqrt{V_{n,k}(Y_0)}} 
\frac{
\big|\sqrt{V_{n,k}(Y_s)}-\sqrt{V_{n,k}(Y_0)}\big|
}{\sqrt{V_{n,k}(Y_s)}}\\[5pt]
&\le&
\Big(1+\sqrt{SN_n^M(0)}\,\Big)
\frac{
B_n^M(s)
}{
n\sqrt{\underline V_n^M(0)}
-\sqrt{n}B_n^M(s)
},
\end{eqnarray*}
which yields the second inequality. 
This completes the proof.

\section{Example in the GARCH(1,1) model}
\label{supp:Sec3}

In this section, we derive the explicit form of $\xi_t$ under AO and IO contamination in the GARCH(1,1) model, and then evaluate $R_n(\xi)$ and $Q_n(\xi)$.

Let $\{X_t(s)\}_{t=1}^n$ be the sample generated from the GARCH(1,1) model 
with true parameter $\T_0=(\omega_0,\alpha_0,\beta_0)$ and contaminated 
by outliers of size $s$ as in \eqref{eq:AOIO}.  Then, the proxy $\tilde\sigma_t^2(\T;s)$ for the conditional variance can be written as
\begin{eqnarray}\label{supp:eq:GARCH}
\tilde\sigma_t^2(\T;s)
=\frac{\omega}{1-\beta}+\alpha\sum_{i=1}^{t-1}\beta^{i-1}X_{t-i}^2(s),\quad t=1,\cdots, n
\end{eqnarray}
(cf. \cite{berkes.et.al:2003}).
To derive the form of $\xi_t$, we assume that
\begin{equation}\label{supp:eq:conv_theta}
   \hat\T_n^\circ(s)\ \rightarrow\ \T_\infty
= (\omega_\infty,\alpha_\infty,\beta_\infty)
\quad\text{as }s\to\infty, 
\end{equation}
where $\alpha_\infty>0$ and $0<\beta_\infty<1$.
By Remark \ref{rmk:AOIO}, it suffices to determine the limits of
$a_{t,A}(s)$ and $a_{t,I}(s)$ at the outlier times.
\subsection{Derivation of $\xi_t$}
\label{supp:subsec:G}
\subsubsection{AO contamination}
Since no outlier occurs before $r_1$, we have $\tilde\sigma_{r_1}^2(\hat\T_n^\circ(s);s)=\tilde\sigma_{r_1}^2(\hat\T_n^\circ(s);0)$, and thus
\[
a_{r_1,A}(s) =
\frac{1}{\tilde\sigma_{r_1}^2(\hat\T_n^\circ(s);s)}
\ \rightarrow\
\frac{1}{\tilde\sigma_{r_1}^2(\T_\infty;0)}\quad \text{as }s\to\infty.
\]
For every $r_j>r_1$, it follows from \eqref{supp:eq:GARCH} that
\[
\tilde\sigma_{r_j}^2(\hat\T_n^\circ(s);s)
\geq
\hat\alpha_n^\circ(s)
\{\hat\beta_n^\circ(s)\}^{r_j-r_1-1}
X_{r_1}^2(s).
\]
Since $\hat\alpha_n^\circ(s)\to\alpha_\infty>0$, $\hat\beta_n^\circ(s)\to\beta_\infty>0$, and  $X^2_{r_1}(s) = (X_{r_1,0} + s\zeta_{r_1}\,\mathrm{sign}(X_{r_1,0}))^2\to \infty$ as $s\to\infty$, $\tilde\sigma_{r_j}^2(\hat\T_n^\circ(s);s)
$ diverges to infinity, and thus we have $a_{r_j,A}(s)\to0$ as $s\to\infty$ for $j=2,\cdots,m$.
Therefore, we obtain
\begin{equation}\label{supp:eq:G_AO}
\xi_t=
\begin{cases}
\displaystyle \frac{\zeta_{r_1}^2}
{\tilde\sigma_{r_1}^2(\T_\infty;0)} &\text{if } t=r_1,\\[15pt]
0  &\text{if } t\neq r_1.
\end{cases}
\end{equation}
Hence, $c_{\s A}$ in \eqref{eq:G_AOIO} is $\zeta_{r_1}^2/\tilde\sigma^2_{r_1}(\theta_\infty;0)$.
\subsubsection{IO contamination}
To deal with the IO case, put $\eta_t(s)=\epsilon_t+s\zeta_t\operatorname{sign}(\epsilon_t)I_t$.
Then, we have
\[
\sigma_{t+1}^2(\T_0;s)
=
\omega_0+(\alpha_0\eta_t^2(s)+\beta_0)\sigma_t^2(\T_0;s).
\]
Note that $\sigma_{r_1}^2(\T_0;s)=\sigma_{r_1}^2(\T_0;0)$ and 
$\eta_{r_1}^2(s)/s^2\to\zeta_{r_1}^2$ as $s\to\infty$. Hence, it follows that
\begin{eqnarray*}
\frac{\sigma_{r_1+1}^2(\T_0;s)}{s^2}
=
\frac{\omega_0}{s^2}+\alpha_0\sigma_{r_1}^2(\T_0;0)\frac{\eta_{r_1}^2(s)}{s^2}
+
\beta_0\frac{\sigma_{r_1}^2(\T_0;0)}{s^2}
\ \rightarrow\
\alpha_0\sigma_{r_1}^2(\T_0;0)\zeta_{r_1}^2\quad \text{as }s\to\infty.
\end{eqnarray*}
Further, since $\eta_t(s)=\epsilon_t$ for $r_1+1 \le t<r_2$, if $\sigma_t^2(\T_0;s)/s^2\to \kappa_{t}$ for some positive finite random variable $\kappa_t$, then we have
\begin{eqnarray*}
\frac{\sigma_{t+1}^2(\T_0;s)}{s^2}
=
\frac{\omega_0}{s^2}
+
(\alpha_0\epsilon_t^2+\beta_0)
\frac{\sigma_t^2(\T_0;s)}{s^2}
\ \rightarrow\
(\alpha_0\epsilon_t^2+\beta_0)\kappa_{t}\quad \text{as }s\to\infty.
\end{eqnarray*}
Therefore, it follows by induction that for $r_1+1 \le t\le r_2$,
\[
\frac{\sigma_t^2(\T_0;s)}{s^2}
\ \rightarrow\ \kappa_t=\alpha_0\sigma_{r_1}^2(\T_0;0)\zeta_{r_1}^2 \prod_{u=r_1+1}^{t-1}(\alpha_0\epsilon_u^2+\beta_0)\quad \text{as }s\to\infty.
\]
Next, using $\sigma_{r_2}^2(\T_0;s)/s^2\to\kappa_{r_2}$ and $\eta_{r_2}^2(s)/s^2\to\zeta_{r_2}^2$ as $s\to\infty$, we have
\begin{eqnarray*}
\frac{\sigma_{r_2+1}^2(\T_0;s)}{s^4}
=
\frac{\omega_0}{s^4}+\alpha_0\frac{\sigma_{r_2}^2(\T_0;s)}{s^2}\frac{\eta_{r_2}^2(s)}{s^2}
+
\beta_0\frac{\sigma_{r_2}^2(\T_0;s)}{s^4}
\ \rightarrow\
\alpha_0\kappa_{r_2}\zeta_{r_2}^2\quad \text{as }s\to\infty.
\end{eqnarray*}
By the same argument above, we have that for $r_2+1 \le t\le r_3$,
\[
\frac{\sigma_t^2(\T_0;s)}{s^4}
\ \rightarrow\ \alpha_0\kappa_{r_2}\zeta_{r_2}^2\prod_{u=r_2+1}^{t-1}(\alpha_0\epsilon_u^2+\beta_0)\quad \text{as }s\to\infty.
\]
Repeating the same argument after each IO, we finally obtain that, for
every fixed $t$,
\begin{equation}\label{supp:eq:kappa}
 \frac{\sigma_t^2(\T_0;s)}{s^{2d_t}}\ 
\rightarrow\ \kappa_t\quad \text{as }s\to\infty,   
\end{equation}
where $d_t=\sum_{j=1}^m\mathbf 1(r_j<t)$ and
\begin{equation*}
\kappa_{t}
=
\begin{cases}
\sigma_t^2(\T_0;0)
&\text{if }  d_t=0,\\[7pt]
\displaystyle
\alpha_0^{d_t}\,\sigma_{r_1}^2(\T_0;0)\,
\prod_{j=1}^{d_t}\zeta_{r_j}^2
\,
\prod_{\substack{r_1<u<t\\u\notin O_n}}
(\alpha_0\epsilon_u^2+\beta_0)
&\text{if } d_t\geq1.
\end{cases}
\end{equation*}

Meanwhile, we can see from \eqref{supp:eq:kappa} that
\begin{eqnarray*}
\frac{X_t^2(s)}{s^{2d_t}}
=
\frac{\sigma_t^2(\T_0;s)}{s^{2d_t}}\epsilon_t^2
\ \rightarrow\
\kappa_t \epsilon_t^2\quad \text{for } t\notin O_n
\end{eqnarray*}
and
\begin{eqnarray*}
\frac{X_t^2(s)}{s^{2d_t+2}}
=
\frac{\sigma_t^2(\T_0;s)}{s^{2d_t}}
\frac{\eta_t^2(s)}{s^2}
\ \rightarrow\ 
\kappa_t\zeta_t^2\quad\text{for } t\in O_n.
\end{eqnarray*}
Recall that $\xi_t=0$ for $t<r_1$. Hence, it suffices to consider $t\ge r_1$. 
Suppose that
$\tilde\sigma_t^2(\hat\T_n^\circ(s);s)/s^{2d_t}
\to\tilde\kappa_t$ as $s\to\infty$ for some positive finite random variable $\tilde \kappa_t$.
Since  $d_{t+1}=d_t$ for $t\notin O_n$ and $d_{t+1}=d_t+1$ for $t\in O_n$, it follows from the preceding limits and \eqref{supp:eq:conv_theta} that
\[
\frac{\tilde\sigma_{t+1}^2(\hat\T_n^\circ(s);s)}
{s^{2d_{t+1}}}
=
\frac{\hat\omega_n^\circ(s)}{s^{2d_{t+1}}}
+\hat\alpha_n^\circ(s)\frac{X_t^2(s)}{s^{2d_{t+1}}}
+\hat\beta_n^\circ(s)
 \frac{\tilde\sigma_t^2(\hat\T_n^\circ(s);s)}{s^{2d_{t+1}}}
\ \rightarrow\
\begin{cases}
\alpha_\infty\kappa_t\epsilon_t^2+\beta_\infty\tilde\kappa_t &\text{if }  t\notin O_n,\\[5pt]
\alpha_\infty\kappa_t\zeta_t^2  &\text{if }  t\in O_n.
\end{cases}
\]
That is, for $t\geq r_1$, $\tilde\kappa_t$  satisfies the following recursion:
\[
\tilde \kappa_{t+1}=
\begin{cases}
\alpha_\infty\kappa_t\epsilon_t^2+\beta_\infty\tilde\kappa_t &\text{if }  t\notin O_n,\\[5pt]
\alpha_\infty\kappa_t\zeta_t^2  &\text{if }  t\in O_n,
\end{cases}
\]
with $\tilde \kappa_{r_1}=\tilde\sigma_{r_1}^2(\T_\infty;0)$.
Iterating this recursion, we can obtain that 
\[
\frac{\tilde \sigma_t^2(\hat\T_n^\circ(s);s)}{s^{2d_t}}
\ \rightarrow\
\tilde \kappa_t
\quad\text{as }s\to\infty,
\]
where
\[
\tilde\kappa_t
=
\begin{cases}
\tilde\sigma_t^2(\T_\infty;0)
&\text{if } d_t=0,\\[7pt]
\displaystyle
\alpha_\infty
\beta_\infty^{t-r_{d_t}-1}
\kappa_{r_{d_t}}\zeta_{r_{d_t}}^2
+
\alpha_\infty
\sum_{u=r_{d_t}+1}^{t-1}
\beta_\infty^{t-u-1}
\kappa_u\epsilon_u^2
&\text{if } d_t\geq1.
\end{cases}
\]
Consequently, since $a_{t,{\s I}}(s)\ \rightarrow\ \kappa_t/\tilde\kappa_t$ as $s\to\infty$, we have by Remark \ref{rmk:AOIO} that
\begin{equation}\label{supp:eq:G_IO}
\xi_t=
\begin{cases}
\displaystyle\frac{\kappa_t}{\tilde\kappa_t} \zeta_t^2 &\text{if } t\in O_n,\\[12pt]
0  &\text{if } t\notin O_n.
\end{cases}
\end{equation}
Hence, $c_{t,{\s I}}$ in \eqref{eq:G_AOIO} is given by $\kappa_t\zeta_t^2/\tilde\kappa_t$ for $t\in\mathcal O_n$.

\subsection{Evaluation of $R_n(\xi)$ and $Q_n(\xi)$}
\label{supp:subsec:Rn}


\subsubsection{AO contamination}
Note from \eqref{supp:eq:G_AO} that $\xi_t$ has a nonzero value $c_{\s A}$ only at $t=r_1$. Then, we have
\begin{equation}\label{sup:eq:D_G}
D_{n,k}(\xi)
=
\begin{cases}
\displaystyle -\frac{k}{n}c_{\s A}
&\text{if }1\leq k<r_1,\\[10pt]
\displaystyle \Big(1-\frac{k}{n}\Big)c_{\s A}
&\text{if } r_1\leq k\leq n,
\end{cases} 
\end{equation}
and thus
\[
\max_{1\leq k\leq n}|D_{n,k}(\xi)|
=
\frac{c_{\s A}}{n}\max\{r_1-1,n-r_1\}.
\]
Since 
\begin{eqnarray*}
v_{\xi}^2
=
\frac1n\Big\{
\Big(c_{\s A}-\frac{c_{\s A}}{n}\Big)^2
+(n-1)\left(\frac{c_{\s A}}{n}\right)^2
\Big\}  
=
\frac{c_{\s A}^2(n-1)}{n^2},
\end{eqnarray*}
we have
\begin{eqnarray*}\label{supp:eq:R_AO}
R_n(\xi)
=
\frac{\max\{r_1-1,n-r_1\}}{\sqrt{n(n-1)}}.
\end{eqnarray*}

We next derive $Q_n(\xi)$. Put $J_\ell:=\sum_{j=1}^{\ell}j^2$ with $J_0=0$.
If $k<r_1$, the first part of $V_{n,k}(\xi)$ is zero, and  thus we have by a direct
calculation that
\[
V_{n,k}(\xi)=\frac{c_{\s A}^2}{(n-k)^2}\big(J_{r_1-k-1}+J_{n-r_1}\big).
\]
For $k\geq r_1$, the second part of $V_{n,k}(\xi)$ is zero, and similarly, we also have
\[
V_{n,k}(\xi)
=
\frac{c_{\s A}^2}{k^2} \big(J_{r_1-1}+J_{k-r_1}\big).
\]
Note that for $1<r_1<n$, $V_{n,k}(\xi)>0$ holds for all $1\le k\le n-1$. In this case, it follows from \eqref{sup:eq:D_G} that

\begin{eqnarray}\label{supp:eq:Q_AO}
Q_n(\xi)
=
\max_{1\leq k\leq n-1}q_{n,k}(r_1),
\end{eqnarray}
where
\[
q_{n,k}(r_1)
=
\begin{cases}
\displaystyle
\frac{k^2(n-k)^2}
{n (J_{r_1-k-1}+J_{n-r_1})}
&\text{if }1\leq k<r_1,\\[15pt]
\displaystyle
\frac{k^2(n-k)^2}
{n(J_{r_1-1}+J_{k-r_1})}
&\text{if }r_1\leq k\leq n-1.
\end{cases}
\]
When $r_1=1$ (resp. $r_1=n$), we have $D_{n,1}(\xi)\neq 0$ and $V_{n,1}(\xi)=0$ (resp. $D_{n,n-1}(\xi)\neq 0$ and $V_{n,n-1}(\xi)=0$), and thus $Q_n(\xi)=\infty$. In this case, it can also be shown from the proof of Proposition 1 that $SN_n(\hat\theta_n^\circ(s))\to\infty$ as $s\to\infty$.


\subsubsection{IO contamination}
 To avoid cumbersome derivations, we consider the simplified case $c_{r_j,{\s I}}=c_{\s I}$ for all $j$. For brevity, we omit some algebraic calculations needed to derive the results in this subsection. 

First, we can see that 
\[
R_n(\xi)=\sqrt{\frac{n}{m(n-m)}}
\max_{1\leq k\leq n}\Big|N_k-\frac{km}{n}\Big|,
\]
where $N_k=\sum_{j=1}^m\mathbf 1(r_j\leq k)$.
Consider the situation in which the IOs are evenly spaced at $r_j=jn/m$, where $n/m$ is assumed to be an integer. Then, we have
\[
R_n(\xi)=\sqrt{\frac{n-m}{mn}}<1.
\]
Hence, when large IOs are distributed nearly evenly  over the sample, the naive CUSUM test may show a tendency to produce a relatively small value. In contrast, suppose that the $m$ IOs are located at consecutive times $r+1,\cdots,r+m$ for some $0\leq r\leq n-m$. Then, we have
\[
R_n(\xi)=\sqrt{\frac{m}{n(n-m)}}\max\{r,n-r-m\}.
\]
Note that $\max\{r,n-r-m\}$ has the minimum
value $(n-m)/2$ at $r=(n-m)/2$, provided that $n-m$ is even,  and has the maximum value $n-m$ at $r=0$ or
$r=n-m$. Hence, if $m/n$ is nearly zero, then $R_n(\xi)$ takes a value close to $\sqrt m/2$  when the outliers
are concentrated around the center of the sample, and close to $\sqrt m$  when they
are clustered near a boundary. Thus, unlike the evenly spaced case, nearly consecutive IOs can inflate  the value of the naive CUSUM statistic as $m$ increases, particularly when they are clustered near a boundary.

Next, to deal with $Q_n(\xi)$, letting
\begin{eqnarray*}
 W_{n,k}
:=
\sum_{t=1}^k
\Big(N_t-\frac{t}{k}N_k\Big)^2
+
\sum_{t=k+1}^n
\Big(
m-N_{t-1}
-\frac{n-t+1}{n-k}(m-N_k)
\Big)^2,
\end{eqnarray*}
we have
\begin{eqnarray}\label{supp:eq:Q_IO}
Q_n(\xi)
=
\max_{1\leq k\leq n-1}
\frac{n\Big(N_k-\dfrac{km}{n}\Big)^2}
{W_{n,k}}.
\end{eqnarray}
For the evenly spaced case above, if $m\geq2$ and $n/m\geq2$, then the maximum in \eqref{supp:eq:Q_IO} is attained at $k=n-1$, and it can be shown that 
\begin{eqnarray*}\label{supp:eq:Q_IO_even}
Q_n(\xi)
=
\frac{6m(n-m)(n-1)}{n^2(m-1)} <12.
\end{eqnarray*}
In the consecutive case where the cluster is located at the center of the sample, the maximum in \eqref{supp:eq:Q_IO} is attained at $k=r$ and $k=r+m$, yielding
\begin{eqnarray}\label{supp:eq:Q_IO_cluster}
Q_n(\xi)
=
\frac{6rm(r+m)}{n(2rm+1)}<3.
\end{eqnarray}
Note that  when $m$ is small relative to $n$, the right side of \eqref{supp:eq:Q_IO_cluster} is close to 3/2. On the other hand, when the cluster is located at the boundary, that is, $r=0$ or $r=n-m$, the denominator in \eqref{supp:eq:Q_IO} is zero for some $k$ while the numerator remains nonzero. Hence, the corresponding ratio is not finite. In sum, one can expect that roughly evenly spaced IOs yield a
relatively small value of $Q_n(\xi)$, whereas for consecutive IOs,
$Q_n(\xi)$ depends strongly on the location of the cluster, taking a small
value when the cluster is near the center and becoming much larger as it
approaches a sample boundary.

\putbib[jun]

\end{bibunit}

\end{document}